\documentclass{aa}  

\usepackage{graphicx}
\usepackage{txfonts}
\usepackage{enumitem}
\usepackage{natbib}
\usepackage{siunitx}
\usepackage{adjustbox}
\usepackage{textcomp}
\usepackage{tablefootnote}
\usepackage{float}
\usepackage{placeins}
\usepackage[flushleft]{threeparttable}
\usepackage{lipsum}
\usepackage{pifont}
\newcommand{\cmark}{\ding{51}}%
\usepackage{dblfloatfix}
\usepackage{makecell}
\usepackage{booktabs}
\usepackage{hyperref}

\hypersetup{
    colorlinks=true,
    linkcolor=black,
    citecolor=blue,
    filecolor=blue,
    urlcolor=blue,
}

\usepackage{lipsum}
\bibpunct{(}{)}{;}{a}{}{,} 

\providecommand{\e}[1]{\ensuremath{\times 10^{#1}}}

\usepackage{enumitem}
\setlist{parsep=0pt,listparindent=\parindent}

\begin{document}

\title{Probing the weak wind phenomenon in Galactic O-type giants}

\author{E. S. G. de Almeida\inst{\ref{inst1},\ref{inst2}} 
 \and W. L. F. Marcolino\inst{\ref{inst2}} 
 \and J.-C. Bouret\inst{\ref{inst3}}
 \and C. B. Pereira\inst{\ref{inst4}}
}

\institute{Université Côte d'Azur, Observatoire de la Côte d'Azur, CNRS, Laboratoire Lagrange, France\\ \email {Elisson.Saldanha@oca.eu}\label{inst1} \and Universidade Federal do Rio de Janeiro, Observatório do Valongo, Rio de Janeiro, Brasil \label{inst2} \and Aix Marseille Univ, CNRS, CNES, LAM, Laboratoire d’Astrophysique de Marseille, Marseille \label{inst3} \and Observatório Nacional/MCTIC, Rio de Janeiro, Brasil \label{inst4}}

\date{}

  \abstract
   {}
   {Analyses of Galactic late O dwarfs (O8-O9.5V stars) raised the ``weak wind problem'': spectroscopic mass-loss rates ($\dot{M}$) are up to two orders of magnitude lower than the theoretical values. We investigated the stellar and wind properties of Galactic late O giants (O8-O9.5III stars). These stars have luminosities $\log(L_\star/L_\odot) \sim 5.2$, which is the critical value (onset of weak winds) proposed in the literature.}
   {We performed a spectroscopic analysis of nine O8-O9.5III stars in the ultraviolet (UV) and optical regions using the model atmosphere code CMFGEN.}
   {Stellar luminosities were adopted using calibrations from the literature. Overall, our model spectral energy distributions agree well with the observed ones considering parallaxes from the latest GAIA data release (DR2). The effective temperature derived from the UV region agrees well with the ones from the optical. As expected, the analysis of the Hertzsprung-Russell (HR) diagram shows that our sample is more evolved than late O dwarfs. From the UV region, we found $\dot{M} \sim 10^{-8}-10^{-9}$ $\mathrm{M_\odot}$ yr\textsuperscript{-1}  overall. This is lower by $\sim 0.9 - 2.3$ dex than predicted values based on the (global) conservation of energy in the wind. The mass-loss rates predicted from first principles, based on the moving reversing layer theory, agree better with our findings, but it fails to match the spectroscopic $\dot{M}$ for the most luminous OB stars. The region of $\log(L_\star/L_\odot) \sim 5.2$ is critical for both sets of predictions in comparison with the spectroscopic mass-loss rates. CMFGEN models with the predicted $\dot{M}$ (the former one) fail to reproduce the UV wind lines for all the stars of our sample. We reproduce the observed H$\alpha$ profiles of four objects with our $\dot{M}$ derived from the UV. Hence, low $\dot{M}$ values (weak winds) are favored to fit the observations (UV + optical), but discrepancies between the UV and H$\alpha$ diagnostics remain for some objects.}
   {Our results indicate weak winds beyond the O8-9.5V class, since the region of $\log(L_\star/L_\odot) \sim 5.2$ is indeed critical to the weak wind phenomenon. Since O8-O9.5III stars are more evolved than O8-9.5V, evolutionary effects do not seem to play a role in the onset of the weak wind phenomenon. These findings support that the $\dot{M}$ (for low luminosity O stars) in use in the majority of modern stellar evolution codes must be severely overestimated up to the end of the H-burning phase. Further investigations must evaluate the consequences of weak winds in terms of physical parameters for massive stars (e.g., angular momentum and CNO surface abundances).}



\maketitle

\hypersetup{
    colorlinks=true,
    linkcolor=blue,
    citecolor=blue,
    filecolor=blue,
    urlcolor=blue,
}


\section{Introduction}

With their high effective temperatures ($\gtrsim 30$ kK) and intense radiation fields ($L_\star \lesssim 10^{6}$ L$_\odot$), massive O-type stars have a huge impact on the interstellar medium through ionizing photons \citep[e.g.,][]{abbott82} and strong line-driven outflows ($\lesssim 10^{-6}$ M$_{\odot}$ yr\textsuperscript{-1}). After they leave the main sequence, O stars evolve to become objects such as red supergiants, luminous blue variables, and Wolf-Rayet stars, depending on the initial mass \citep{meynet11}. Therefore, they are also progenitors of type II-Ib-Ic supernovae, neutron stars, black holes, long gamma-ray bursts \citep[e.g.,][]{gehrels13}, and possibly gravitational wave events at low metallicity \citep{abbott16}.\par

Despite having been studied for several decades, exciting findings regarding O stars have been reported in the literature in recent years. For example, it is still unclear why about 10$\%$ of O stars have been found to possess surface magnetic fields \citep{grunhut17}. Moreover, it has been argued that O stars form almost exclusively in multiple systems \citep{sana14}. As is the case for magnetic fields, the effects of binary interactions (e.g., tides, mass transfer, and mergers) on stellar evolution are complex and constitute a hot topic in the literature \citep[see, e.g.,][]{keszthelyi17,demink14}. The instability inherent in line-driven winds is another critical issue; it induces the far from smooth density and velocity structures --  inhomogeneities -- that present a challenge to hydrodynamics and to implementation in atmosphere codes \citep[e.g.,][]{sundqvist14, sundqvist18}. Such inhomogeneities directly affect the mass-loss rate obtained for these stars, and with likely evolutionary consequences.\par

Another open question that has been called urgent in the massive star community \citep[e.g.,][]{puls09} is the so-called weak wind problem\footnote{We use the term "problem" throughout the text to state the disagreement between spectroscopic and predicted mass-loss rates.}. It is characterized by a large discrepancy between theoretical predictions for the mass-loss rates ($\dot{M}$) of O8-O9V stars -- provided by \citet{vink00, vink01} -- and the results obtained from spectroscopic analyses using atmosphere models \citep[e.g.,][]{martins05_weakwinds, marcolino09}. The "measured" (i.e., derived using atmosphere models) mass-loss rates of late-type O dwarfs (O8-9.5V stars) are up to two orders of magnitude lower than the predicted ones. \par
 
The first results regarding weak winds were presented almost three decades ago, showing a significant difference (approximately a factor of 5-10) between spectroscopic and predicted $\dot{M}$ for the faintest O stars \citep[e.g.,][]{chlebowski91}. However, one of the first quantitative findings about the weak wind problem with modern photosphere-wind unified models came from \citet{bouret03} and \citet{martins04}. These authors investigated O dwarfs in the Small Magellanic Cloud and found weak mass-loss on the order of $10^{-9} - 10^{-8}$ $\mathrm{M_\sun}$ yr\textsuperscript{-1} (in O6-O9.5V stars). An important question raised by these results was the role played by the low metallicity environment (like the Small Magellanic Cloud) and the youth of the sample stars. Later, \citet{martins05_weakwinds} and \citet{marcolino09} presented spectroscopic analyses using optical and ultraviolet data for a total of 17 Galactic O dwarfs and inferred much lower mass-loss rates than predicted for the late-type objects (O8-9V). The discrepancies found reached up to two orders of magnitude. This ruled out an environmental effect due to metallicity. Overall, these results indicated that weak winds concern a particular range of stellar luminosity, corresponding to late O dwarfs, namely $\log(L/\mathrm{L_\sun}) \sim 4.5-5.0$.\par

Interestingly, one of the most recent hydrodynamical results regarding O stars \citep{muijres12} fails to drive winds for low luminosity O stars (O6.5V and later). \citet{muijres12} identified the absence of physical solutions for their hydrodynamical approach as resulting from the lack of \ion{Fe}{V} at the basis of the wind for this spectral range. This is notable since P-Cygni profiles are observed in the UV spectra of these stars. It exposes our lack of knowledge about the force that maintains the wind at this luminosity range. Moreover, such discrepancy between the hydrodynamical\footnote{We note that the mass-loss rates predicted by \citet{vink00} are not rigorously hydrodynamical since the authors do not explicitly solve the wind equation through simulations. Nevertheless, \citet{vink00} provide $\dot{M}$ using a global energy argument, where the conservation of energy is globally satisfied in the wind for solely one value of $\dot{M}$. For more details, see \protect\citet{vink99}.} mass-loss rates of \citet{vink00} and results from atmosphere models is currently a serious question because the most modern evolutionary models use the predicted values of \citet{vink00} during the main sequence phase \citep[e.g.,][]{ekstrom12,meynet15}. Stated differently, the mass-loss rate of the majority of massive stars (low luminosity) at the main sequence is severely overestimated in most evolution models. The mass range affected is around 20-25 $\mathrm{M_{\sun}}$, that is, the mass range of late O dwarfs. The evolutionary consequences of this fact up to the end of the main sequence are as yet unknown.\par

We aim to investigate the weak wind phenomenon among more evolved O stars for the first time. We are particularly interested in late O giants because of their luminosity -- $\log(L_\star/L_\sun) \sim 5.2$ -- which seems to define the outset of the weak wind problem. So far, this problem has only been reported in O8-9.5V stars. We performed a quantitative analysis of nine Galactic late-type O giants (O8-9.5III) using ultraviolet and optical data to derive their main stellar and wind physical parameters. Our analysis increases the number of O8-9.5III stars analyzed in the literature through a quantitative approach in the UV and visible spectral regions. Up to date, \citet{mahy15} is one of the most comprehensive works deriving the stellar and wind parameters of Galactic late O giants, combining UV and visible spectra: five objects in total, with UV data for two out of five stars.\par
  
This paper is organized as follows. In Sect. \ref{sec_observations}, we present the observational data used in our analysis. In Sect. \ref{sec_stellar_atmosphere_code}, we present the code to generate the non-LTE expanding atmosphere models (CMFGEN). The code assumptions and our methodology for the analysis of the UV and optical data are discussed. In Sect. \ref{sec_physical_properties_evolutionary_analysis}, we first present the derived stellar and wind parameters and our fits (Sect. \ref{sec_spectral}). Then, we analyze the energy spectral distribution, the photospheric parameters, and the evolutionary status of our sample in Sect. \ref{sec_stellar}. The wind parameters and the weak wind phenomenon are analyzed in detail in Sect. \ref{sec_wind}. After, we discuss the derivation of $\dot{M}$ from the UV in comparison with the optical region (from H$\alpha$) in Sect. \ref{sec_mass_loss_rates_uv_visible}. In Sect. \ref{sec_degeneracies}, we evaluate the effect of different parameters (e.g., CNO abundances) on the $\dot{M}$ diagnostics. Finally, our conclusions are presented in Sect. \ref{conclusions}.\par


\section{Observations}\label{sec_observations}

\begin{sidewaystable*}
\caption{Star sample and photometric/spectroscopic data.}\label{sample_data} 
\centering
\renewcommand{\arraystretch}{1.2}
\begin{adjustbox}{width=0.85\textwidth}
\begin{tabular}{lccccccccc}
\toprule
\toprule
Star &HD 156292 &HD 24431 &HD 105627 &HD 116852 &HD 153426 &HD 218195 & \makecell{ HD 36861 A \\ ($\lambda$ Orionis A)} &HD 115455 &HD 135591\\

\midrule
Spec. type &O9.7III &O9III &O9III &O8.5II-III((f)) &O8.5III &O8.5III Nstr &O8III  &O8III((f)) &O8IV((f))\\

\midrule
Spec. bin. &SB2 &--- &--- &--- &SB2 &--- &--- &SB2\\

\midrule
Member. &NGC 6322 &\makecell{Sh 2-205 \\ Cam OB1}&--- &--- &Sh 2-2 &Cep OB1 &\makecell{Sh 2-264 \\ Collinder 69 \\ Ori OB1} &\makecell{RCW 75 \\ Cen OB1} &ASCC 79\\

\midrule
Johnson U &7.101 &6.507 &7.270 &8.380 &6.650 &7.657 &2.196 &7.400\,\tablefootmark{a} &4.442 \\

\midrule
Johnson B &7.773 &7.117 &8.182 &8.380 &7.610 &8.650 &3.218 &8.170\,\tablefootmark{a} &5.372 \\

\midrule
Johnson V &7.509 &6.745 &8.140 &8.470 &7.470 &8.357 &3.405 &7.970\,\tablefootmark{a} &5.457 \\

\midrule
Johnson R &--- &6.360\,\tablefootmark{a} &--- &--- &--- &--- &---  &--- &--- \\

\midrule
Johnson I &--- &6.100\,\tablefootmark{a} &--- &--- &--- &--- &---  &--- &--- \\

\midrule
2MASS J &6.944 &5.917 &7.985 &8.720 &7.057 &7.755 &3.735\,\tablefootmark{b}  &7.469 &5.554 \\

\midrule
2MASS H &6.886 &5.826 &8.030 &8.789 &7.056 &7.708 &3.769\,\tablefootmark{b}  &7.442 &5.566 \\

\midrule
2MASS Ks &6.855 &5.839 &8.069 &8.795 &7.027 &7.735 &3.876\,\tablefootmark{b}  &7.454 &5.616 \\

\midrule
$E(B-V)$ &0.52 &0.63 &0.30 &0.17 &0.40 &0.55 &0.07 &0.46 &0.17 \\

\midrule
Dist. (pc) &$1833^{+270}_{-208}$ &$823^{+62}_{-54}$ &$2541^{+287}_{-234}$ & \makecell{ $22726_{-12766}$ \\ \addlinespace[0.1cm] $2857_{-1956} \; c$ } &$2163^{+269}_{-215}$ &$1588^{+276}_{-204}$ &$417^{+10}_{-10}$ &$2266^{+271}_{-219}$ & $836^{+137}_{-103}$ \\

\midrule
SWP \textnumero  &16218 &30166  &20623 &09332  &\makecell{01517 \\ 07827-28 \\ 07828} &26975 &\makecell{46234 \\ 46237 \\ 46241 \\ 46245 \\ 46247}  &16087 &48294 \\

\midrule
SWP date &1982-Jan-31 &1987-Jan-26 &1983-Aug-06 &1980-Jun-20 &\makecell{1978-May-10 \\ 1980-Jan-31 \\ 1980-Jan-31} &1985-Oct-24 &\makecell{1992-Nov-12 \\ 1992-Nov-12 \\ 1992-Nov-12 \\ 1992-Nov-12 \\ 1992-Nov-12} &1982-Jan-21 &1993-Aug-03 \\

\midrule
LWP \textnumero &--- &10024 &16537 &15610 &06841 &06987 &15311 &--- &25759 \\

\midrule
LWP date &--- &1987-Jan-26 &1983-Aug-06 &1989-May-27 &1980-Feb-02 &1985-Oct-25 &1989-Apr-05 &--- &1993-Jun-17 \\

\midrule
Vis. inst. &FEROS &ESPADONS &FEROS &FEROS &FEROS &ESPADONS &NARVAL &FEROS &FEROS \\

\midrule
Vis. date &2016-Mar-23 &2011-Nov-07 &2016-Mar-18 &2016-Mar-18 &2016-Mar-18 &2011-Jul-05 &2007-Sep-21 &2016-Mar-18 &2016-Mar-18 \\

\bottomrule

\end{tabular}
\end{adjustbox}

\tablefoot{Spectral types, spectroscopic binary classification, and memberships are from \citet{sota14}. Photometric data are from \citet{maiz04} with exception to the specific references. Color excesses are calculated considering intrinsic colors calibrated by spectral type from \citet{martins06}. Distances are from \citet{gaia18}. For HD36861, we adopted the distance from the mean Gaia DR2 parallaxes for the components C and D, as in \citet{gordon18}. Distance from \citet{hipparcos07} for HD 116852 is shown too. All distances are obtained from the direct inversion of the measured parallaxes. We list information about the analyzed data in the UV (SWP and LWP number and observation date) and in the visible (instrument name/observation date) regions.\\
\tablefoottext{a}{From \citet{ducati02}.}
\tablefoottext{b}{From \citet{cutri03}.}
\tablefoottext{c}{From \citet{hipparcos07}.}
}

\end{sidewaystable*}

Our sample contains nine Galactic late O giants. They belong to the O8-9.5III spectral types, according to the classification provided by the Galactic O-Star Catalog \citep[GOSC,][]{maiz13}. We present them in Table \ref{sample_data}. Three stars of our sample are double-lined spectroscopic binaries (HD 156292, HD 153426, and HD 115455), and we discuss in Sect. \ref{sec_mass_loss_rates_binary} possible binary effects on the analysis. Two stars of our sample have a classification outside the luminosity class III: HD 116852 (O8.5II-III) and HD 135591 (O8IV). We initially included these objects in the analysis based on an outdated GOSC classification, O9III for HD 116852 and O7.5III for HD 135591 \citep{maiz04}. Nevertheless, HD 116852 and HD 135591 have expected stellar parameters (such as bolometric luminosity and effective temperature) consistent with the rest of our sample of giants. The known population of O8-9.5 giants is close to sixty stars \citep{maiz13}. From an observational point of view, our analyzed spectra have a fairly good morphological homogeneity. We consider our sample representative and it corresponds to  $\sim 25 \%$ of the late O giants with available spectra from the IUE/Short-Wavelength Prime (SWP) instrument.\par

We used high-resolution ($\Delta\lambda \sim 0.2$ {\AA}) data from the International Ultraviolet Explorer (IUE) satellite\footnote{Public data available in the Barbara A. Mikulski Archive for Space Telescopes (MAST): https://archive.stsci.edu/iue/.}. We focused on the $\sim 1200-1975$ {\AA} interval (SWP instrument). This region contains the most important photospheric and wind lines useful to our purposes (e.g., iron forest, \ion{N}{V} $\lambda$1240, \ion{Si}{IV} $\lambda$$\lambda$1394,1403, \ion{C}{IV} $\lambda$$\lambda$1548,1551, and \ion{N}{IV} $\lambda$1718). When available, we co-added different observations for the same target in order to achieve a better signal-to-noise ratio (SNR). In addition, we used IUE/ Long-Wavelength Prime (LWP) data ($\sim 1900-3125$ {\AA}) and UBVJHK photometry in the analysis of the spectral energy distribution of our sample.\par

We also acquired high-resolution optical spectroscopic data for six stars of the sample: HD 156292, HD 105627, HD 116852, HD 153426, HD 115455, and HD 135591. The observations were done with the FEROS spectrograph (resolving power R = 48000) at the ESO/MPG 2.2 m telescope in La Silla (Chile) from 17 to 22 March of 2016. Technical details about FEROS can be found in \citet{kaufer99,kaufer00}. The exposure times varied from 80 to 420 seconds. The spectra cover the interval $\sim 3500-9200$ {\AA}, and the SNR achieved is about 100 for all the stars. Furthermore, we used high-resolution optical data for HD 36861 from the NARVAL spectrograph (R = 75000) at the 2 m Télescope Bernard Lyot on Pic du Midi (France), covering $\sim 3700-6800$ {\AA}. For HD 24431 and HD 218195, we used public data from the Magnetism in Massive Stars (MiMeS) survey \citep{wade16} obtained using the ESPADONS instrument (R = 68000) at the Canada-France-Hawaii Telescope (USA). The optical data of our sample were analyzed in a second step in our methodology, thus allowing us to check the consistency of the results derived from a pure ultraviolet analysis (e.g., effective temperature and mass-loss rate).\par 


\section{Atmosphere models}\label{sec_stellar_atmosphere_code}

\subsection{Code}\label{sec_code}

We used the code CMFGEN \citep{hillier98} to derive the stellar and wind properties of the late-type O giant stars of our sample. It allows us to solve the radiative transfer, statistical and radiative equilibrium equations in a spherically symmetric outflow. It includes, for example, the effects of line blanketing, clumping, and Auger ionization by X-rays, and provides realistic spectra from the UV to the middle infrared.\par

The code requires an initial estimate of the hydrostatic structure. For this purpose, we used a grid of non-LTE plane-parallel models computed with the code TLUSTY \citep{hubeny95}, 
based on the OSTAR2002 grid \citep{lanz03}. The sampling steps of our grid are $\sim$ 500 K in effective temperature and $\sim$ 0.25 dex in surface gravity. When necessary, we interpolated on $T_\mathrm{eff}$ and $\log(g)$. For the wind, we used a standard $\beta$ velocity law, in the form $v(r) = v_\infty\left(1 - \frac{R_{\star}}{r}\right)^{\beta}$, which is smoothly connected to the hydrostatic density structure just above the sonic point.\par 

Initially, we used the following assumptions in the modeling for all the stars of our sample:

\begin{enumerate}[label=(\roman*)]

\item We adopted standard solar abundances \citep{grevesse10} for all the chemical elements. Later, we performed tests concerning the effects of CNO abundance changes on the derivation of the mass-loss rate (Sect. \ref{sec_deg_CNO_mdot}). The atomic species included in each model and their number of energy levels are shown in Table \ref{atomic_species}, together with the total number of computed bound-bound transitions.

\item We assumed $\beta = 1.0$ for the wind velocity structure. Values of $\beta = 0.8-1.0$ are recognized as typical for O stars since they are supported both from spectroscopic modeling \citep[e.g.,][]{bouret13} and hydrodynamical predictions \citep[e.g.,][]{muijres12}. As we will show later (Sect. \ref{sec_betalaw}), lower values for this parameter provide a better fit to the observed \ion{C}{IV} $\lambda$$\lambda$1548,1551 profiles in late O giants.

\item All models include the effects of X-rays (energy interval of $0.1-1.0$ keV) produced in the wind with the canonical value for O-type stars of 
$\log(L_{\textrm{X}}/L_{\textrm{BOL}}) \sim -7.0$ \citep[e.g.,][]{sana06, rauw15}. We adopted the value of $\log(L_{\textrm{X}}/L_{\textrm{BOL}}) = -7.0 \pm 0.1$. In fact, two objects of our sample have observed values for $\log(L_{\textrm{X}}/L_{\textrm{BOL}})$ in the literature: HD 36861 ($\lambda$ Orionis A) and HD 135591. The first has determinations of -6.96 \citep{berghoefer96} and -6.81 \citep{naze09}, while the latter one shows -7.14 \citep{berghoefer96}. 

\begin{table}
\caption{\label{atomic_species} Number of levels, super-levels \tablefootnote{CMFGEN approach for a faster computational treatment. For more details see, for example, \citet{hillier98}.}, and bound-bound transitions for each atomic species included in our basic models.}
\centering
\renewcommand{\arraystretch}{1.1}
\begin{adjustbox}{width=0.4\textwidth}
\begin{tabular}{lcccc}
\toprule
\toprule
Ion & Full levels & Super-levels & b-b transitions \\ 
\midrule

\ion{H}{I} & 30 & 30  & 435\\ 

\ion{He}{I} & 69 & 69 & 905 \\ 
 
\ion{He}{II} & 30 & 30 & 435 \\ 

\ion{C}{III} & 243 & 99 & 5528 \\  

\ion{C}{IV}  & 64 & 64 & 1446 \\  

\ion{N}{III} & 287 & 57 & 6223 \\ 

\ion{N}{IV} & 70 & 44 & 440 \\ 

\ion{N}{V} & 49  & 41 & 519 \\ 

\ion{O}{III} & 104 & 36 & 761 \\ 

\ion{O}{IV} & 64 & 30 & 359\\ 

\ion{O}{V} & 56 & 32 & 314 \\ 

\ion{Mg}{II} & 44 & 36 & 348 \\ 

\ion{Si}{III} & 50  & 50 & 232 \\ 

\ion{Si}{IV} & 66  & 66 & 1090 \\ 

\ion{S}{V} & 144  & 37 & 1673 \\ 

\ion{Fe}{III} & 607  & 65 & 5482 \\ 

\ion{Fe}{IV} & 1000  & 100 & 25241\\ 

\ion{Fe}{V} & 1000  & 139 & 25173 \\ 

\ion{Fe}{VI} &1000  & 59 & 24798 \\ 
\bottomrule

\end{tabular}
\end{adjustbox}

\end{table}

\item Wind clumping was included by default in the models. In CMFGEN, a volume filling factor is used according to the formula $f(r) = f_{\infty} + (1 - f_{\infty})\mathrm{e}^{-\frac{v(r)}{v_{\mathrm{initial}}}}$ (microclumping approximation). The free parameters $v_{\mathrm{initial}}$ and $f_{\infty}$ are the onset velocity of clumping and the filling factor value at $r\to\infty$, respectively. We adopted $v_{\mathrm{initial}}$ = 30 km s$^{-1}$ and $f_{\infty}$ = 0.1 \citep[e.g.,][]{bouret03, martins05_weakwinds}. It is important to note that "clumped models" imply underestimation of the derived mass-loss rates in comparison with "unclumped models" \citep[see, e.g.,][]{martins11}. When needed, we scaled our mass-loss rates by a factor of $1/\sqrt{f_{\infty}}$ ($\sim 3.16$) to compare them with unclumped results from the literature.\par

\item In CMFGEN, a depth-dependent microturbulence velocity is used to compute the emergent spectrum (formal solution). It is parameterized as $\xi_{t}(r) = \xi^{min}_{t} + (\xi^{max}_{t} - \xi^{min}_{t})\frac{v(r)}{v_\infty}$, where $\xi^{min}_{t}$ and $\xi^{max}_{t}$ are the minimum and maximum microturbulence velocities. As in \citet{mahy15}, we fixed $\xi^{min}_{t}$ = 10 km s\textsuperscript{-1} and $\xi^{max}_{t}$ is set to 0.1$v_\infty$. In Sect. \ref{sec_spectral}, we discuss the effect of this assumption particularly on the analysis of the effective temperature.

\end{enumerate}


\begin{figure}[!h]
\centerline{\resizebox{0.42\textwidth}{!}{\input{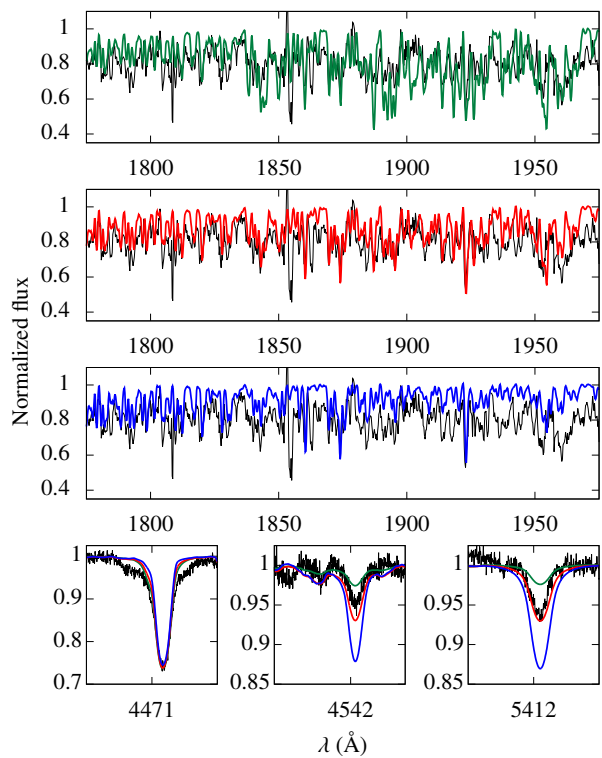}}}
\vspace*{1cm}
\caption{Determination of $T_{\mathrm{eff}}^{\mathrm{UV}}$ for HD 156292 (O9.7III). The IUE spectrum is in black. In the UV, most features are due to \ion{Fe}{IV} ($\sim 1775-1840$ {\AA}) and \ion{Fe}{III} ($\gtrsim 1840$ {\AA}). Models are shown in green ($T_{\mathrm{eff}}^{\mathrm{UV}}$ lower limit), red (final model), and blue ($T_{\mathrm{eff}}^{\mathrm{UV}}$ upper limit). In the bottom panels, these models are shown for $T_{\mathrm{eff}}$ diagnostics in the optical (\ion{He}{I} $\lambda$4471, \ion{He}{II} $\lambda$4542, and \ion{He}{II} $\lambda$5412).}
\label{hd156292_teff_uv_optical}
\end{figure}

\subsection{Diagnostics for the photospheric and wind parameters}\label{sec_photospheric_and_wind_parameters}

First, we performed the analysis of all objects using the UV data set alone. Ultraviolet spectroscopy is suitable to investigate the weak wind phenomenon since the most traditional mass-loss diagnostic in the optical (H$\alpha$ line) is found to be insensitive for the analysis of mass losses lower than $\sim 10^{-8}-10^{-7} \mathrm{M_{\sun}}$ yr\textsuperscript{-1} \citep[e.g.,][]{puls08, martins11}. Thereafter, we extended the analysis to the visible region, comparing with the results derived from the UV  (e.g., for $T_{\mathrm{eff}}$ and $\dot{M}$).\par

We have used typical line diagnostics in the optical and UV for the determination of the photospheric and wind properties --- the effective temperature $T_{\mathrm{eff}}$, surface gravity $\log(g)$, projected rotational velocity $v\sin i$, mass-loss rate ($\dot{M}$), and terminal velocity $v_{\infty}$ \citep[e.g.,][]{martins04,marcolino09,mahy15}. The spectroscopic parameters are obtained through a direct comparison between the synthetic spectrum and the data. The uncertainty for each parameter is inferred in a conservative way: the upper and lower limits for each parameter do not provide an acceptable ``by eye'' fit to the data \citep[as, e.g., in][]{marcolino09}. We provide examples for the derivation of $T_{\mathrm{eff}}$ and of $\dot{M}$ below.\par

In the following we summarize our methodology in more detail:

\begin{enumerate}[label=(\roman*)]

\item The bolometric luminosity $\log(L_\star/\mathrm{L_{\odot}})$ was adopted according to the spectral type of each star of our sample. We used the calibrations for Galactic O stars provided by \citet{martins05_calibration} adopting conservative error bars, namely, $\pm$ 0.2 dex in $\log(L_\star/\mathrm{L_{\odot}})$. Since \citet{martins05_calibration} provide results for luminosity classes V, III, and I, we use mean values among these classes for the stars with intermediate classification, such as HD 116852 (II-III) and HD 135591 (IV). We initially chose this assumption because the astrometric distances derived from the Hipparcos parallaxes are highly uncertain for most of our sample. For O stars, the discrepancy between the Hipparcos distances and the ones predicted from the spectral type is notorious \citep[e.g.,][]{schroder04}. The total amplitude of 0.4 dex in $\log(L_\star/\mathrm{L_{\odot}})$ uncertainty covers a significant deviation in terms of spectral types from O9.5III to O6.5III \citep{martins05_calibration}. In Sect. \ref{sec_sed}, we discuss this assumption for the stellar luminosity, using recent astrometric results from \citet{gaia18}.\par

\begin{figure}[!h]
\centerline{\resizebox{0.42\textwidth}{!}{\input{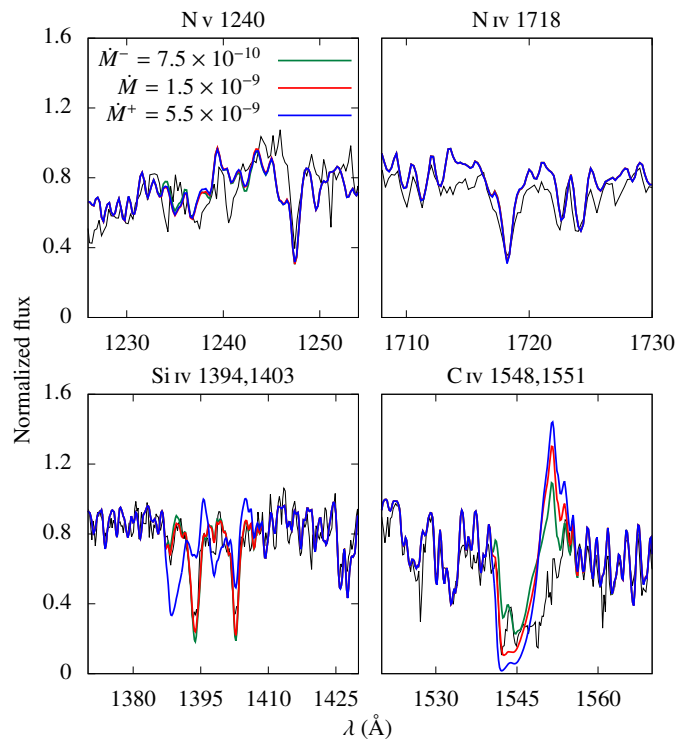}}}
\vspace*{0.9cm}
\caption{Determination of $\dot{M}$ for HD 156292 from the UV lines. The IUE spectrum is in black. We show the models in green (lower limit on $\dot{M}$), red (final model), and blue (upper limit on $\dot{M}$). Mass-loss rate unit is in $\mathrm{M_{\sun}}$ yr\textsuperscript{-1}. We note that \ion{Si}{IV} $\lambda$$\lambda$1394,1403 and \ion{C}{IV} $\lambda$$\lambda$1548,1551 are the most useful lines for the analysis of the wind mass loss in late O giants. All the models have $v_{\infty}$ fixed in 1300 km s\textsuperscript{-1}.}
\label{hd156292_mdot_uv}
\end{figure}

\item The effective temperature $T_{\mathrm{eff}}$ was derived by the ionization balance method through the intensity of the \ion{Fe}{III-IV-V} lines in the ultraviolet, especially \ion{Fe}{III} and \ion{Fe}{V}. In the optical, it was derived by the relative intensity of the \ion{He}{I} and \ion{He}{II} profiles, mainly \ion{He}{I} $\lambda$4471 and \ion{He}{II} $\lambda$4542. Additional lines for consistency checking include \ion{He}{I} $\lambda$4144, \ion{He}{II} $\lambda$4200, \ion{He}{I} $\lambda$4713, \ion{He}{I} $\lambda$4920, \ion{He}{I} $\lambda$5016, and \ion{He}{II} $\lambda$5412. \par
We emphasize that we have inferred independent values for the temperature using the ultraviolet and optical separately. In Fig. \ref{hd156292_teff_uv_optical}, we illustrate the derivation of $T_{\mathrm{eff}}$ for one of the stars of our sample, HD 156292. The first three panels show the models for the determination of the effective temperature (considering the error bars) in the ultraviolet. The bottom panels show the same three models in the optical region for \ion{He}{I} $\lambda$4471, \ion{He}{II} $\lambda$4542, and  \ion{He}{II} $\lambda$5412. It is conspicuous that the same effective temperature fits both the UV and the visible spectra.\par

\item The surface gravity $\log(g)$ was initially adopted according to the spectral type using the calibrations of \citet{martins05_calibration}. After the UV analysis, we checked the fits for the wings of the Balmer lines, mainly H$\gamma$ and H$\beta$, for all our sample.\par

\item The stellar radius $R_{\star}$ of each object follows from the Stefan-Boltzmann equation for a specific $T_{\mathrm{eff}}$ and $\log(L_\star/\mathrm{L_\sun})$,

\begin{equation}
R_\star = \sqrt{\frac{L_\star}{4\pi\sigma T^{4}_{\mathrm{eff}}}}
,\end{equation}
where $\sigma \equiv$ Stefan-Bolztmann constant.\par

The spectroscopic $M_{\star}$ is found from the gravity law

\begin{equation}
M_\star = \frac{gR^{2}_{\star}}{\mathrm{G}}
,\end{equation}
where G is the universal gravity constant.\par

The error bars for $R_{\star}$ are calculated from the uncertainties in $\log(L_\star/\mathrm{L_{\odot}})$ (highest contribution to the error propagation on the radius) and for $M_{\star}$ from the uncertainties in $R_{\star}$ (highest contribution to the error propagation on the mass), being thus underestimated values.

\begin{sidewaystable*}
\caption{\label{resumo_resultados}Summary of the results for the stellar and wind parameters.}
\centering
\renewcommand{\arraystretch}{2.0}
\begin{adjustbox}{width=1.0\textwidth}
\begin{tabular}{lccccccccc}
\toprule
\toprule
Parameter & HD 156292 & HD 24431 & HD 105627 & HD 116852 & HD 153426 & HD 218195 & HD 36861 & HD 115455 & HD 135591\\
\midrule

Spectral type & O9.7III & O9III & O9III & O8.5II-III((f)) & O8.5III & O8.5III Nstr & O8III & O8III((f)) & O8IV((f)) \\

log($L_\star/\mathrm{L_\odot}$) & 5.12 $\pm$ 0.20  & 5.17 $\pm$ 0.20 & 5.17 $\pm$ 0.20 & 5.33 $\pm$ 0.20 & 5.24 $\pm$ 0.20 & 5.24 $\pm$ 0.20 & 5.30 $\pm$ 0.20 & 5.30 $\pm$ 0.20 & 5.10 $\pm$ 0.20 \\

$T_{\mathrm{eff}}^{\mathrm{UV}}$ (kK)\,\tablefootmark{a} & $31.0^{+2.0}_{-3.0}$ & 33.0 $\pm$ 3.0 & $33.0^{+1.5}_{-2.0}$ & $32.5^{+2.0}_{-2.5}$ & 32.0 $\pm$ 2.0 & 33.0 $\pm$ 2.0 & 33.5 $\pm$ 2.5 & $34.0^{+3.5}_{-1.5}$ & 35.0 $\pm$ 2.5 \\ 

$T_{\mathrm{eff}}^{\mathrm{opt}}$ (kK)\,\tablefootmark{b} & 30.0 $\pm$ 2.0 & 32.5 $\pm$ 1.5 & $33.0^{+1.5}_{-2.0}$ & $33.0^{+2.0}_{-2.5}$ & 35.0 $\pm$ 1.0 & 35.0 $\pm$ 1.5 & $35.0^{+2.0}_{-1.5}$ & $34.0^{+2.0}_{-1.5}$ & 36.0 $\pm$ 1.5 \\ 

$\log(g)$ & 3.50 $\pm$ 0.10 & 3.75 $\pm$ 0.10 & 3.50 $\pm$ 0.10 & 3.50 $\pm$ 0.10 & 3.55 $\pm$ 0.10 & 3.55 $\pm$ 0.10 & 3.60 $\pm$ 0.10 & 3.57 $\pm$ 0.10 & 3.75 $\pm$ 0.10 \\ 

$R_\star$ ($\mathrm{R_\odot}$) & $13.0^{+3.7}_{-2.3}$ & $11.9^{+3.5}_{-2.2}$ & $11.9^{+3.5}_{-2.2}$ & $14.7^{+4.3}_{-2.7}$ & $13.7^{+4.0}_{-2.5}$ & $12.9^{+3.7}_{-2.4}$  & $13.4^{+3.9}_{-2.5}$  & $13.0^{+3.8}_{-2.4}$ & $9.7^{+2.8}_{-1.8}$ \\

$M_\star$ ($\mathrm{M_\odot}$) & $19.6^{+11.1}_{-7.0}$ & $28.9^{+16.8}_{-10.6}$ & $16.3^{+9.5}_{-6.0}$ & $25.0^{+14.5}_{-9.1}$ & $24.3^{+14.1}_{-8.9}$ & $21.5^{+12.5}_{-7.9}$  & $26.1^{+15.1}_{-9.5}$ & $22.9^{+13.3}_{-8.4}$ & $19.4^{+11.2}_{-7.1}$ \\ 

$v\sin i$ (km s\textsuperscript{-1}) & 100 & 80 & 160 & 120 & 110 & 80 & 75 & 70 & 80\\

$v_\infty$ (km s\textsuperscript{-1}) & $1300 \pm 200$ & $2300 \pm 300$ & $2100 \pm 300$ & $2100 \pm 300$ & $2400 \pm 300$ & $2000 \pm 200$ & $2000 \pm 200$ & $2300 \pm 300$ & $2100 \pm 300$ \\

\midrule

$\dot{M}_{\mathrm{derived}}$ ($\mathrm{M_\odot}$ yr\textsuperscript{-1})\,\tablefootmark{c} & $1.50^{+4.00}_{-0.75}\e{-9}$ & $2.5^{+7.5}_{-1.5}\e{-9}$ & $4.0^{+16.0}_{-2.5}\e{-9}$ & $6.0^{+19.0}_{-4.5}\e{-8}$ & $4.5^{+10.5}_{-2.5}\e{-9}$ & $1.00^{+2.50}_{-0.75}\e{-8}$ & $2.5^{+3.5}_{-2.2}\e{-8}$ & $5.0^{+10.0}_{-3.0}\e{-9}$ & $2.00^{+6.00}_{-1.85}\e{-8}$ \\

$\dot{M}_{\mathrm{unclumped}}$ ($\mathrm{M_\odot}$ yr\textsuperscript{-1})\,\tablefootmark{d} & $4.8^{+12.6}_{-2.5}\e{-9}$ & $7.9^{+23.7}_{-4.7}\e{-9}$ & $1.3^{+5.0}_{-0.8}\e{-8}$ & $1.90^{+6.00}_{-1.43}\e{-7}$ & $1.4^{+3.4}_{-0.8}\e{-8}$ & $3.2^{+7.8}_{-2.4}\e{-8}$ & $7.9^{+11.1}_{-7.0}\e{-8}$ & $1.6^{+3.2}_{-1.0}\e{-8}$ & $6.30^{+18.70}_{-5.83}\e{-8}$ \\

$\dot{M}_{\mathrm{Vink}}$ ($\mathrm{M_\odot}$ yr\textsuperscript{-1})\,\tablefootmark{e} & $2.6^{+0.5}_{-0.3}\e{-7}$ & $5.4^{+0.5}_{-0.3}\e{-7}$ & $5.4^{+0.8}_{-0.7}\e{-7}$ & $6.6^{+1.1}_{-1.0}\e{-7}$ & $4.2^{+0.7}_{-0.6}\e{-7}$ & $5.4^{+0.8}_{-0.8}\e{-7}$ & $5.8^{+1.0}_{-0.7}\e{-7}$ & $7.1^{+1.2}_{-0.9}\e{-7}$ & $3.5^{+0.5}_{-0.5}\e{-7}$ \\ 

$\dot{M}_{\mathrm{Lucy}}$ ($\mathrm{M_\odot}$ yr\textsuperscript{-1})\,\tablefootmark{f} & $2.1^{+1.4}_{-0.7}\e{-8}$ & $1.2^{+0.8}_{-0.4}\e{-8}$ & $7.1^{+4.9}_{-2.4}\e{-8}$ & $9.8^{+6.5}_{-3.3}\e{-8}$ & $4.2^{+2.9}_{-1.4}\e{-8}$ & $5.9^{+3.9}_{-2.0}\e{-8}$ & $5.0^{+3.3}_{-1.7}\e{-8}$ & $6.5^{+4.3}_{-2.2}\e{-8}$ & $1.3^{+0.8}_{-0.5}\e{-8}$ \\ 

\midrule

$\log(D_{mom}^{\mathrm{unclumped}})$\,\tablefootmark{g}  & $26.15^{+0.56}_{-0.48}$ & $26.60^{+0.60}_{-0.40}$ & $26.74^{+0.72}_{-0.40}$ & $27.98^{+0.62}_{-0.60}$ & $26.90^{+0.52}_{-0.35}$ & $27.16^{+0.52}_{-0.61}$ & $27.56^{+0.38}_{-0.92}$ & $26.92^{+0.47}_{-0.40}$ & $27.41^{+0.60}_{-1.12}$ \\ 

$\log(D_{mom}^{\mathrm{Vink}})$\,\tablefootmark{e} & $28.03^{+0.49}_{-0.49}$ & $28.12^{+0.49}_{-0.49}$ & $28.12^{+0.49}_{-0.49}$ & $28.32^{+0.49}_{-0.49}$ & $28.25^{+0.49}_{-0.49}$ & $28.25^{+0.49}_{-0.49}$ & $28.36^{+0.49}_{-0.49}$ & $28.36^{+0.49}_{-0.49}$ & $28.10^{+0.49}_{-0.49}$ \\ 

\bottomrule

\end{tabular}
\end{adjustbox}
\tablefoot{
Solar units and $g$ unit are in the cgs system. Modified wind momenta ($D_{mom}$) are calculated using the mass-loss rate and terminal velocity in the cgs system and the radius in solar unit. \\
\tablefoottext{a}{Determined from the UV region (\ion{Fe}{III-IV-V} lines).}
\tablefoottext{b}{Determined from the optical region (\ion{He}{I-II} lines).} 
\tablefoottext{c}{Mass-loss rate using $f_{\infty}$ = 0.1.}
\tablefoottext{d}{$\dot{M}_{\mathrm{unclumped}}$ = $\frac{\dot{M}_{\mathrm{derived}}}{\sqrt{f_{\infty} = 0.1}}$ $\sim 3.16 \, \dot{M}_{\mathrm{derived}}$.}
\tablefoottext{e}{Calculated from \citet{vink00}.}
\tablefoottext{f}{Calculated from \citet{lucy10_grid}.}
\tablefoottext{g}{$D_{mom}^{\mathrm{unclumped}} = v_{\infty}\sqrt{R_\star}\dot{M}_{\mathrm{unclumped}}$.}
}

\end{sidewaystable*}

\begin{figure*}
\centerline{\resizebox{1.0\textwidth}{!}{\input{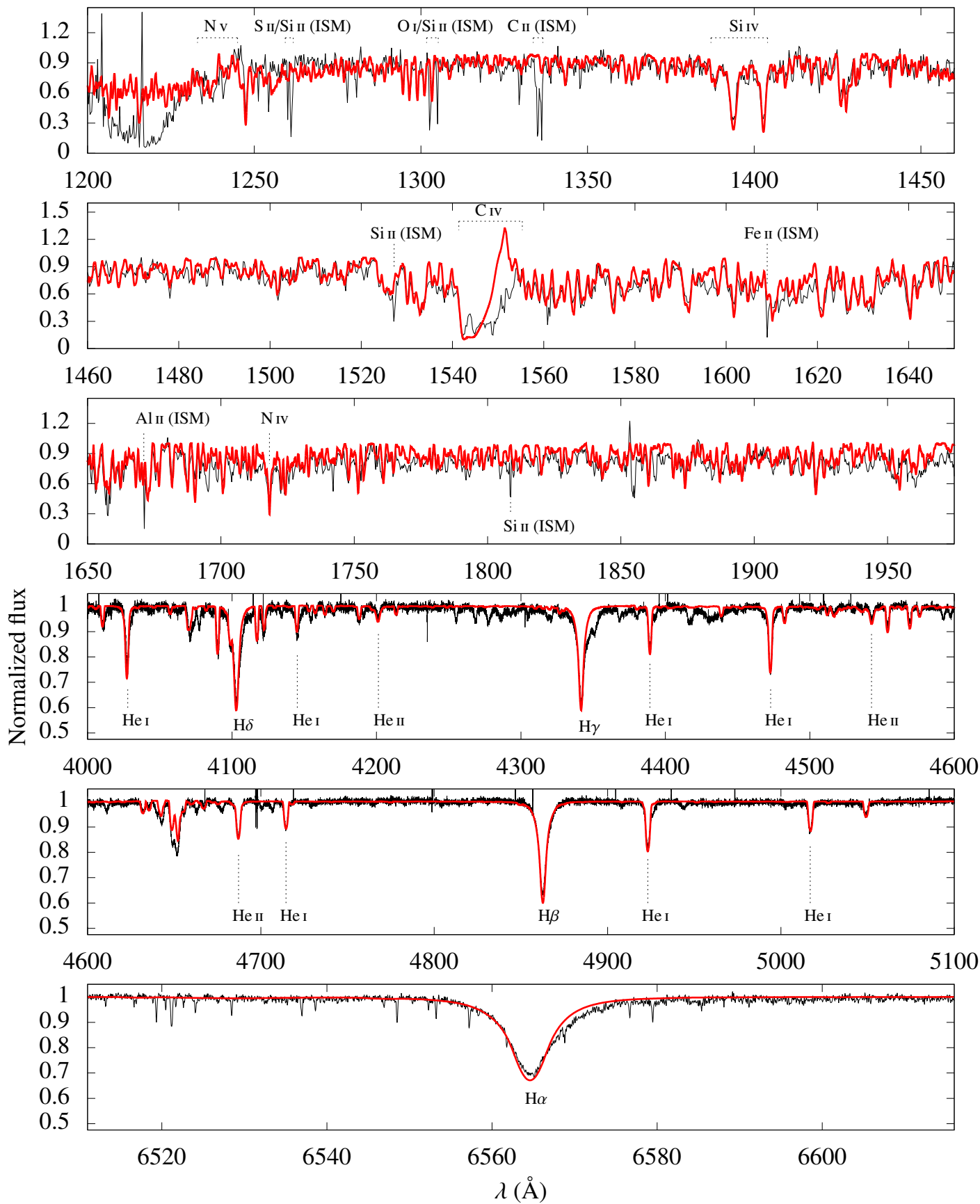}}}
\vspace*{1cm}
\caption{Modeling (red) of HD 156292 (O9.7III) in the UV and optical. The IUE and FEROS data are shown in black. The effective temperature is derived from fitting the UV region (see $T_{\mathrm{eff}}^{\mathrm{UV}}$ in Table \ref{resumo_resultados}). We obtain a good fit to both the UV and optical observed spectra. In addition to the stellar and wind diagnostics, we list some interstellar (ISM) lines. In this case, H$\alpha$ is well reproduced with our $\dot{M}$ derived from the UV. Discussion can be found in the text. }
\label{hd156292_uv_plus_optical}
\end{figure*}

\item The projected rotational velocity $v\sin i$  was initially adopted from \citet{howarth97}, and modified when needed in order to provide a better fitting to the observed broadening. We analyzed the broadening of UV \ion{Fe}{III-IV-V} transitions, as well as of weak metal lines and \ion{He}{I} transitions in the optical. We stress here that macroturbulence is not accounted for in our models. We are aware that the inclusion of macroturbulence must provide a better overall fit to the optical data, but it does not have a significant impact on the wind parameters. Thus, our values of $v\sin i$, in fact, express the total line broadening and they must be seen as upper limits.\par

\item The wind terminal velocity $v_{\infty}$ is derived from fitting the blueward extension (formed up to $v_{\infty}$+$\xi^{max}_{t}$) of the absorption component of the \ion{C}{IV} $\lambda$$\lambda$1548,1551 profile. Overall, we are able to provide a very reasonable fit to the observed blueward extension of \ion{C}{IV} $\lambda$$\lambda$1548,1551 with our adopted value of $\xi^{max}_{t}$ = 0.1$v_{\infty}$.\par

\item The mass-loss rate $\dot{M}$ was determined by fitting the intensity of the ultraviolet P-Cygni profiles  \ion{Si}{IV} $\lambda$$\lambda$1394,1403 and \ion{C}{IV} $\lambda$$\lambda$1548,1551. The H$\alpha$ profile was also used to infer $\dot{M}$, allowing us to compare with the values derived from the UV. In Fig. \ref{hd156292_mdot_uv}, we illustrate the determination of the wind mass loss of HD 156292 from the UV lines. The model parameters are fixed except the mass-loss rate. \ion{The lines N}{V} $\lambda$1240 and \ion{N}{IV} $\lambda$1718 are much less sensitive to the variation in $\dot{M}$ than the lines due to \ion{Si}{IV} and \ion{C}{IV}. Nevertheless, they provide at most constraints on the mass-loss rate. For example, models with $\dot{M} \sim 10^{-7}$ $\mathrm{M_{\sun}}$ yr\textsuperscript{-1} provide stronger nitrogen lines than the observed ones in our sample. For HD 156292, the modeling provided by our lower limit on $\dot{M}$ is quite close to our final model. Such uncertainty is due to the discrete absorption components in the observed \ion{C}{IV} $\lambda$$\lambda$1548,1551 of HD 156292, which are not included in our modeling. In any case, it would imply an overestimated $\dot{M}$ and thus provides a proper comparison with the theoretical values for this star.\par

\end{enumerate}


\section{Results}
\label{sec_physical_properties_evolutionary_analysis}

We present the stellar and wind parameters derived for our sample in Table \ref{resumo_resultados}. Effective temperature determined through the analysis of \ion{Fe}{III-IV-V} (ultraviolet) is denoted as $T_{\mathrm{eff}}^{\mathrm{UV}}$, while the values obtained by \ion{He}{I-II} (optical) are denoted by $T_{\mathrm{eff}}^{\mathrm{opt}}$. For a proper comparison with the theoretical values, we list our unclumped mass-loss rates ($\dot{M}_{\mathrm{unclumped}}$). Unclumped modified wind momenta $\log(D_{mom}^{\mathrm{unclumped}})$ are calculated using $\dot{M}_{\mathrm{unclumped}}$.\par

In Table \ref{resumo_resultados}, we denote $\dot{M}_{\mathrm{derived}}$ as our mass-loss rate derived with the inclusion of clumping (adopted value of $f_{\infty}$ = 0.1), while $\dot{M}_{\mathrm{unclumped}}$ is calculated from $\dot{M}_{\mathrm{derived}}/\sqrt{f_{\infty} = 0.1} \sim 3.16 \, \dot{M}_{\mathrm{derived}}$. In the rest of this paper, we will keep referring to the clumped values as $\dot{M}_{\mathrm{derived}}$ and to the unclumped ones as $\dot{M}_{\mathrm{unclumped}}$.\par

The mass-loss rate $\dot{M}_{\mathrm{Vink}}$ is the theoretical rate from the mass-loss recipe of \citet{vink00}. It was calculated considering $T_{\mathrm{eff}}^{\mathrm{UV}}$, the derived $M_\star$, and adopting the ratio $v_\infty$/$v_\mathrm{esc}$ = 2.6. Accordingly, we provide values  for $\log(D_{mom}^{\mathrm{Vink}})$ that were calculated from log($L_\star/\mathrm{L_\odot}$). We find that our mass-loss rates ($\dot{M}_{\mathrm{derived}}$) are systematically lower than the predictions of \citet{vink00} by $\sim 0.9-2.2$ dex. The discrepancy is reduced to $\sim 0.4-1.7$ dex considering the unclumped values for the mass-loss rate ($\dot{M}_{\mathrm{unclumped}}$).\par

We also present in Table \ref{resumo_resultados} the mass-loss rates predicted by the hydrodynamical approach of \citet{lucy10_grid}, namely, the moving reversing layer theory \citep{lucy70}. We performed bivariate linear interpolation in the model grid provided by Lucy and we computed mass fluxes $J$ for the sample taking $\log(g)$ and $T_{\mathrm{eff}}^{\mathrm{UV}}$ into account. Values for $\dot{M}_{\mathrm{Lucy}}$ were then obtained from our values for the stellar radii \citep[see Eq. 3 in][]{lucy10}. We see a significant reduction in the discrepancy regarding the predicted mass-loss rates from \citet{lucy10_grid}. The values for $\dot{M}_{\mathrm{derived}}$ are lower\footnote{However, note that $\dot{M}_{\mathrm{derived}}$ is higher than $\dot{M}_{\mathrm{Lucy}}$ by $\sim$ 0.1 dex for HD 135591.} by $\sim 0.2-1.5$ dex. However, we note that $\dot{M}_{\mathrm{Lucy}}$ is overestimated (up to $\sim$ 1.0 dex) in comparison with $\dot{M}_{\mathrm{unclumped}}$ for most of our sample. More details will be discussed later in the paper. \par

\begin{figure}
\centerline{\resizebox{0.42\textwidth}{!}{\input{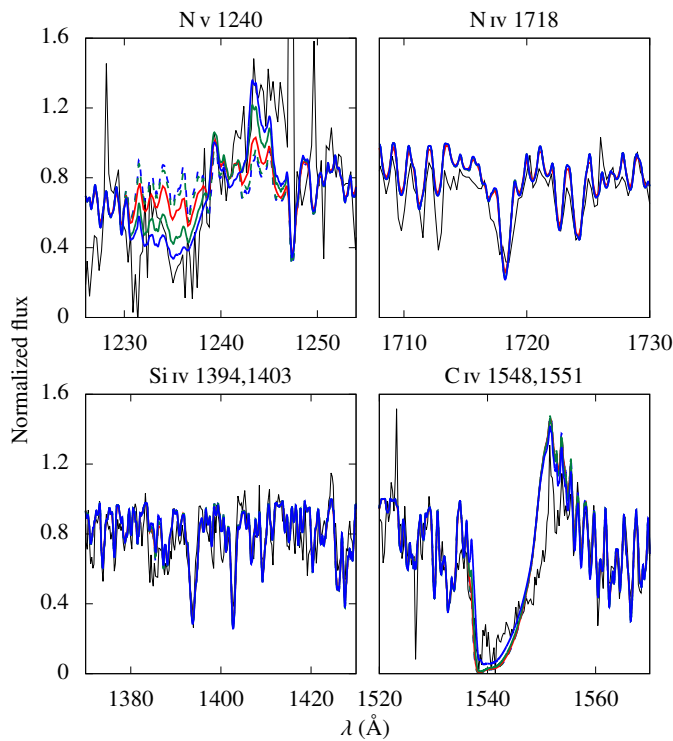}}}
\vspace*{0.8cm}
\caption{Models with different X-ray fluxes compared to the IUE spectrum (black line) of HD 218195. All the other parameters are fixed. These models have the following $\log(L_{\textrm{X}}/L_{\textrm{BOL}})$: -7.96 (blue dashed), -7.49 (green dashed), -7.00 (solid red), -6.49 (solid green), and -6.00 (solid blue). Our final model for HD 218195 is shown in red line (typical X-ray luminosity for O stars). We note how the modeling of \ion{N}{V} $\lambda$1240 is sensitive to the inclusion of X-Rays, while \ion{Si}{IV} $\lambda$$\lambda$1394,1403 and \ion{C}{IV} $\lambda$$\lambda$1548,1551 are almost unchanged.}
\label{hd218195_xray_uv_lines}
\end{figure}

\subsection{Spectral modelling}
\label{sec_spectral}

\begin{figure*}
\centerline{\resizebox{0.85\textwidth}{!}{\input{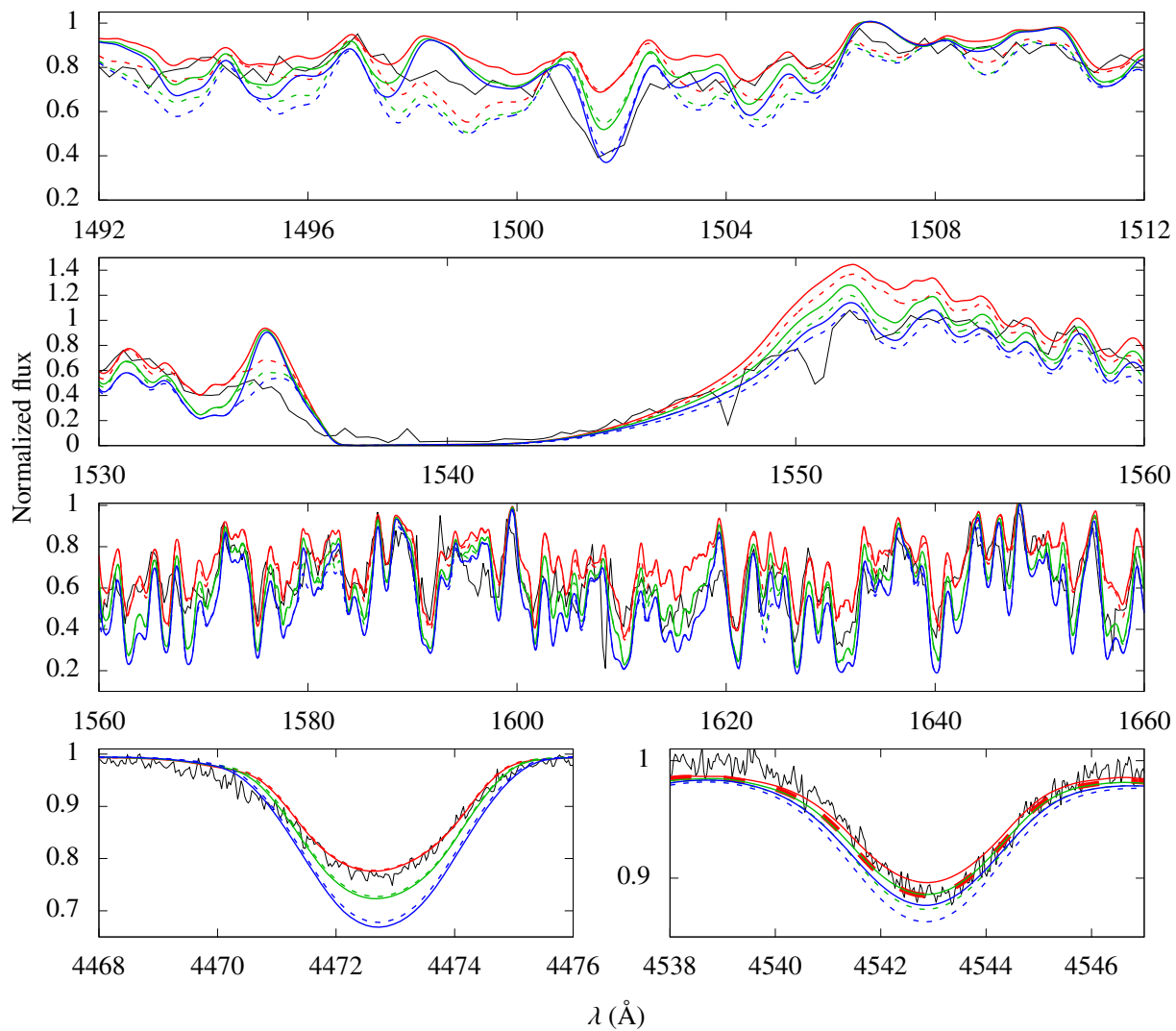}}}
\vspace*{1cm}
\caption{Effect of $\xi^{min}_{t}$ in the $T_{\mathrm{eff}}$ analysis (HD 116852). Models with $\xi^{min}_{t}$ = 10, 20, and 30 km s\textsuperscript{-1} are shown respectively in red, green, and blue. The IUE and FEROS data are shown in black. Models with a fuller account of atomic species are shown by dashed lines. The model with $\xi^{min}_{t}$ = 10 km s\textsuperscript{-1} and a fuller account of species (dashed red line) is shown with thicker lines in \ion{He}{II} $\lambda$4542 since it is overlapped with the basic model $\xi^{min}_{t}$ = 20 km s\textsuperscript{-1} (solid green). \ion{Line S}{V} $\lambda$1502 is reproduced by our model with the highest photospheric microturbulence velocity. The emission component of \ion{C}{IV} $\lambda$$\lambda$1548,1551 is also better reproduced with a higher $\xi^{min}_{t}$ up to 20-30 km s\textsuperscript{-1}. However, $T_{\mathrm{eff}}$ diagnostics are misfitted in this case. It is not possible to fit simultaneously \ion{He}{I} $\lambda$4471 and \ion{He}{II} $\lambda$4542 considering $\xi^{min}_{t}$ = 20-30 km s\textsuperscript{-1}. Our conclusions are unchanged regardless of the improved atomic in the modeling.} 
\label{hd116852_microturb_uv_optical}
\end{figure*}

We present the fits to the UV and optical spectra of each object of our sample in Appendices \ref{appendix_uv} and \ref{appendix_optical}, respectively. All the fits presented in the appendices use the UV mass-loss rate and $T_{\mathrm{eff}}^{\mathrm{UV}}$ (see Table \ref{resumo_resultados}). In the rest of this paper, we also only present models with the effective temperature derived from the UV region. This approach is followed in this paper, since we extensively used our final models with $T_{\mathrm{eff}}^{\mathrm{UV}}$ in the analysis of degeneracy tests for the $\dot{M}$ derivation. Our principal tests (such as for $T_{\mathrm{eff}}$) will be discussed in detail in Sect. \ref{sec_degeneracies}. Here, as an example, we present the final model for HD 156292 in Fig. \ref{hd156292_uv_plus_optical}. Overall, we achieve a very reasonable fit to the UV and optical data simultaneously. Additional details and observed discrepancies are discussed below.\par

Despite our efforts, a perfect fit to the observed spectra is elusive. In the UV (see Figs. \ref{final_model_uv_hd156292} to \ref{final_model_uv_hd135591}, Appendix \ref{appendix_uv}), the spectrum below $\sim 1240$ {\AA} is affected by geocoronal emission and severe interstellar \ion{H}{I} absorption that is not taken into account here. We also note the presence of interstellar lines created by low ionized metals, neglected in our models. From the atlas of \citet{dean85}, the most common ones found in the IUE (SWP) spectra are: \ion{S}{II} $\lambda$1259, \ion{Si}{II} $\lambda$1260, \ion{O}{I} $\lambda$1302, \ion{Si}{II} $\lambda$1304, \ion{C}{II} $\lambda$1334, \ion{C}{II} $\lambda$1306, \ion{Si}{II} $\lambda$1527, \ion{Fe}{II} $\lambda$1608, \ion{C}{I} $\lambda$1656, \ion{C}{I} $\lambda$1657, 
\ion{C}{I} $\lambda$1658, \ion{Al}{II} $\lambda$1657, \ion{Si}{II} $\lambda$1808, \ion{Al}{III} $\lambda$1855, and \ion{Al}{III} $\lambda$1863. Several of them can be identified in our stars.\par

The \ion{N}{V} $\lambda$1240 wind profile is not reproduced in some stars of our sample (e.g., see Fig. \ref{final_model_uv_hd218195} for HD 218195). However, this transition is known to be very sensitive to the X-ray luminosity from the wind and to the N/H abundance. In Fig. \ref{hd218195_xray_uv_lines}, we show the behavior of the UV wind lines due to the variation in X-ray luminosity ($\pm 1.0$ dex) in the modeling of HD 218195. We found that our mass-loss diagnostics (\ion{Si}{IV} $\lambda$$\lambda$1394,1403 and \ion{C}{IV} $\lambda$$\lambda$1548,1551) are not strongly affected by such variation in X-ray in the parameter space of O8-9.5III. Thus, our results for $\dot{M}$ are unlikely biased by X-ray effects. On the other hand, it is clear that \ion{N}{V} $\lambda$1240 is much more affected. Therefore, we did not consider it for the mass-loss rate determination. 

\begin{table*}[b]
\caption{Comparison between our adopted ISM extinction parameters ($R_V$ = 3.1) to the total to selective extinction ratio derived by \citet{wegner03} for stars in common with our sample.}\label{extinction} 
\centering
\renewcommand{\arraystretch}{1.8}
\begin{adjustbox}{angle=0, width=0.8\textwidth}
\begin{tabular}{lcccccc}
\toprule
\toprule
Star &HD 24431 &HD 105627 &HD 153426 &\makecell{ HD 36861 A \\ ($\lambda$ Orionis A)}  &HD 115455 &HD 135591\\

\midrule

$E(B-V)$\,\tablefootmark{a} &0.63 &0.30 &0.40 &0.07 &0.46 &0.17\\

\midrule

$E(B-V)$\,\tablefootmark{b} &0.65 &0.31 &0.43 &0.09 &0.47 &0.22\\

$R_V$\,\tablefootmark{b} & 3.46 $\pm$ 0.06  & 3.24 $\pm$ 0.16  & 3.19 $\pm$ 0.16  & 2.46 $\pm$ 0.60 & 3.29 $\pm$ 0.13  & 3.57 $\pm$ 0.18\\

\bottomrule

\end{tabular}
\end{adjustbox}

\tablefoot{\tablefoottext{a}{Adopted in this work (see Table \ref{sample_data}).}
\tablefoottext{b}{From \citet{wegner03}.}}

\end{table*}

Overall, our synthetic profiles of \ion{S}{V} $\lambda$1502 (in absorption) are stronger than the observations. Since this line is very sensitive to the microturbulence velocity at the photosphere, we tested different set of values for $\xi^{min}_{t}$ from our assumption of 10 km s\textsuperscript{-1} to 30 km s\textsuperscript{-1}.  For example, we show in Fig. \ref{hd116852_microturb_uv_optical} our model for HD 116852 computed with $\xi^{min}_{t}$ = 10, 20, and 30 km s\textsuperscript{-1}. It is necessary to increase $\xi^{min}_{t}$ from 10 km s\textsuperscript{-1} up to 30 km s\textsuperscript{-1} to reproduce the observed \ion{S}{V} $\lambda$1502. On the other hand, $T_{\mathrm{eff}}$ diagnostic lines in the UV and in the visible are misfitted considering a microturbulence velocity higher than 10 km s\textsuperscript{-1}. Thus, it is not possible to obtain a consistent fit simultaneously to the \ion{Fe}{III-IV-V} lines and to the \ion{He}{I-II} lines with this suggested higher $\xi^{min}_{t}$.\par

In addition to our basic model (atomic species shown in Table \ref{atomic_species}), we also compare in Fig. \ref{hd116852_microturb_uv_optical} models computed with the inclusion of the following species in order to test possible effects due to line blanketing: \ion{C}{II}, \ion{N}{II}, \ion{O}{II}, \ion{Ne}{II}, \ion{Ne}{III}, \ion{Ne}{IV}, \ion{Ne}{V}, \ion{P}{IV}, \ion{P}{V}, \ion{S}{III}, \ion{S}{IV}, \ion{Ar}{III}, \ion{Ar}{IV}, \ion{Ar}{V}, \ion{Ar}{VI}, \ion{Cr}{IV}, \ion{Cr}{V}, \ion{Cr}{VI}, \ion{Ni}{III}, \ion{Ni}{IV}, \ion{Ni}{V}, and \ion{Ni}{VI}. Our results regarding the modeling of \ion{S}{V} $\lambda$1502 are unchanged. Still from Fig. \ref{hd116852_microturb_uv_optical}, one sees that the \ion{C}{IV} $\lambda$$\lambda$1548,1551 profile from our final model (solid red line) has an emission component stronger than observed. In advance of the discussion, this issue is systematic in our sample. We are not able to reproduce the observed emission component by just considering our models with a fuller account of species, we need a higher $\xi^{min}_{t}$ up to 20-30 km s\textsuperscript{-1} to better reproduce the observed emission. As discussed above, despite being able to fit the \ion{S}{V} $\lambda$1502 line, such high photospheric microturbulence prevents a self-consistent analysis of the effective temperature both from the UV and the visible for our sample. Therefore, we present our results with the default value of $\xi^{min}_{t}$ = 10 km s\textsuperscript{-1}.\par

We point out that \citet{holgado18} provide limits on the photospheric microturbulence from optical spectroscopic analysis to four stars of our sample: HD 24431 ($\xi^{min}_{t}$ > 18 km s\textsuperscript{-1}), HD 218195 ($\xi^{min}_{t}$ > 18 km s\textsuperscript{-1}), HD 36861 ($\xi^{min}_{t}$ > 11 km s\textsuperscript{-1}), and HD 135591 ($\xi^{min}_{t}$ < 8 km s\textsuperscript{-1}). From Figs. \ref{final_model_uv_hd24431}, \ref{final_model_uv_hd218195}, \ref{final_model_uv_hd36861}, and \ref{final_model_uv_hd135591}, our final models for HD 24431 and HD 218195 (high $\xi^{min}_{t}$) overestimate the observed emission component of \ion{C}{IV} $\lambda$$\lambda$1548,1551 practically as much as in the cases of HD 36861 and HD 135591 (low $\xi^{min}_{t}$). Thus, even considering these estimations for the microturbulence, we are not able to explain our systematic overestimation of the emission component in \ion{C}{IV} $\lambda$$\lambda$1548,1551 by just regarding $\xi^{min}_{t}$. This issue concerning \ion{C}{IV} $\lambda$$\lambda$1548,1551 will be discussed in terms of the wind velocity in Sect \ref{sec_betalaw}.\par

In the optical, it is conspicuous that our models do not reproduce the features of \ion{C}{III} $\lambda 4647-4650-4651$ (see Figs. \ref{final_model_optical_hd156292} to \ref{final_model_optical_hd135591}, Appendix \ref{appendix_optical}). For HD 105627, HD 116852, and HD 115455, they are barely produced by our models. In contrast, the final models for HD 36861 and HD 135591 show these profiles in emission, while the data reveal them in absorption. For HD 156292 and HD 153426, the synthetic lines are in absorption but weaker than observed. We note, however, that these lines are quite sensitive to radiative transfer details in the extreme UV -- such as the lack of robust atomic data for these transitions -- as already pointed out by \citet{martins12}. Recent efforts on a better carbon atomic model, using the code FASTWIND \citep{puls05}, were presented by \citet{carneiro18}. Thus, despite being sensitive to $\dot{M}$, \ion{C}{III} $\lambda 4647-4650-4651$ must not be used as diagnostics for this parameter.\par

\subsection{Stellar properties}
\label{sec_stellar}

\subsubsection{Spectral energy distribution}
\label{sec_sed}

\begin{sidewaysfigure*}
\centering
\centerline{\resizebox{0.850\textwidth}{!}{\includegraphics[angle=0]{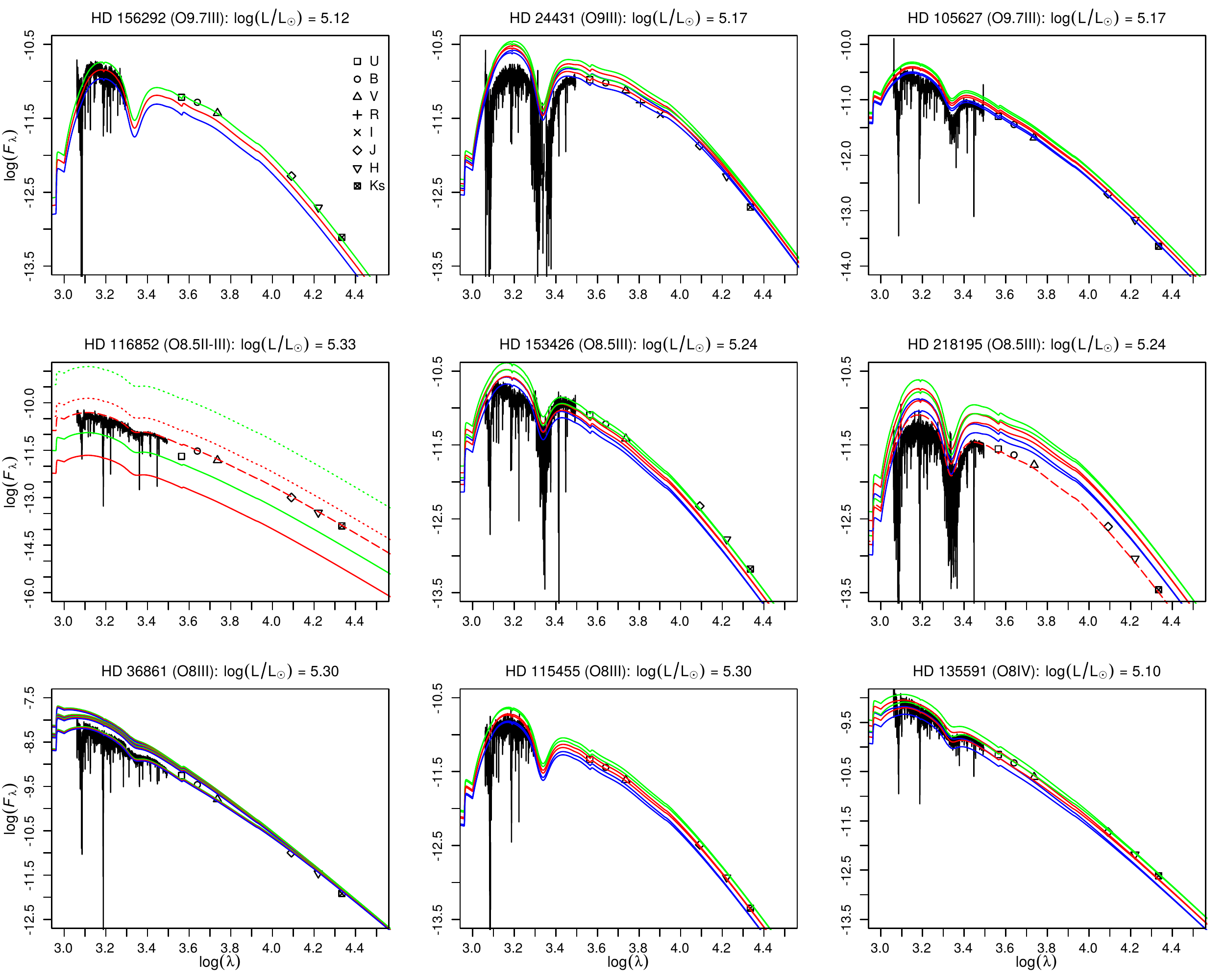}}}
\caption{Model SEDs (color lines) compared to the observed ones (black). The IUE/SWP+LWP and photometric data are listed in Table \ref{sample_data}. Flux unit is in erg cm\textsuperscript{-2} s\textsuperscript{-1} {\AA} \textsuperscript{-1} and wavelength is in {\AA}. Model SEDs in solid lines are computed with distances from GAIA DR2: $1/(\pi+\sigma_{\pi})$ (green), $1/\pi$ (red), $1/(\pi-\sigma_{\pi})$ (blue). Model SEDs taking into account Hipparcos distances are shown in dotted lines (HD 116852). For the stars listed in Table \ref{extinction}, we show two set of models with different values of $E(B-V)$ and $R_{\mathrm{V}}$ from our assumption and from \citet{wegner03}. For HD 36861, there are shown four sets of models with different extinction parameters, including one with  $E(B-V)$ and $R_{\mathrm{V}}$ from \citet{wegner03}. For HD 218195, we compare two sets of models with different values of $E(B-V)$. For HD 116852 and HD 218195, SED models considering the distances of 4.8 kpc and 2.5 kpc are shown in red dashed line. See text for discussion.}
\label{sed_gaia_all_sample}
\end{sidewaysfigure*}

The spectral energy distribution (SED) for all the stars of our sample are presented in Fig. \ref{sed_gaia_all_sample}. We included the effect of interstellar medium (ISM) extinction in the synthetic SEDs using the reddening law from \citet{cardelli89} with $R_{V}$ = 3.1. The color excess $E(B-V)$ (Table \ref{sample_data}) was assumed according to the calibrated intrinsic colors $(B-V)_{0}$ from \citet{martins06}. We compare the data with our synthetic SEDs scaled to take into account the GAIA DR2 parallaxes \citep{gaia16, gaia18}: with $1/(\pi+\sigma_{\pi})$ in solid green, $1/\pi$ in solid red, and $1/(\pi-\sigma_{\pi})$ in solid blue. Synthetic SEDs taking into account Hipparcos parallaxes \citep{hipparcos07} are shown in dotted lines for HD 116852. For HD 36861 ($\lambda$ Ori A), the distance from the GAIA DR2 parallaxes is $271^{+47}_{-35}$ pc. As in \citet{gordon18}, we adopted the distance of $417 \pm 10$ pc from the mean of the parallaxes for components C and D, since the GAIA DR2 parallaxes for HD 36861 have large error bars. Furthermore, different methods in the literature provide a distance estimation for this star up to $\sim$ 400 pc \citep[e.g.,][]{hipparcos07, maiz08, mayne08, maiz18}.\par

From Fig. \ref{sed_gaia_all_sample}, we verify that our models provide a very reasonable fit to the observed SEDs overall (e.g, for HD 156292). Again, these luminosity values (Table \ref{resumo_resultados}) are adopted given the spectral type using the calibrations of \citet{martins05_calibration}. Log($L_\star/\mathrm{L_\odot}$) is fixed here for each star, thus we are not taking the error bar in log($L_\star/\mathrm{L_\odot}$) into account in this analysis. We tested possible effects on the SED fit due to our adoptions on the color excess (Table \ref{sample_data}) and on the total to selective extinction ratio ($R_{V}$ = 3.1). In Table \ref{extinction}, we compare this assumption on $R_{\mathrm{V}}$ with the values derived from \citet{wegner03} since our sample has six objects in common with this work: HD 24431, HD 105627, HD 153426, HD 36861, HD 115455, and HD 135591. The color excess $E(B-V)$ in \citet{wegner03} is adopted considering intrinsic colors from \citet{wegner94}. There are no large discrepancies between these literature results and the adopted value of $R_{\mathrm{V}}$ in our analysis. One of the highest discrepancies is found for HD 36861 ($R_{V} \sim$ 2.5), but with a large error bar compatible with $R_{\mathrm{V}} \sim$ 3.1. For these six stars, we present two sets of model SEDs in Fig \ref{sed_gaia_all_sample}: one with our adopted values for the extinction parameters and another one with the parameters (without the error bars) from \citet{wegner03}. For HD 36861, we show four different sets of models, including the one with the extinction parameters from \citet{wegner03}, as discussed below. Both sets of extinction parameters provide very reasonable fits to the observed SEDs, in particular to the shape of the 2200 {\AA} bump for the targets with IUE/LWP data. Thus, the analysis of the stellar luminosity is unlikely biased by our adoption of $R_{V}$ = 3.1. Despite individual departures from this value, other studies in the literature support that $R_{V} \sim$ 3.1 is a reasonable assumption for galactic O-type stars \citep[e.g.,][]{majaess16}.

The highest discrepancy in Fig. \ref{sed_gaia_all_sample} is seen for HD 116852: we underestimate the data in $\sim$ 1.5 dex (solid red line). Taking distances from \citet{hipparcos07} into account, our model overestimates the observations in $\sim$ 0.5 dex (dashed red line). From both \citet{gaia18} and \citet{hipparcos07}, the parallax $\pi$ has the same order of magnitude of $\sigma_{\pi}$. There is no model shown with distance $1/(\pi-\sigma_{\pi})$ in both cases due to negative parallax values. We stress that the direct inversion of the GAIA DR2 parallax is a reasonable distance estimator for stars with $\sigma_{\pi}/\pi \lesssim 0.2$ \citep{bailer18}. Eight out of nine stars of our sample have $\sigma_{\pi}/\pi \lesssim 0.2$ from the GAIA DR2 release. HD 116852 is the only exception with a high ratio $\sigma_{\pi}/\pi \sim 1.3$. Therefore, this discrepancy for HD 116852 is more likely due to an unreliable distance estimation, using the direct inversion of $\pi$, than due to our adopted luminosity of log($L_\star/\mathrm{L_\odot}$) = 5.33 for this star. Still from Fig. \ref{sed_gaia_all_sample}, the distance needed to fit the SED is $\sim$ 4.8 kpc with log($L_\star/\mathrm{L_\odot}$) = 5.33 (dashed red line). This result is in agreement with the spectroscopic distance of 4.8 kpc derived by \citet{sembach94} for HD 116852. The closest astrometric result to this distance is provided by the lower limit on $\pi$ from \citet{hipparcos97}, giving an upper limit on the distance of $\sim$ 3.6 kpc.\par

In the case of HD 36861 and HD 218195, the difference between our model and the observations is stronger in the UV than in the near-infrared. For HD 218195, it reaches up to $\sim$ 1.0 dex in the UV continuum. For example, we show in Fig. \ref{sed_gaia_all_sample} two sets of models (solid lines) for HD 218195 computed with the same distance and with slightly different values of $E(B-V)$: 0.55 and 0.60. The latter corresponds to the selective extinction adopted by \citet{patriarchi01} for this star, using intrinsic colors from \citet{wegner94}. Considering this color excess different from our assumption, we are able to reproduce better the SED shape in the continuum UV and to diminish the discrepancy in the UBV-bands. In addition, we are able to improve significantly our fit (red dashed line) taking into account the spectroscopic distance of $\sim$ 2.5 kpc found by \citet{maiz18} for HD 218195. This distance is somewhat larger than the value from GAIA DR2 parallaxes ($\sim$ 1.6 kpc). In this case, we use the extinction parameters from \citet{maiz18} for this star ($E(B-V)=0.54$ and $R_{V}$=3.2), but they are very close to our adopted values. For HD 36861, we show four sets of SED models with different extinction parameters: our adopted values, derived from \citet{wegner03}, from \citet{gordon18}, and from \citet{maiz18}. In this case, our SED models encompass the observed one by just considering different values for $E(B-V)$ and $R_{V}$. An analysis of ISM reddening is beyond the scope of this paper, nevertheless we point out that uncertainties in our adopted values for $E(B-V)$ can explain certain differences between our models and the observed SED.

Therefore, despite uncertainties regarding the implementation of ISM reddening in the models and the distance estimations, we conclude that the luminosities provided in \citet{martins05_calibration} are in fair agreement with the observations of O8-9.5III stars. Considering our adopted $E(B-V)$, it is necessary to decrease $\log (L_\star/\mathrm{L_\odot})$ in $\sim 0.5$ dex for HD 218195. This reduction in luminosity would place this star in the late O dwarfs' loci in the HR diagram. However, no evidence supports such uncertainty in the spectral classification. Nevertheless, a lower $\log (L_\star/\mathrm{L_\odot})$ implies downward revision of $\dot{M}$ for HD 218195 to re-fit the observed \ion{Si}{IV} $\lambda$$\lambda$1394,1403. In this case, our inferred mass-loss rates for this star are at most overestimated due to the adopted luminosity.\par

\subsubsection{Photospheric parameters}
\label{sec_photospheric}

\begin{figure}
\centering
\centerline{\resizebox{0.5\textwidth}{!}{\includegraphics[angle=90]{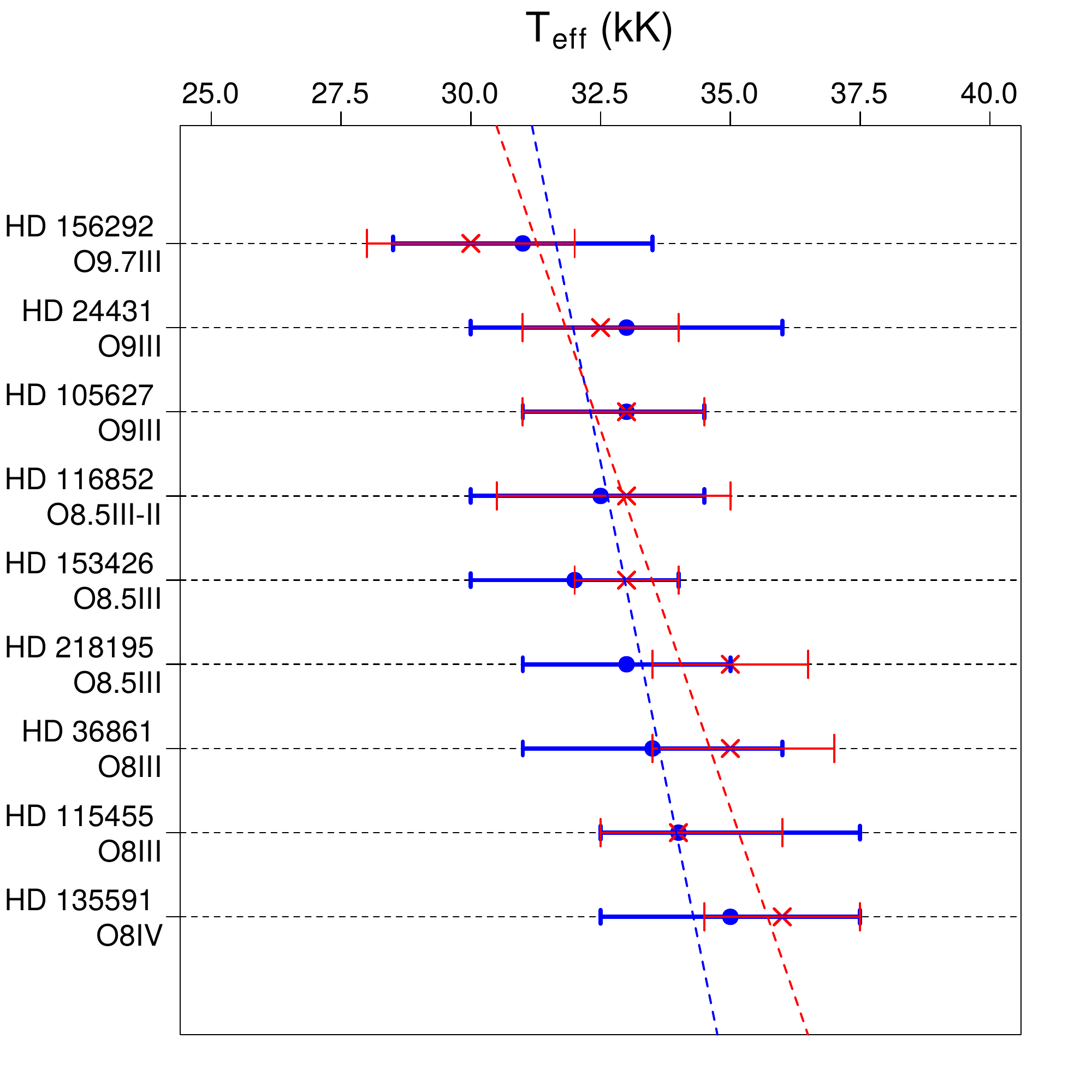}}}
\caption{Comparison between effective temperatures obtained from the UV versus optical for all our sample. The stars are ordered from the later to the earlier types. The effective temperature derived from the UV and the optical regions are shown, respectively, by circles (blue) and crosses (red). We show weighted least squares fits to the UV $T_{\mathrm{eff}}$ (dashed blue line) and to the optical $T_{\mathrm{eff}}$ (dashed red line). We note the good agreement between them and the trend of higher $T_{\mathrm{eff}}$ towards earlier spectral types.}
\label{diagrama_teff_uv_optico}
\end{figure}

In the following, we analyze the UV and optical effective temperatures inferred for all objects of our sample in Fig. \ref{diagrama_teff_uv_optico}. We find good agreement between the effective temperatures derived from the iron forest lines in the ultraviolet and from the helium lines in the visible region. The highest discrepancy (2000K) is seen for HD 218195. However, even in this case, the ultraviolet and optical results are consistent within the error bars. The expected trend of higher temperatures towards earlier spectral classes (from O9.7III to O8IV) is confirmed: for a better visualization, we provide two linear regressions in Fig. \ref{diagrama_teff_uv_optico} to the UV and optical $T_{\mathrm{eff}}$ in function of the spectral type. We find only two objects (HD 156292 and HD 24431) with $T_{\mathrm{eff}}^{\mathrm{opt}}$ lower than $T_{\mathrm{eff}}^{\mathrm{UV}}$. Others results in the literature find good agreement for $T_{\mathrm{eff}}$ derived from the UV and the optical spectra using the code CMFGEN \citep[e.g.,][]{hillier03,martins05_weakwinds}. Therefore, our results confirm the viability of the determination of the effective temperature for O giants solely through the ultraviolet, despite its relatively high error bars.\par

We compare our photospheric parameters with the ones found by \citet{martins_herve15}\footnote{\citet{martins_herve15} do not provide results for the wind parameters as their focus is on surface abundances for O stars.} as our sample shows four objects in common with them: HD 24431, HD 153426, HD 218195, and HD 36861. We verify a good agreement for the effective temperature. These authors derived the following values for $T_{\mathrm{eff}}$, respectively: 33500 K, 34000 K, 34000 K, and 35000 K. Our values ($T_{\mathrm{eff}}^{\mathrm{opt}}$) differ in 1000 K for all these stars. Such differences are inside our error bars on $T_{\mathrm{eff}}^{\mathrm{opt}}$ and it is also the typical uncertainty from \citet{martins_herve15}.\par

For $\log(g)$, we derived the same value for HD 24431, but overall our values are lower (up to 0.25 dex) than the ones found in \citet{martins_herve15}. Here, the lowest discrepancy is 0.15 dex for HD 36861 \citep[$\log(g)$ = 3.75 from][]{martins_herve15} and the highest one is 0.25 dex for HD 218195 \citep[$\log(g)$ = 3.80 from][]{martins_herve15}. This discrepancy for HD 218195 is explained considering our different values between $T_{\mathrm{eff}}^{\mathrm{UV}}$ (33000 K) and $T_{\mathrm{eff}}^{\mathrm{opt}}$ (35000 K). From our tests using $T_{\mathrm{eff}}^{\mathrm{opt}}$, it is necessary to increase $\log(g)$ up to $\sim$ 3.8 to re-fit the wings of the Balmer lines. We are aware that the effective temperature derived from the UV lines is less precise than the ones derived from the optical analysis. Nonetheless, as discussed above, these independent determinations of $T_{\mathrm{eff}}$ are in overall good agreement, attesting that our measured $T_{\mathrm{eff}}$ from the UV are reliable. Thus, such discrepancies must not impact the derivation of the mass-loss rate for the stars of our sample.\par

Regarding $v\sin i$, our values are systematically larger in comparison with \citet{martins_herve15}. These discrepancies are expected, as we do not include macroturbulence in the modeling and these authors include it. In any case, we stress that the effective temperature has the highest potential of affecting our mass-loss analysis.\par

\subsubsection{HR diagram}
\label{sec_HR_diagram}

\begin{figure}
\centerline{\resizebox{0.5\textwidth}{!}{\includegraphics{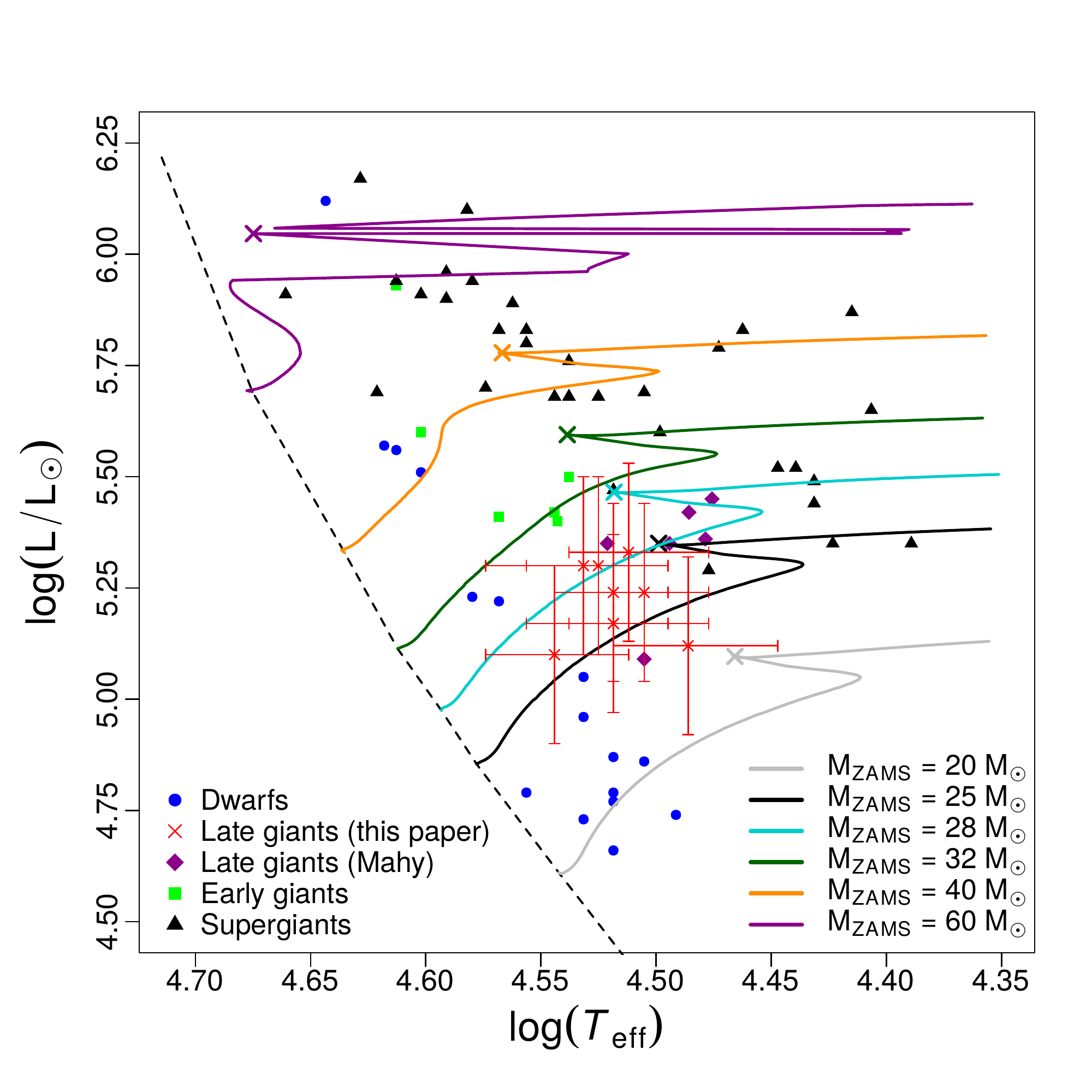}}}
\caption{Evolutionary tracks (in color lines) for samples of O dwarfs, giants, and supergiants. For each model, the central H exhaustion is indicated by crosses. The zero-age line is in dashed black. Stars are grouped by different symbols and colors. Our sample (red crosses) has initial masses ($M_{\textrm{ZAMS}}$) ranging around 25-28 $\mathrm{M_\odot}$. Late O giants are closer to the end of the main sequence phase than the dwarfs.}
\label{HR_diagram_tracks}
\end{figure}

\begin{figure}
\centerline{\resizebox{0.5\textwidth}{!}{\includegraphics{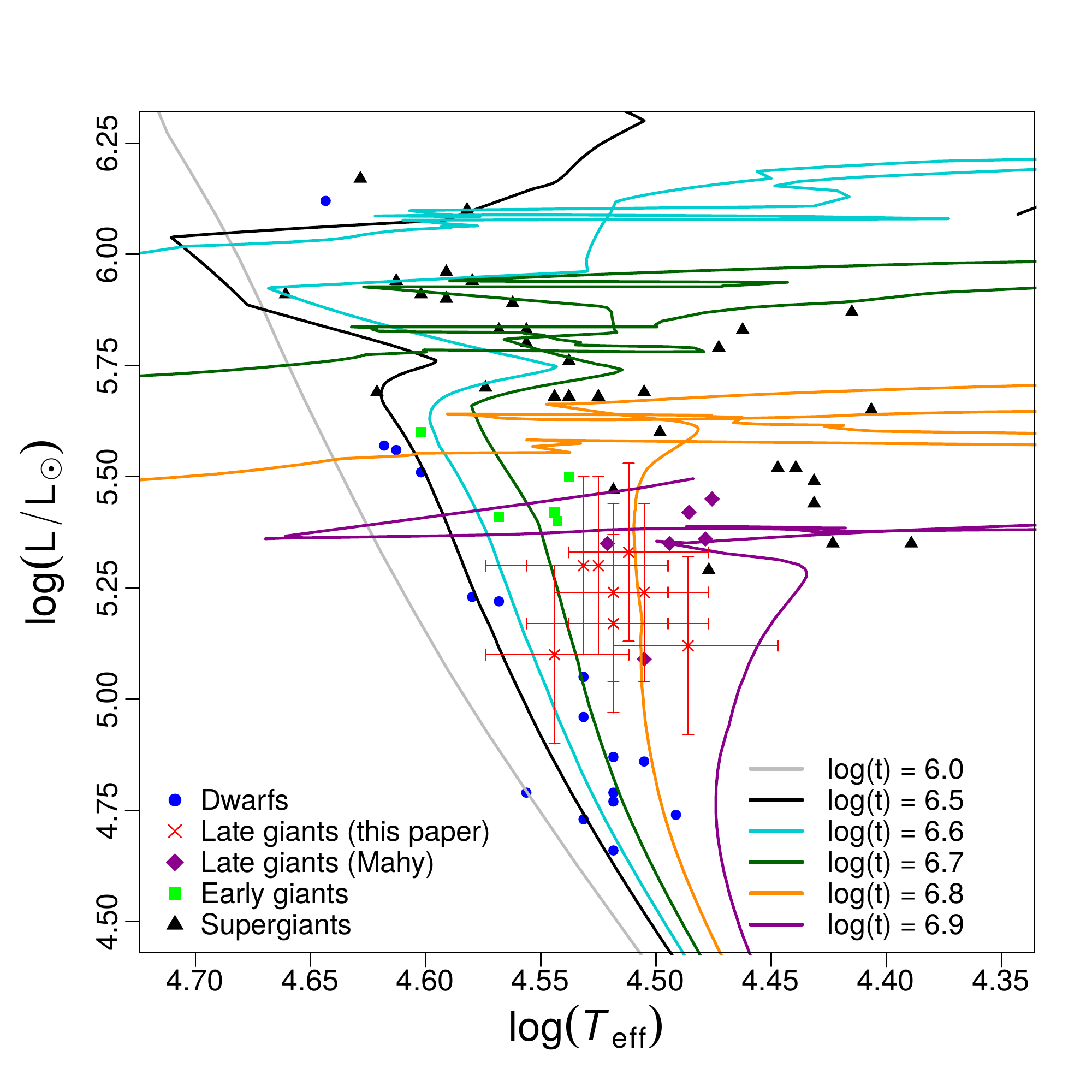}}}
\caption{Same as in Fig. \ref{HR_diagram_tracks}, but showing the isochrones. The bulk of the late O giants show ages ranging around $10^{6.7}-10^{6.8}$ yr.}
\label{HR_diagram_isochrones}
\end{figure}

After deriving the stellar and wind parameters, we analyzed our sample in the HR diagram along with results from the literature for different classes of O-type stars. We used evolutionary tracks and isochrones from \citet{ekstrom12}. The tracks were computed considering solar metallicity (Z = 0.014) and $v_{\mathrm{initial}}$/$v_{\mathrm{critical}}$ = 0.4 \citep{ekstrom12}.\par

We present the results in Figs. \ref{HR_diagram_tracks} and \ref{HR_diagram_isochrones}. We show evolutionary tracks for the initial masses ($M_{\textrm{ZAMS}}$) of 20, 25, 28, 32, 40, and 60 $\mathrm{M_\odot}$, as well as isochrones for the ages ($t$) of $10^{6.0}$, $10^{6.5}$, $10^{6.6}$, $10^{6.7}$, $10^{6.8}$, $10^{6.9}$ yr. Results concerning dwarfs (O3.5-9.5V) are from \citet{martins05_weakwinds} and \citet{marcolino09}. The OB supergiants (O3-9.7I and B0-0.5I) are from \citet{repolust04}, \citet{mokiem05}, \citet{crowther06}, and \citet{bouret12}. The early-type giants (O5-7.5III) are from \citet{repolust04} and \citet{mokiem05}. Additionally, results for late O giants (six stars O8-9.5III in total, excluding giants earlier than O7) from \citet{mahy15} are shown too. There are no objects in common between \citet{mahy15} and our sample.\par

\begin{figure*}
\centerline{\resizebox{0.85\textwidth}{!}{\input{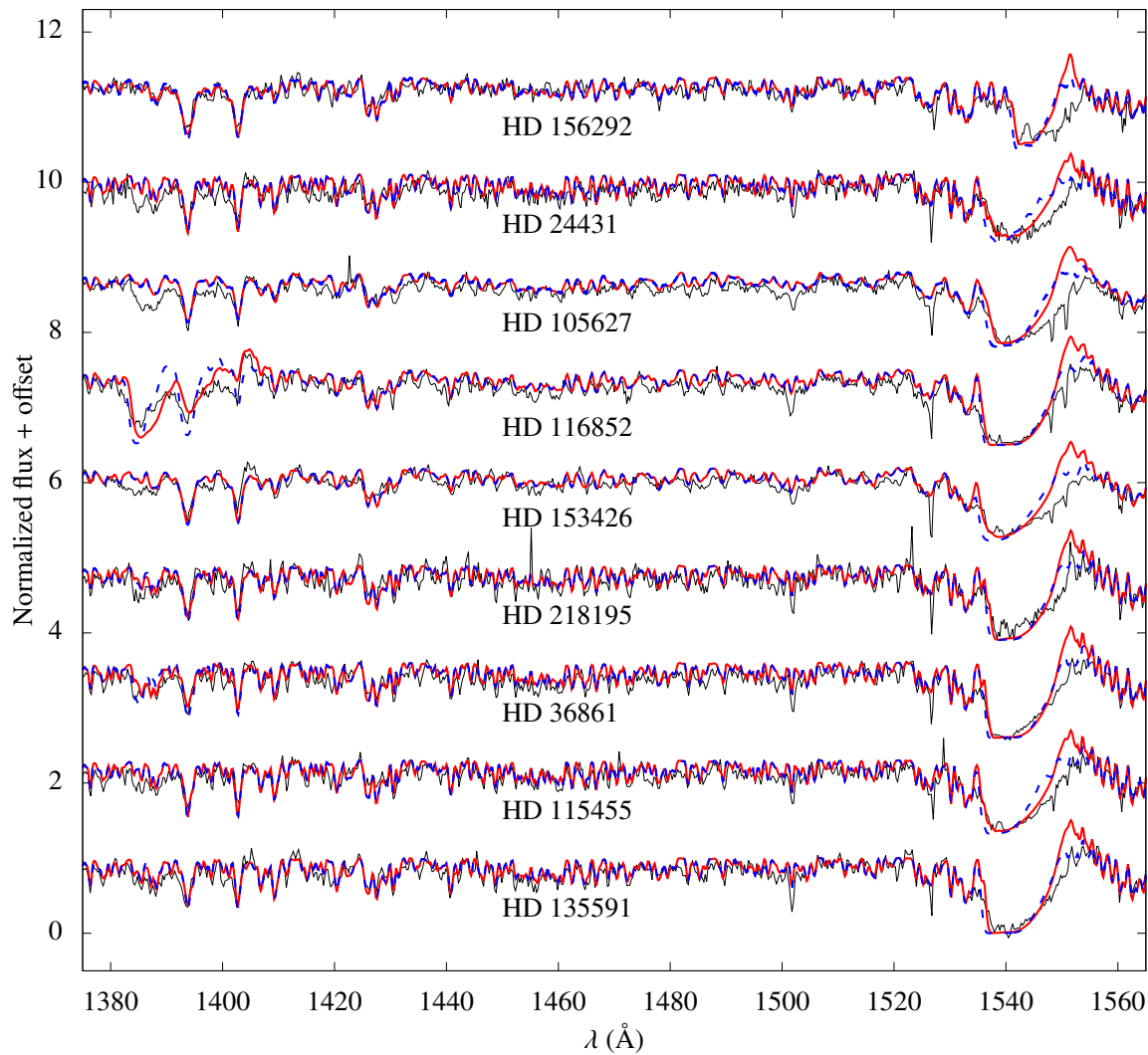}}}
\vspace*{1cm}
\caption{Final models for $\beta = 1.0$ (red) and tests with $\beta \sim 0.3$ (dashed blue) for each star in the UV. Here, $\beta = 0.35$ for HD 24431 and HD 115455 due to model convergence issues with $\beta = 0.3$. For the other stars, $\beta = 0.3$. The IUE spectra are in black. The emission component of \ion{C}{IV} $\lambda$$\lambda$1548,1551 is better modeled with $\beta \sim 0.3$.}
\label{beta_law}
\end{figure*}

As expected, O dwarfs, giants, and supergiants occupy different loci in the HR diagram. In particular, our sample of late O giants populate a narrow region due to the low dispersion in luminosity (adopted) and effective temperature (from UV): $\log(L_\star/\mathrm{L_\odot}) \sim 5.1-5.3$ and $\log(T_{\mathrm{eff}}) \sim 4.50-4.55$. The bulk of our sample shows initial (evolutionary) masses of $\sim 25-28$ $\mathrm{M_\odot}$ and ages of $\sim 10^{6.7}-10^{6.8}$ yr. The star of our sample with the lowest $T_{\mathrm{eff}}$, HD 156292, has initial mass between 20 and 25 $\mathrm{M_\odot}$ (being closer to the latter) and age between $10^{6.8}$ and $10^{6.9}$ yr. In contrast, the O5-7.5III stars correspond to different intervals of mass and age, with $M_{\textrm{ZAMS}} \sim 32-60$ $\mathrm{M_\odot}$ and $t \sim 10^{6.6}-10^{6.7}$ yr, and hence they are more massive and younger than our sample, as expected from the spectral classification. We see that the late giants of \citet{mahy15} populate the region around our sample despite the two stars that are close to the edge of our upper limits on $\log(L_\star/\mathrm{L_\odot})$. Indeed, \citet{mahy15} noted the discrepancies between their luminosities and the spectral-type calibration of \citet{martins05_calibration}. They argue that this trend is related to their methodology for the derivation of the luminosity, and thus slightly affecting the analysis on the HR diagram.\par

The dwarfs considered here present a larger interval in mass and age ($\sim 25-60$ $\mathrm{M_\odot}$ and t = $\sim 10^{6.0}-10^{6.8}$), since they encompass a larger range of spectral types (from O9.5V to O3.5V). This is the same as for the OB supergiants that spread over the whole diagram in $T_{\mathrm{eff}}$, implying $M_{\textrm{ZAMS}} \sim 25-60$ $\mathrm{M_\odot}$ and t = $\sim 10^{6.5}-10^{6.9}$ yr. We recall here that the O dwarfs with $\log(L_\star/\mathrm{L_\odot}) < 5.2$ present the weak wind problem and correspond to the O8-9.5V spectral types. For these stars, we observe masses of $\sim 20-25$ $\mathrm{M_\odot}$ and ages around $10^{6.7}$ yr. In fact, as expected, we can perceive a clear division in ages between dwarfs, giants, and supergiants from Fig. \ref{HR_diagram_isochrones}.\par

In conclusion, we corroborate the literature results showing that O giants are slightly more evolved objects than the dwarfs, being closer to the end of the main sequence phase \citep[e.g.,][]{mahy15, martins_herve15}. Our sample is described on the HR diagram as a descent of O dwarfs with $\log(L_\star/\mathrm{L_\odot}) \sim 5.0$, corresponding to the spectral types O6.5-8V. These O dwarfs are the onset of the weak wind problem. Thus, weak winds in O giants would imply that this phenomenon is not exclusively associated to younger stars on the main sequence. The bulk of our sample is halfway between the O dwarfs' loci and the end of the H-burning phase, thus weak winds could persist up to end of the main sequence before undergoing the supergiant phase. We stress that OB supergiants do not present the weak wind phenomenon \citep[e.g.,][]{bouret12}.

\subsection{Wind properties}
\label{sec_wind}

\subsubsection{Wind velocity law}
\label{sec_betalaw}

As previously mentioned, the emission component of the \ion{C}{IV} $\lambda$$\lambda$1548,1551 P-Cygni profile is overestimated in our models. Different parameters can affect this profile, for example, the carbon abundance, mass-loss rate, X-ray flux, and wind velocity structure. However, we only found better fits by changing this last, more specifically, the $\beta$ parameter. Tests performed with other parameters did not change the profile in the desired way and/or produced undesired effects in other parts of the spectrum. It is possible to decrease the emission to the observed level by decreasing the mass-loss rate or the carbon abundance\footnote{A lower carbon abundance is actually expected as O stars evolve \citep[e.g.,][]{martins_herve15}.}. On the other hand, the absorption component of the P-Cygni decreases too much in comparison with the observations. It is beyond the scope of this paper to derive CNO abundances for late O giants. Nevertheless, we discuss in Sect. \ref{sec_deg_CNO_mdot} the effects of CNO abundances on the determination of $\dot{M}$ from the UV.\par

We have tried different values for the $\beta$ parameter in the velocity law. In Fig. \ref{beta_law}, we compare our final models ($\beta$ = 1.0) to models recalculated with lower $\beta$ values around 0.3. Our tests are limited to this value because we could not reach model convergence below $\beta \lesssim 0.3$\footnote{Indeed, we could not reach convergence even with $\beta$ = 0.3 for some stars (HD 24431 and HD 115455), but it worked with a slightly higher value $\beta$ = 0.35-0.40.}. Overall, the fit to the observed profiles is improved with a $\beta \sim 0.3$. The emission component of the profiles decreases in comparison with $\beta = 1.0$ models and provides a better match to the observations.  We see that the effects of varying $\beta$ on the \ion{Si}{IV} $\lambda$$\lambda$1394,1403 profiles are not significant. The exception is for HD 116852, but this modification of the spectral lines due to $\beta$ is much smaller than the changes created by the limits on $\dot{M}$ of this star.\par

In the framework of the Sobolev approximation, the variation in $\beta$ impacts differently on line formation in the inner and outer regions of the wind. In the inner wind, the Sobolev length is proportional to $ \left(\frac{dv}{dr}\right)^{-1} $. That is, a lower $\beta$ (higher gradient) implies a smaller interaction region: we have less absorption and emission at low velocities (close to the line center). This can be seen in the \ion{C}{IV} $\lambda$$\lambda$1548,1551 profiles in Fig. \ref{beta_law}. On the other hand, the Sobolev length is proportional to $ \frac{r^{2}}{v} \beta^{-1}$ in the outer wind. By decreasing $\beta$, we have a larger interaction region at high velocities (far from the line center). This is also observed in Fig. \ref{beta_law} (more absorption), but the effect is lower compared to the decrease in emission.\par

Low values for $\beta$, as suggested by our fits, are uncommon from the spectroscopic modelling of O stars: most O stars have $\beta$ close to unity (dwarfs) or even higher, up to $\sim 2.0-3.0$ in supergiants \citep[see, e.g.,][]{crowther06,martins_marcolino15}. Moreover, there are hydrodynamical results showing $\beta$ $\sim$ 1.0-0.9 for O8-9.5 giants \citep{muijres12}. Therefore, our tests suggesting very low values of $\beta$ are an artifact of our modeling assumptions, they do not represent a viable solution to the wind velocity structure of O-type stars. We recall that we assumed a standard $\beta$ velocity law to describe the wind region. One possibility relies on less simple parameterizations for wind velocity structure, for example, a two-component $\beta$ velocity describing separately the inner and outer regions of the wind \citep[e.g.,][]{hillier99}. Thus, a deeper investigation is needed, but it is beyond the scope of this paper.\par

\subsubsection{Mass-loss rates: Weak winds}
\label{sec_weak_wind}

\begin{figure}
\centerline{\resizebox{0.5\textwidth}{!}{\includegraphics{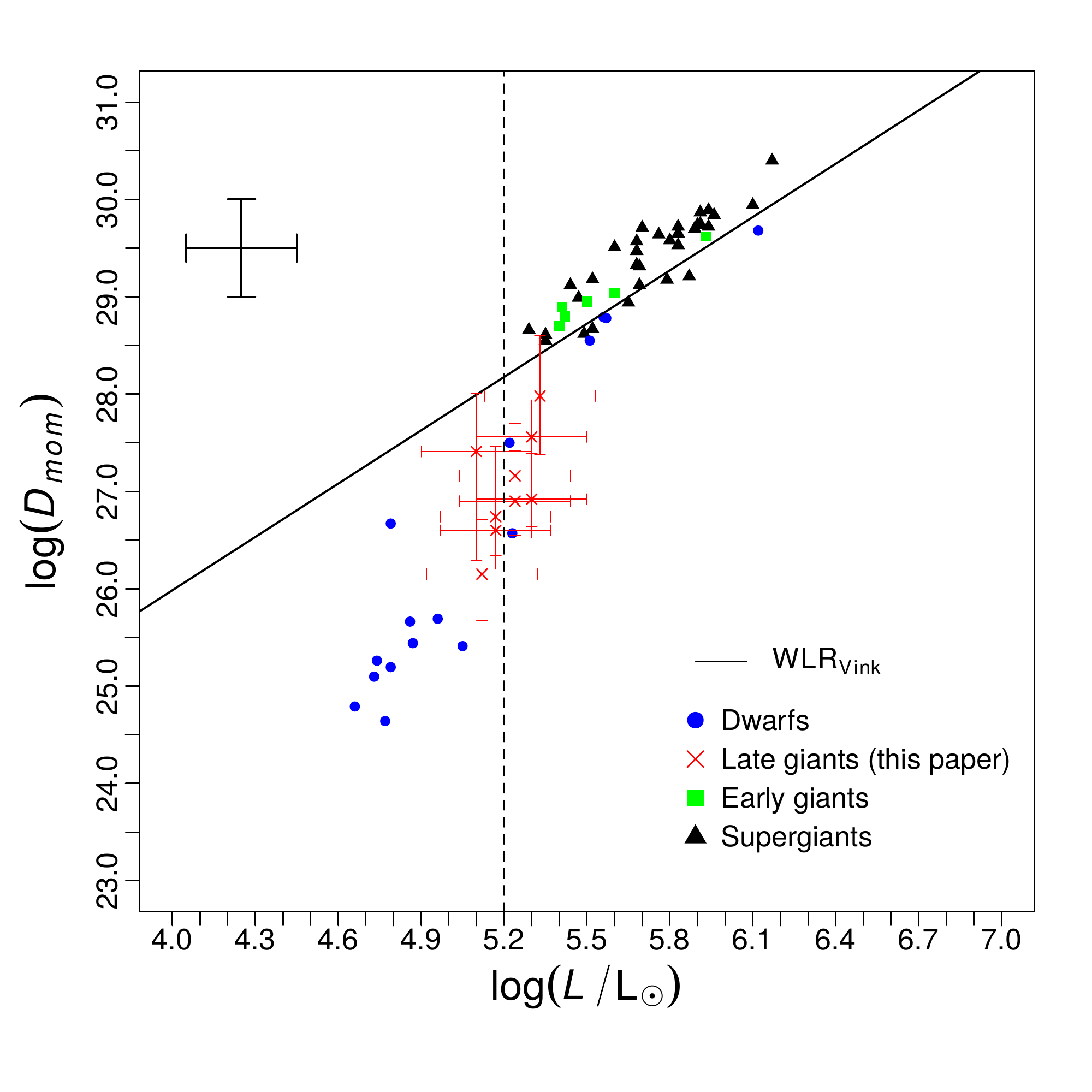}}}
\caption{Wind momentum-luminosity diagram for O dwarfs, giants, and supergiants. Colors and geometric figures stand as in Fig. \ref{HR_diagram_tracks}, our results are shown in red crosses. Our results are derived from the UV analysis. All the spectroscopic results consider (or are scaled to) unclumped $\dot{M}$. The theoretical relation of \citet{vink00} is in solid black. We mark $\log (L_\star/\mathrm{L_\odot}) = 5.2$ in dashed black line, and representative error bars for the literature results are shown in the top left.}
\label{diagrama_dmom_anas_giants_super}
\end{figure}

\begin{figure}
\centerline{\resizebox{0.5\textwidth}{!}{\includegraphics{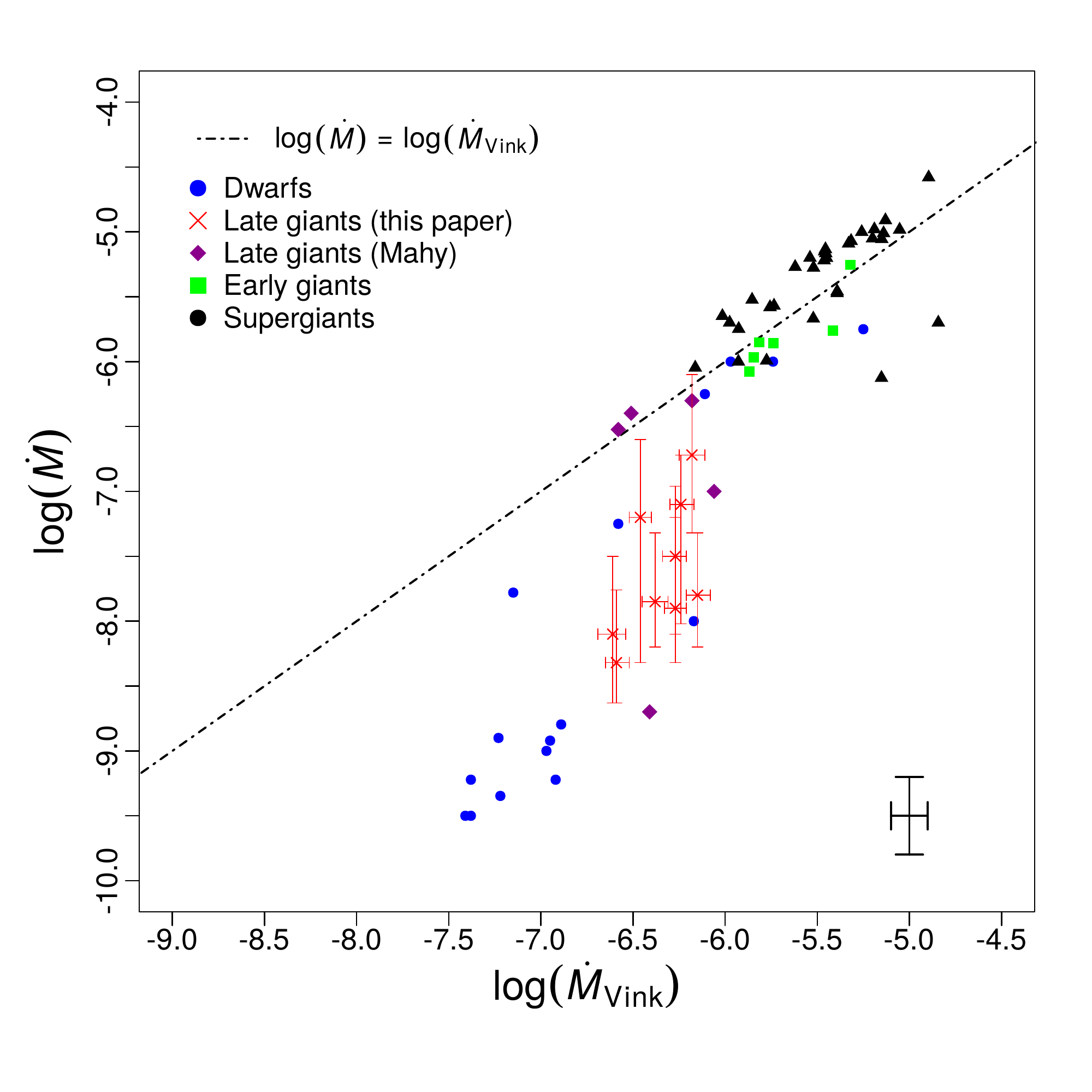}}}
\caption{Comparison between the spectroscopic $\dot{M}$ and the ones predicted from \citet{vink00} for O dwarfs, giants, and supergiants. Colors and geometric figures stand as in Fig. \ref{HR_diagram_tracks}, our results are shown in red crosses. One-to-one relation is shown in dotted-dashed line, and representative error bars for the literature results are shown in the bottom right. O8-9 dwarfs (weak winds) are shown in blue points. We see that late O giants also present weak winds.}
\label{diagrama_mdot_anas_giants_super}
\end{figure}

\begin{figure}[!ht]
\centerline{\resizebox{0.5\textwidth}{!}{\includegraphics{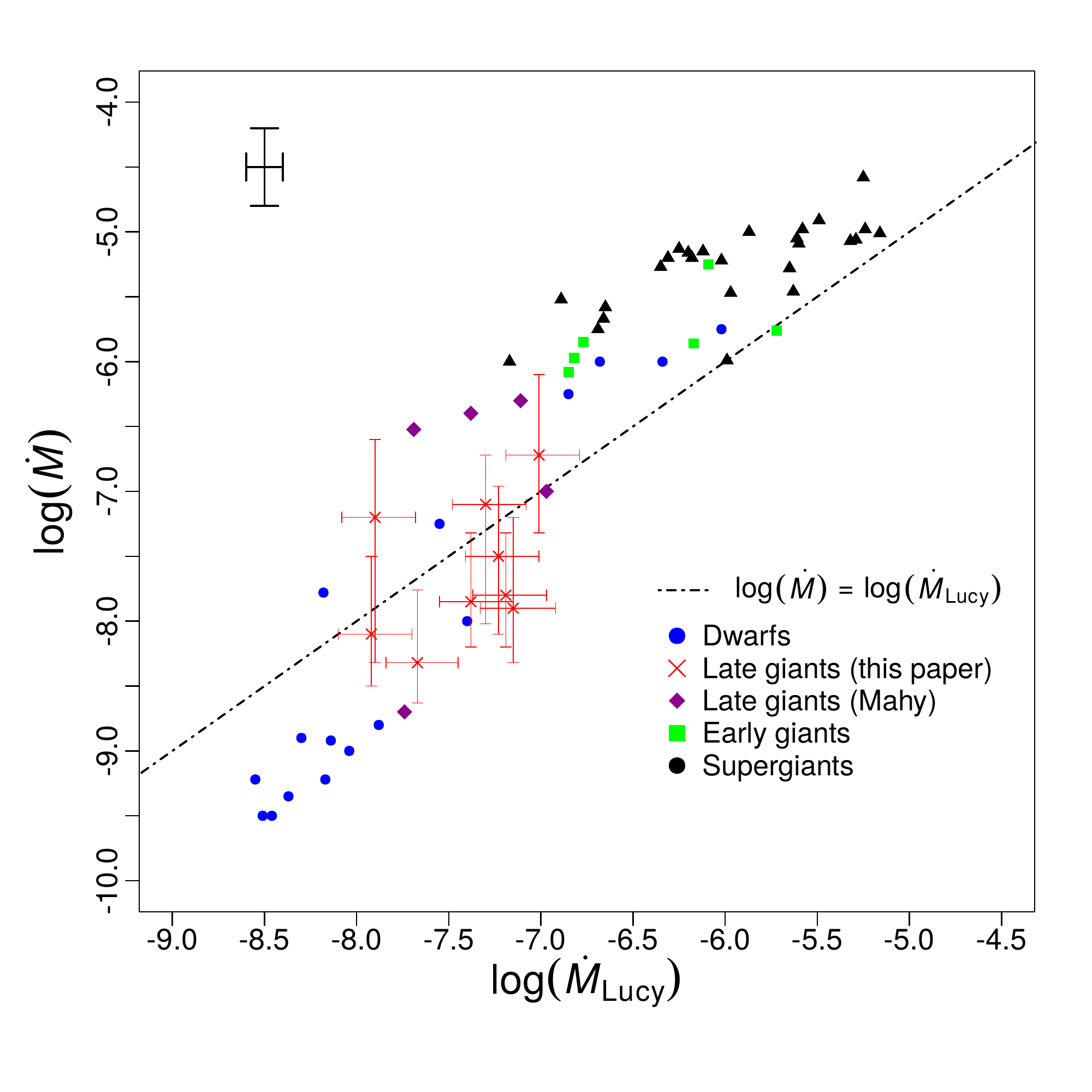}}}
\caption{Same as Fig. \ref{diagrama_mdot_anas_giants_super}, but comparing with theoretical $\dot{M}$ from \citet{lucy10_grid}. The weak wind problem is significantly lessened to about one order of magnitude. On the other hand, the discrepancy here for the luminous OB stars (shown in triangles) is increased in comparison with \citet{vink00}.}
\label{diagrama_mdot_lucy_anas_giants_super}
\end{figure}

\begin{figure}[!ht]
\centerline{\resizebox{0.50\textwidth}{!}{\includegraphics[angle=90]{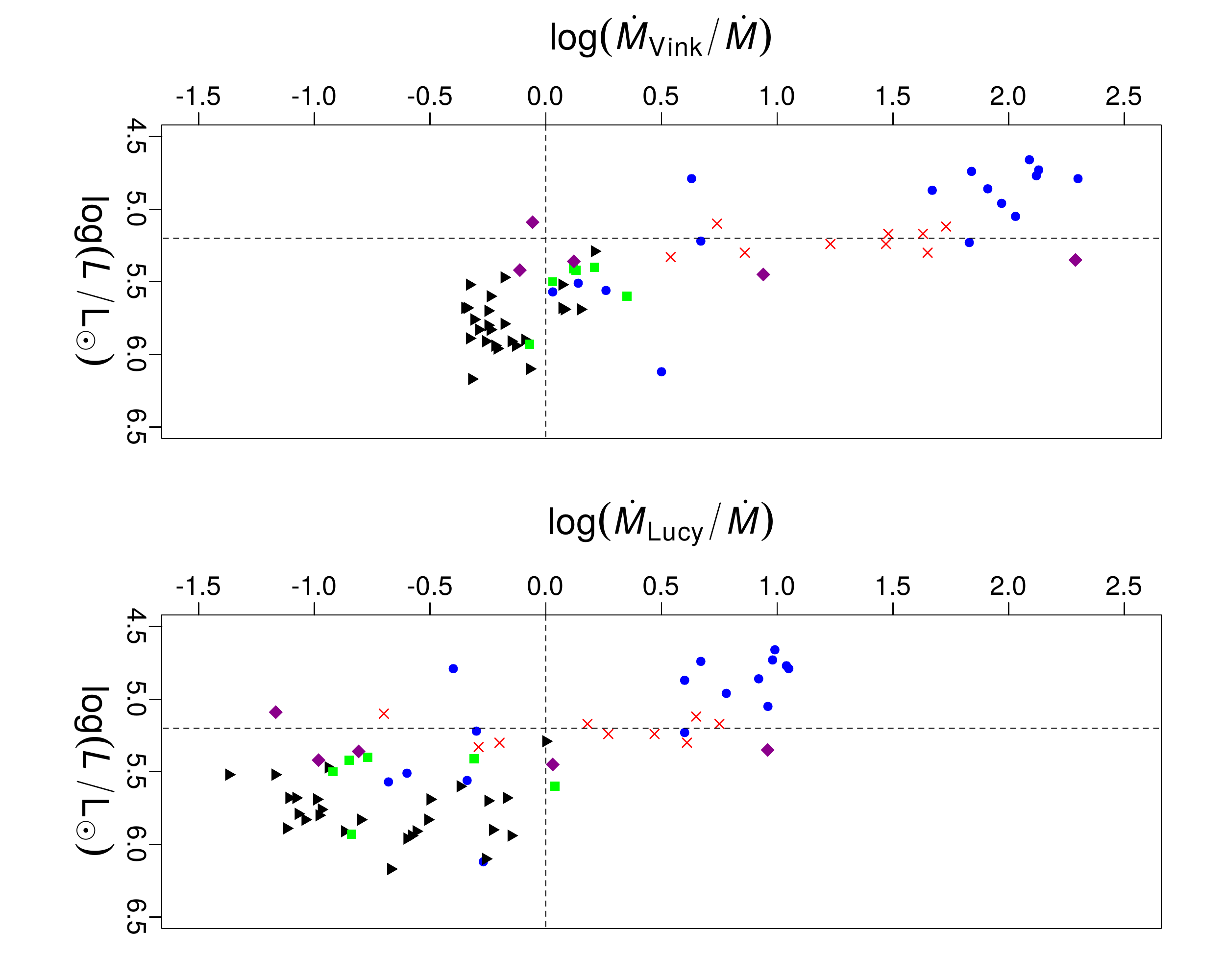}}}
\caption{Difference (as a function of luminosity) between the measured $\dot{M}$ (clumped) and their theoretical values by \citet{vink00} on the left, and by \citet{lucy10_grid} on the right. Symbols stand as presented in Fig. \ref{HR_diagram_tracks}, our results are shown in red crosses. The luminosity value of $\log (L_\star/\mathrm{L_\odot}) = 5.2$ and the match between the spectroscopic and theoretical $\dot{M}$ are indicated by dashed black lines. We note that $\dot{M}_{\mathrm{Lucy}}$ attenuates the weak wind problem, but it increases the discrepancy to the spectroscopic $\dot{M}$ in $\log (L_\star/\mathrm{L_\odot}) \gtrsim 5.2$.}
\label{diagrama_ratio_mdots_luminosity}
\end{figure}

In this section, we compare the mass-loss rates determined from atmospheric models with the ones predicted by \citet{vink00} and \citet{lucy10_grid}. First, we consider the $\log(D_{mom})$ versus $\log (L_\star/\mathrm{L_\odot})$ diagram in Fig. \ref{diagrama_dmom_anas_giants_super}. Our results for late O giants are presented along with dwarfs, giants, and supergiants of different spectral classes from the literature. We do not include here the results of \citet{mahy15} since they derived $v_\infty$ for just two objects out of six late O giants. All spectroscopic results in Figs. \ref{diagrama_dmom_anas_giants_super}, \ref{diagrama_mdot_anas_giants_super}, \ref{diagrama_mdot_lucy_anas_giants_super}, and \ref{diagrama_ratio_mdots_luminosity} consider homogeneous wind models: the literature results with clumping were scaled by a factor of 1/$\sqrt{f_{\infty}}$.\par

The weak wind phenomenon is seen for the late O dwarfs (O8-9.5V) with $\log (L_\star/\mathrm{L_\odot}) \lesssim 5.2$. Their modified momentum are up to two orders of magnitude lower than the theoretical relation of \citet{vink00}. The late O dwarf closest to the predicted value -- $\log (L_\star/\mathrm{L_\odot}) \sim 4.8$ and log($D_{mom}) \sim 26.6$ -- had its mass-loss rate derived by \citet{martins05_weakwinds} as a conservative upper limit value. For the most luminous stars with $\log (L_\star/\mathrm{L_\odot}) \gtrsim 5.2$, there is a good agreement between the measured and predicted values. Our results fall below the values expected from theory, even considering the error bars. Only one object of our sample (HD 116852) marginally agrees with the wind momentum-luminosity relation from \citet{vink00}. Hence, we conclude that late O giants also present winds weaker than predicted by theory. The discrepancy is more severe for O8 giants (HD 156292, O9.7III, lowest $D_{mom}$) and is attenuated towards O9 giants (HD 116852, O8.5II-III, highest $D_{mom}$). It suggests a gradual change from ``weak'' to ``normal'' winds (agreement with predictions) for the stars of our sample. \par

In Fig. \ref{diagrama_mdot_anas_giants_super}, we present a direct comparison between the spectroscopic $\dot{M}$ and the predicted ones using \citet{vink00}. Stars are divided by colors and geometric symbols as in Fig. \ref{diagrama_dmom_anas_giants_super}. In addition, we include here the results of \citet{mahy15} for O8-9.5III stars for which mass-loss rates were determined (five out of six stars). It reflects the same basic conclusions obtained from the wind momentum-luminosity diagram in Fig. \ref{diagrama_dmom_anas_giants_super}. Nevertheless, the mass-loss range and the types of O stars for which the radiative wind theory is successful are conspicuous. We note that the sample of late O giants from \citet{mahy15} tends to agree better with the predictions of \citet{vink00}, but we still observe the weak wind problem here: three out of five stars in good agreement and two stars presenting significant deviations (with one clear weak wind star). Since \citet{mahy15} derived mass-loss rates using UV and H$\alpha$, we will discuss this question in more detail in Sect. \ref{sec_mass_loss_rates_uv_visible}.\par

Furthermore, we performed the same comparison but with the hydrodynamical predictions of \citet{lucy10_grid} for Galactic O stars. These predictions are made in the framework of the most recent updated version of the moving reversing layer theory \citep{lucy70}. In short, for given stellar parameters, the equation of motion has physical solution for a certain value of mass flux $J$ that satisfies null effective gravity surface at the critical point of the wind. In a previous work, \citet{lucy10} found that the discrepancies between the measured $\dot{M}$ for late O dwarfs \citep{marcolino09} and their predictions are significantly reduced up to about one order of magnitude. In Fig. \ref{diagrama_mdot_lucy_anas_giants_super}, we present a comparison between the derived $\dot{M}$ (by atmosphere models) and the ones calculated using the predicted mass fluxes given by \citet{lucy10_grid} for almost the same sample presented in Fig. \ref{diagrama_mdot_anas_giants_super}.\par

The grid of \citet{lucy10_grid} provides mass fluxes for stars with $3.00 \leq  \log(g) \leq 4.50$. Thus, we excluded some OB supergiants (six objects) that were analyzed in the previous comparison with $\dot{M}_{\mathrm{Vink}}$. From the literature sample presented in Fig. \ref{diagrama_mdot_anas_giants_super}, we excluded stars with $\log(g)$ < 2.95. For stars with $2.95 \leq  \log(g) \leq 3.00$ (three objects), we calculated the mass fluxes (and then $\dot{M}_\mathrm{Lucy}$) considering $\log(g) = 3.00$. Interestingly, we observe a better agreement between the spectroscopic and predicted values for the mass-loss rates of low-luminosity objects (late O dwarfs and giants). However, the most part still have $\dot{M}$ values about 0.5-1.0 dex lower than $\dot{M}_\mathrm{Lucy}$. In contrast to the previous comparison with \citet{vink00}, the predictions of \citet{lucy10_grid} for high-luminosity OB stars -- $\log (L_\star/\mathrm{L_\odot}) \gtrsim 5.2$ -- are lower than the mass-loss rates obtained by atmosphere models. For a better visualization, we present again these results in an alternative form in Fig. \ref{diagrama_ratio_mdots_luminosity}. We see that $\dot{M}_\mathrm{Lucy}$ underestimates the mass loss of OB stars with $\log (L_\star/\mathrm{L_\odot}) \gtrsim 5.2$ practically as much as it overestimates for objects with $\log (L_\star/\mathrm{L_\odot}) \lesssim 5.2$.\par

In conclusion, Figs. \ref{diagrama_dmom_anas_giants_super} and \ref{diagrama_mdot_anas_giants_super} indicate that late O giants exhibit weak winds. As O8-9.7III objects are more evolved than late O dwarfs, we naturally exclude evolutionary effects as the reason for weak winds. Put differently, O stars with luminosities lower than $\log (L_\star/\mathrm{L_\odot}) \sim 5.2$ must have weak winds through the H-burning phase. Furthermore, the predictions from \citet{lucy10_grid} attenuate the weak wind problem both for late O dwarfs and for late giants. However, these theoretical values clearly fail \citep[in comparison with][]{vink00} to predict the mass-loss rates for more luminous OB stars, such as OB supergiants, early dwarfs, and early giants. We stress here that the predictions of \citet{vink00} are in good agreement with the hydrodynamical simulations of \citet{muijres12} for O stars with $\log (L_\star/\mathrm{L_\odot}) \gtrsim 5.2$, while the latter fails to predict $\dot{M}$ for objects below this luminosity region. It is hard to compare the predictions of \citet{vink00} with the ones from \citet{lucy10_grid} because they employ different approaches: the first find $\dot{M}$ that is globally (in the wind) consistent with the conservation of energy, while \citet{lucy10_grid} predicts the mass loss from first principles (i.e., solving the equation of motion). Nevertheless, it is remarkable that the region of $\log (L_\star/\mathrm{L_\odot}) \sim 5.2$ shows to be critical for both of them (in comparison with the spectroscopic $\dot{M}$).\par

\begin{figure*}
\centerline{\resizebox{0.85\textwidth}{!}{\input{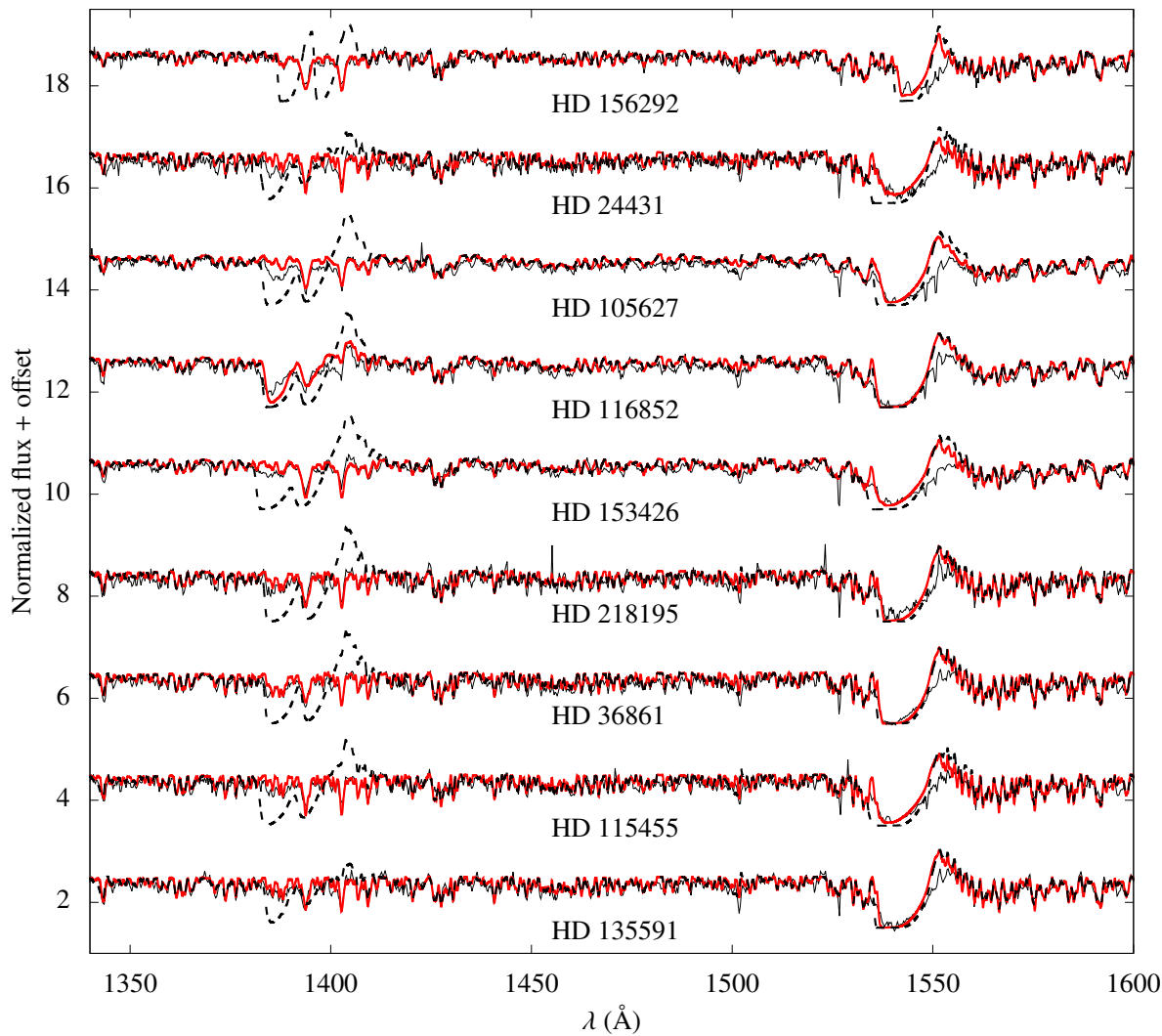}}}
\vspace*{1.0cm}
\caption{Comparison between the final models and the ones computed using the hydrodynamical mass-loss rates of \citet{vink00} in the UV region. The IUE spectra are in solid black, and the star name is indicated right below its spectrum. All the final models ($\dot{M}$ from the UV) are in red, while $\dot{M}_{\mathrm{Vink}}$ is in dashed black. Our final models have clumping ($f_{\infty} = 0.1$), while the models with $\dot{M}_{\mathrm{Vink}}$ are unclumped. We note how $\dot{M}_{\mathrm{Vink}}$ overestimates the intensity in \ion{Si}{IV} $\lambda$$\lambda$1394,1403 for all our sample. \ion{C}{IV} lines also become saturated in a few cases, in contrast to the observations.}
\label{mdot_mdotvink_uv_lines}
\end{figure*}


\section{Discussion}

\subsection{Mass-loss rates: UV versus visible}
\label{sec_mass_loss_rates_uv_visible}

In this section, we compare our final models to the ones computed using $\dot{M}_{\mathrm{Vink}}$, regarding the spectral modeling in the ultraviolet and optical regions. Throughout this section, we only compare our results with the predictions from \citet{vink00} because they are currently used in most modern stellar evolution codes. In the previous discussion, all $\dot{M}$ for the objects of our sample were derived from the UV analysis. Overall, our synthetic H$\alpha$ profiles have deeper cores than the observations, indicating the need to increase the $\dot{M}$ parameter in our models.\par

Regarding O8-9.5V stars, \citet{marcolino09} found that their UV $\dot{M}$ produce H$\alpha$ profiles in absorption, in relatively good agreement with observations. Moreover, they show that in three (out of five) objects the predicted $\dot{M}$ (Vink) implies a shallower H$\alpha$ line, in contrast to the data. For the other two stars, the difference between the final models and $\dot{M}_{\mathrm{Vink}}$ is minor against the observations. We show below that such discrepancies in H$\alpha$ are higher for O8-9.5III stars.\par

First, in Fig. \ref{mdot_mdotvink_uv_lines}, we compare our final models with models using the mass-loss rates from \citet{vink00} in the UV region. All models with $\dot{M}_{\mathrm{Vink}}$ are computed with $f_{\infty}$ = 1.0 because \citet{vink00} do not take clumping into account. These values of $\dot{M}_{\mathrm{Vink}}$ are higher than our unclumped $\dot{M}$ from UV up to about two orders of magnitudes (Table \ref{resumo_resultados}). Our best fits to the observations consider clumping ($f_{\infty} = 0.1$, see Sect. \ref{sec_stellar_atmosphere_code}). We recall, however, that Fig. \ref{mdot_mdotvink_uv_lines} would be virtually identical by preserving $\dot{M}/\sqrt{f_{\infty}}$ constant for each star. All the other physical parameters are fixed. The synthetic profiles of \ion{Si}{IV} $\lambda$$\lambda$1394,1403 using $\dot{M}_{\mathrm{Vink}}$ are systematically more intense than the data for all objects. Regarding \ion{C}{IV} $\lambda$$\lambda$1548,1551, it is difficult to distinguish between our final mass-loss rates and the ones predicted by Vink for stars with saturated profiles (e.g., HD 116852). On the other hand, in HD 24431, HD 105627, and HD 153426 the predicted rates saturate the profiles in contrast to the observations. Hence, we conclude that models considering $\dot{M}_{\mathrm{Vink}}$ are not able to fit the UV mass-loss diagnostics of late O giants.\par

\begin{figure*}
\centerline{\resizebox{0.85\textwidth}{!}{\input{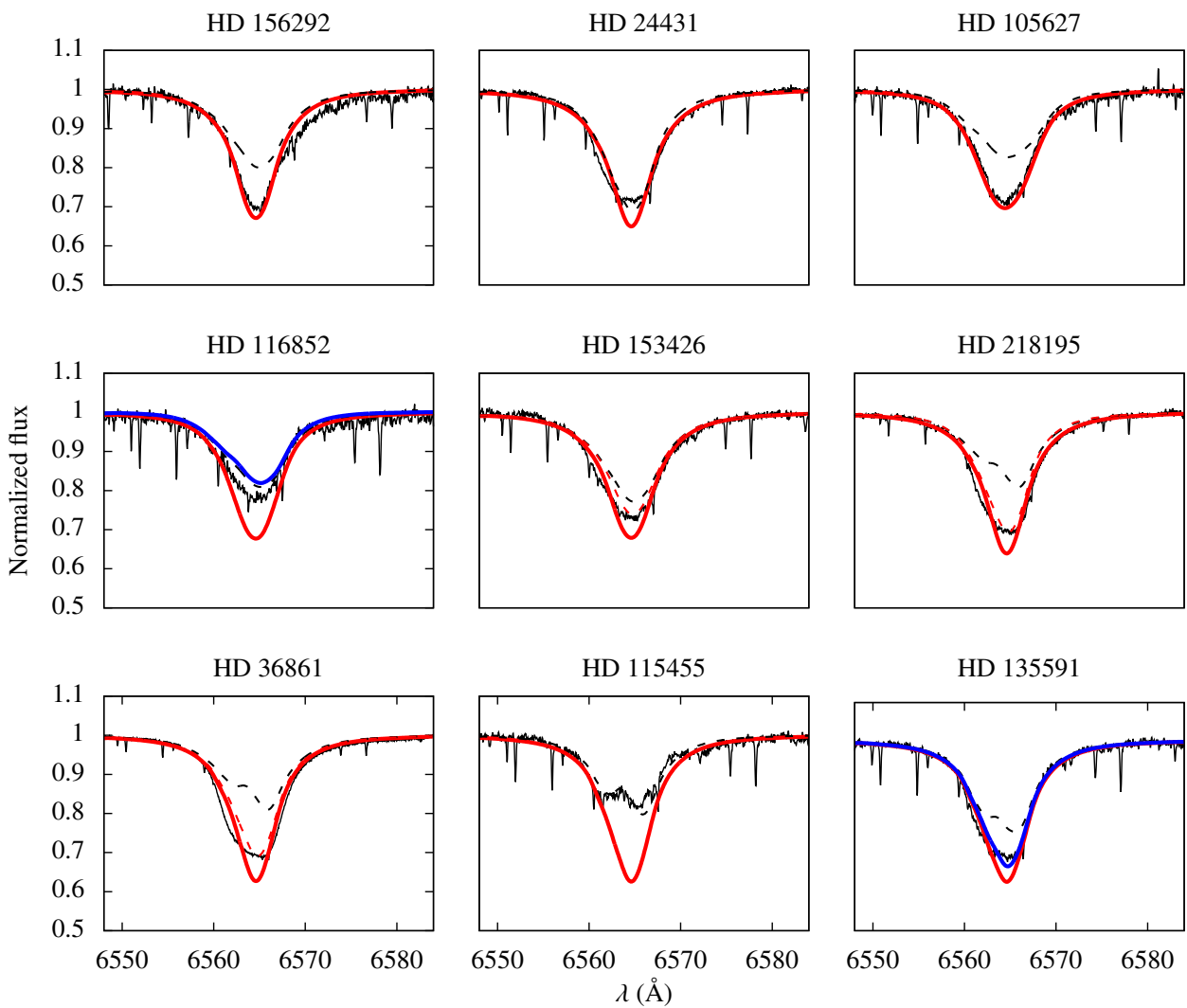}}}
\vspace*{1cm}
\caption{Mass-loss rates from fitting H$\alpha$. Our final models (UV $\dot{M}$) are shown in solid red: ``models (i)''. Upper limits on UV $\dot{M}$ that encompass the observed H$\alpha$ line are shown in blue for some stars: ``models (ii)''. Models with $\dot{M}_{\mathrm{Vink}}$ are presented in dashed black: ``models (iii)''. Again, only the models with $\dot{M}_{\mathrm{Vink}}$ are homogeneous. When none of the previous models are able to adjust the H$\alpha$ intensity, we provide a new $\dot{M}$ determination from fitting H$\alpha$ (dashed-red): ``models (iv)''. The text gives further details concerning the notations ``models (i-iv)''.}
\label{mdot_mdotvink_halpha}
\end{figure*}

Our analysis of the H$\alpha$ profile is presented in Fig. \ref{mdot_mdotvink_halpha}. Again, models with $\dot{M}_{\mathrm{Vink}}$ are computed without clumping. However, in this case, we have four ``types'' of models:

\begin{enumerate}[label=(\roman*)]

\item with UV mass loss (solid red). We present ``models (i)'' for all the stars of our sample. These models have $f_{\infty}$ = 0.1. \par

\item with UV upper mass loss (solid blue). We show ``models (ii)'' only for HD 116852 and HD 135591 because we are able to reproduce (or to overestimate) with them the observed H$\alpha$. We do not present our UV upper limit on $\dot{M}$ for the other stars, since they produce practically the same H$\alpha$ profile as ``models (i)'' in this case. These models have $f_{\infty}$ = 0.1.\par

\item with Vink's mass loss (dashed black). As for ``models (i)'', ``models (iii)'' are shown for all the stars of our sample. We use unclumped models because \citet{vink00} do not take clumping into account.\par

\item with mass loss derived from fitting the H$\alpha$ profile (dashed red). ``Models (iv)'' are shown only for those stars for which we do not fit H$\alpha$ in any of the above cases. For example, we present this type of model for HD 218195, since neither models with our UV $\dot{M}$, our UV upper $\dot{M}$, nor $\dot{M}_{\mathrm{Vink}}$ are able to fit the observed H$\alpha$ profile. These models have $f_{\infty}$ = 0.1.\par 

\end{enumerate}

We note that the synthetic profiles calculated with $\dot{M}_{\mathrm{Vink}}$ produce H$\alpha$ somewhat more strongly than the observed profiles for five stars of our sample: HD 156292, HD 105627, HD 218195, HD 36861, and HD 135591. We observe that the discrepancies for O8-9III are higher than the ones found by \citet{marcolino09} for late dwarfs. This can be explained in terms of a higher H$\alpha$ sensitivity for $\dot{M} \gtrsim 10^{-7}$ $\mathrm{M_\odot}$ yr\textsuperscript{-1}. In fact, our sample has an average $\dot{M}_{\mathrm{Vink}}$ of $\sim 5.0\e{-7}$ $\mathrm{M_\odot}$ yr\textsuperscript{-1}, while the O8-9V star sample in \citet{marcolino09} has an average value of $\sim 9.0\e{-8}$ $\mathrm{M_\odot}$ yr\textsuperscript{-1} for the predicted $\dot{M}$.\par

From Fig. \ref{mdot_mdotvink_halpha}, the H$\alpha$ profiles of HD 156292 and HD 105627 are well fitted by our final models: $\dot{M}$ derived from fitting the UV resonance lines of \ion{Si}{IV} and \ion{C}{IV}. For the other seven stars, our UV mass-loss rates show a deeper core in H$\alpha$. We see that the profiles of HD 24431, HD 116852, and HD 115455 are fitted considering the mass-loss rate from \citet{vink00}. However, for HD 116852, our UV upper limit on $\dot{M}$ (solid blue line) also reproduces H$\alpha$. It happens because all our models -- used to derive $\dot{M}$ -- have the inclusion of clumping, while the models with $\dot{M}_{\mathrm{Vink}}$ are unclumped. Still regarding the mass-loss upper limit from UV, we are also able to fit the H$\alpha$ data of HD 135591. Thus, our $\dot{M}$ derived from the UV (``models (i)'' plus ``models (ii)'') are consistent with the observed H$\alpha$ profile of four stars out of nine.\par

For the other three objects (HD 153426, HD 218195, HD 36861), we need to increase $\dot{M}$ up to $\sim 10^{-7}$ $\mathrm{M_\odot}$ yr\textsuperscript{-1} to fill their core.  Our models show H$\alpha$ insensitive for $\dot{M} \sim 10^{-8}-10^{-7}$ $\mathrm{M_\odot}$ yr\textsuperscript{-1}, similarly to results found in the literature for late O dwarfs \citep[e.g.,][]{martins_mahy12}. Since the H$\alpha$ data for these three stars tend to be reproduced by just varying $\dot{M}$ in CMFGEN, we consider that these deeper observed profiles are unlikely to be due to circumstellar or interstellar contamination. Nonetheless, such cases of contamination have been reported in the literature for early and late O dwarfs \citep[see, e.g.,][]{martins05_weakwinds}. Another observational issue in this analysis could be due to H$\alpha$ variability for the stars in our sample, potentially impacting the determination of $\dot{M}$ from this transition. For example, \citet{martins_marcolino15} investigated the spectral variability in the optical region in early OB supergiants and late O dwarfs. They found strong profile variability in H$\alpha$ for the supergiants, while the dwarfs do not exhibit any sign of variability. Hence, it would be necessary to investigate this issue in detail for an intermediate luminosity class such as the giants. Moreover, we performed different tests (e.g., changing the number of depth points and including additional ions) to solve this discrepancy between the models with $\dot{M}$ from fitting the UV and the H$\alpha$ data, but the situation was not improved at all. Thus, we conclude that our models cannot fit simultaneously the UV and optical wind signatures in about half of our sample.\par

As mentioned in Sect. \ref{sec_weak_wind}, $\dot{M}$ found by \citet{mahy15} tend to be closer to the predicted values using the mass-loss recipe of \citet{vink00}. This can be explained since their mass-loss analysis is only complete concerning the visible spectra: they have IUE/SWP data only for two out of the six O8-9.5III stars in their sample. Even so, we still see one unequivocal late giant in their sample that shows the weak wind phenomenon: HD 191878 (type O8III). For this object, \citet{mahy15} derived $\dot{M} = 2.0\e{-9}$ $\mathrm{M_\odot}$ yr\textsuperscript{-1} (unclumped CMFGEN model) by simultaneously fitting the UV spectrum and the H$\alpha$ line. Regarding Galactic O3-9.5V stars, \citet{martins_mahy12} also found a disagreement using CMFGEN between the UV mass-loss rates and the ones derived from the fitting of H$\alpha$. They found the most severe disagreements for the O8-9V stars. 
Their $\dot{M}$ derived from the UV region are up to two orders of magnitude lower than $\dot{M}$ from H$\alpha$, being this latter closer to $\dot{M}_{\mathrm{Vink}}$. Thus, we verify a similar trend in our sample. One of the possibilities stressed by \citet{martins_mahy12} to explain this issue is the neglect of macroclumping in the modeling with CMFGEN. The literature shows that accounting for macroclumping reduces more significantly the intensity in the UV lines than in H$\alpha$ \citep[e.g.,][]{oskinova07, sundqvist11, sundqvist18}. \citet{martins_mahy12} pointed out that the inclusion of macroclumping could lead to a better agreement between $\dot{M}$ from the UV and H$\alpha$ fittings, since the UV values would be reduced in this case. On the other hand, it also implies that $\dot{M}$ predicted neglecting clumping (such as the Vink's value for HD 116852 in Fig. \ref{mdot_mdotvink_halpha}) must overestimate the real rates.\par

In short, $\dot{M}$ computed using the recipe of \citet{vink00} are not able to fit the UV resonance lines for any of the stars of our sample. Lower $\dot{M}$ are supported in four out of nine stars considering simultaneously the fitting of the UV and the optical regions, so, in this sense, favoring the weak wind phenomenon in late O giants. Besides possible effects resulting from our physical assumptions in the modeling, environmental contamination, and spectroscopic variability, we need to increase the UV mass loss of about half of our sample to find a better modeling of H$\alpha$. These higher $\dot{M}$ values from H$\alpha$ are incompatible with the UV modeling. Hence, we have a partial agreement between $\dot{M}$ derived from the fitting to the UV resonance lines and to the H$\alpha$ line. Again, this issue between the UV and the visible analyses is also present in the literature for O dwarfs and deserves further study.\par

\begin{table}
\caption{Observation log for the FEROS data of HD 156292, HD 153426, and HD 115455 analyzed in Fig. \ref{hd156292_hd153426_hd115455_binary_halpha}. Our observed spectra are shown in bold. Orbital phases ($\phi$) are calculated considering the epoch time $T_{0}$ ($\phi$ = 0) from our observation.}\label{espectros_feros_binary}
\centering
\renewcommand{\arraystretch}{1.2}
\begin{adjustbox}{width=0.44\textwidth}
\begin{tabular}{cccc}

\toprule
\toprule

UTC (start time) &  MJD (start time) & ESO program &$\phi$ \\
\midrule

\noalign{\vskip 1mm}  
\multicolumn{4}{c}{HD 156292}\\
\noalign{\vskip 1mm} 

\textbf{2016-03-23T09:00:47.612} &\textbf{57470.37555106} &\textbf{096.A-9027(A)} & \textbf{$\equiv$ 0}\\
2015-04-04T06:35:39.030 &57116.27475729 &089.D-0975(A) & -0.680\\   
2012-06-20T07:17:03.206 &56098.30350933 &089.D-0975(A) & -0.747\\   
2012-05-21T06:52:16.748 &56068.28630495 &089.D-0975(A) & -0.824\\ 
2009-05-05T05:48:32.002 &54956.24203706 &083.D-0589(A) & -0.934\\     
2009-05-04T04:45:08.265 &54955.19801233 &083.D-0589(A) & -0.145\\   
2009-05-03T06:19:11.026 &54954.26332206 &083.D-0589(A) & -0.335\\   
2009-05-02T06:58:08.837 &54953.29038006 &083.D-0589(A) & -0.531\\   
2008-05-14T08:54:46.852 &54600.3713756  &081.D-2008(A) & -0.973\\   
2005-06-25T08:24:03.531 &53546.35004087 &075.D-0061(A) & -0.337\\

\midrule

\noalign{\vskip 1mm}  
\multicolumn{4}{c}{HD 153426}\\
\noalign{\vskip 1mm} 

\textbf{2016-03-18T09:15:25.182} &\textbf{57465.38570813} &\textbf{096.A-9027(A)} & \textbf{$\equiv$ 0}\\

2015-04-04T06:07:00.909 &57116.25487163 &089.D-0975(A) & -0.586\\   
2012-06-19T07:36:40.433 &56097.31713464 &089.D-0975(A) & -0.075\\   
2012-05-21T06:08:58.157 &56068.25622867 &089.D-0975(A) & -0.372\\   
2011-05-18T06:42:56.653 &55699.27982237 &087.D-0946(A) & -0.844\\   
2009-05-25T09:24:12.238 &54976.39180831 &083.D-0589(B) & -0.116\\   
2008-05-14T08:02:44.235 &54600.3352342  &081.D-2008(A) & -0.904\\   
2005-06-26T08:28:32.329 &53547.35315196 &075.D-0061(A) & -0.912\\   

\midrule

\noalign{\vskip 1mm}    
\multicolumn{4}{c}{HD 115455}\\
\noalign{\vskip 1mm}    

\textbf{2016-03-18T04:28:43.671} & \textbf{57465.18661656} & \textbf{096.A-9027(A)} & \textbf{$\equiv$ 0}\\

2007-06-15T23:58:03.854 &54266.99865572 &079.D-0564(B) & -0.081\\   
2007-05-27T04:31:13.627 &54247.18835217 &079.D-0564(C) & -0.395\\   
2007-05-24T03:49:08.750 &54244.15912905 &079.D-0564(B) & -0.596\\   
2007-04-22T06:11:41.302 &54212.25811692 &079.D-0564(B) & -0.711\\   
2007-04-21T05:39:19.139 &54211.23563818 &079.D-0564(A) & -0.779\\
2007-04-19T05:24:50.083 &54209.22557966 &079.D-0564(A) & -0.913\\   
2007-04-19T05:05:31.858 &54209.21217428 &079.D-0564(A) & -0.913\\   
2007-04-01T04:56:14.188 &54191.20571977 &079.D-0564(B) & -0.107\\   
2006-08-22T23:42:49.633 &53969.98807446 &077.B-0348(A) & -0.777\\  
2006-08-21T23:24:36.329 &53968.97542048 &077.B-0348(A) & -0.844\\   
2006-08-20T23:22:12.614 &53967.97375711 &077.B-0348(A) & -0.911\\ 
2006-08-19T23:22:41.721 &53966.97409399 &077.B-0348(A) & -0.977\\  
2006-08-18T23:36:52.296 &53965.98393861 &077.B-0348(A) & -0.043\\  
2005-01-03T07:28:56.719 &53373.31176758 &074.D-0300(A) & -0.344\\

\bottomrule
\end{tabular}
\end{adjustbox}
\end{table}

\subsection{Mass-loss rates: Binary effects}
\label{sec_mass_loss_rates_binary}

\begin{figure}
\centerline{\resizebox{0.45\textwidth}{!}{\input{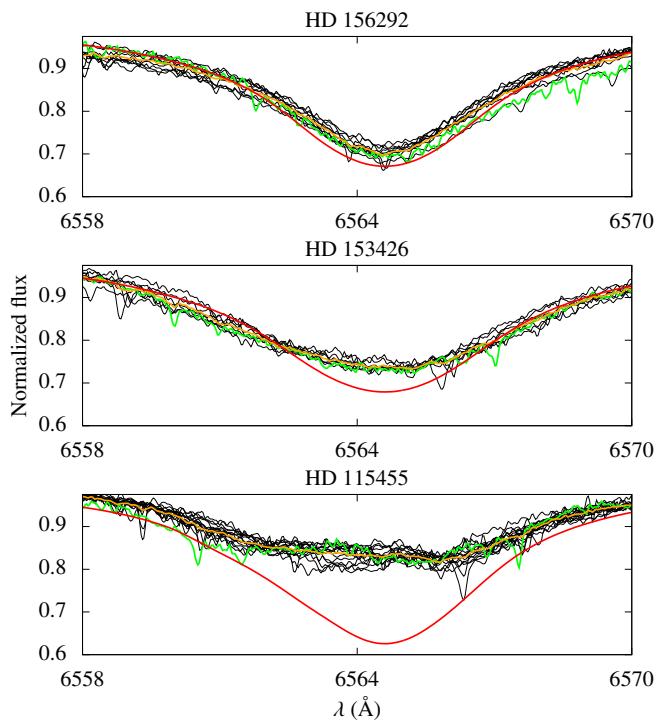}}}
\vspace*{0.8cm}
\caption{Effect of binarity on the H$\alpha$ profile of the SB2 systems in our sample: HD 156292, HD 153426, and HD 115455. Archival FEROS data are shown in black. Our observed spectrum for each star is shown in green. Best-fit CMFGEN model derived from the UV is shown in red. The observed spectra are shifted in $\lambda$ to match the line core of the model. The mean spectra among all the observations is shown in orange. The observed spectra are on average more intense from HD 156292 towards HD 115455.}
\label{hd156292_hd153426_hd115455_binary_halpha}
\end{figure}

As previously commented, three stars of our sample are spectroscopic double-lined binary systems: HD 156292, HD 153426, and HD 156292 (see Table \ref{sample_data}). In Fig. \ref{hd156292_hd153426_hd115455_binary_halpha}, we compare our models (derived from fitting the UV region) to multi-epoch FEROS spectra of HD 156292, HD 153426, and HD 115455 in the H$\alpha$ line. Apart from the spectra that are modeled in this paper for each of these stars (listed in Table \ref{sample_data}), the observed spectra in this figure are retrieved from the ESO Science Archive Facility (Phase 3). The dates of the observations are shown in Table \ref{espectros_feros_binary} with orbital phases calculated considering $T_{0}$ from our observations. The orbital periods $P_{orb}$ of HD 156292, HD 153426, and HD 115455 are, respectively, 4.94 days, 22.40 days, and 15.08 days \citep{sota14}. The orbital phases are fairly uniformly distributed from about 0.1 to 1.0.\par

As discussed in the previous section, we are able to provide an acceptable fit to H$\alpha$ in HD 156292, considering our parameters derived from the UV. On the other hand, we need to increase our UV mass-loss rates for HD 153426 and HD 115455 to fit their H$\alpha$ spectra. From Fig. \ref{hd156292_hd153426_hd115455_binary_halpha}, we see how the morphology of the H$\alpha$ profiles changes due to different orbital configuration of these binary systems. Our failure to reproduce the shape of both H$\alpha$ wings, in particular HD 156292, can be explained due to the binary nature of the system. Furthermore, the line core is also affected with the intensity varying by about 2-3\%. For HD 153426 and HD 115455, our model with UV $\dot{M}$ fails to reproduce the observed H$\alpha$ profiles regardless of the epoch time.\par

Still from Fig. \ref{hd156292_hd153426_hd115455_binary_halpha}, one sees that the averaged spectra, among all the observations for each of these three stars, and ours observed spectra are very similar. Therefore, our H$\alpha$ analysis (considering just our observed spectra) and conclusions presented in Sect. \ref{sec_mass_loss_rates_uv_visible} are unchanged. Binarity is not affecting the  H$\alpha$ line cores in a significant way in the SB2 stars of our sample.\par 

\begin{figure*}
\centerline{\resizebox{0.85\textwidth}{!}{\input{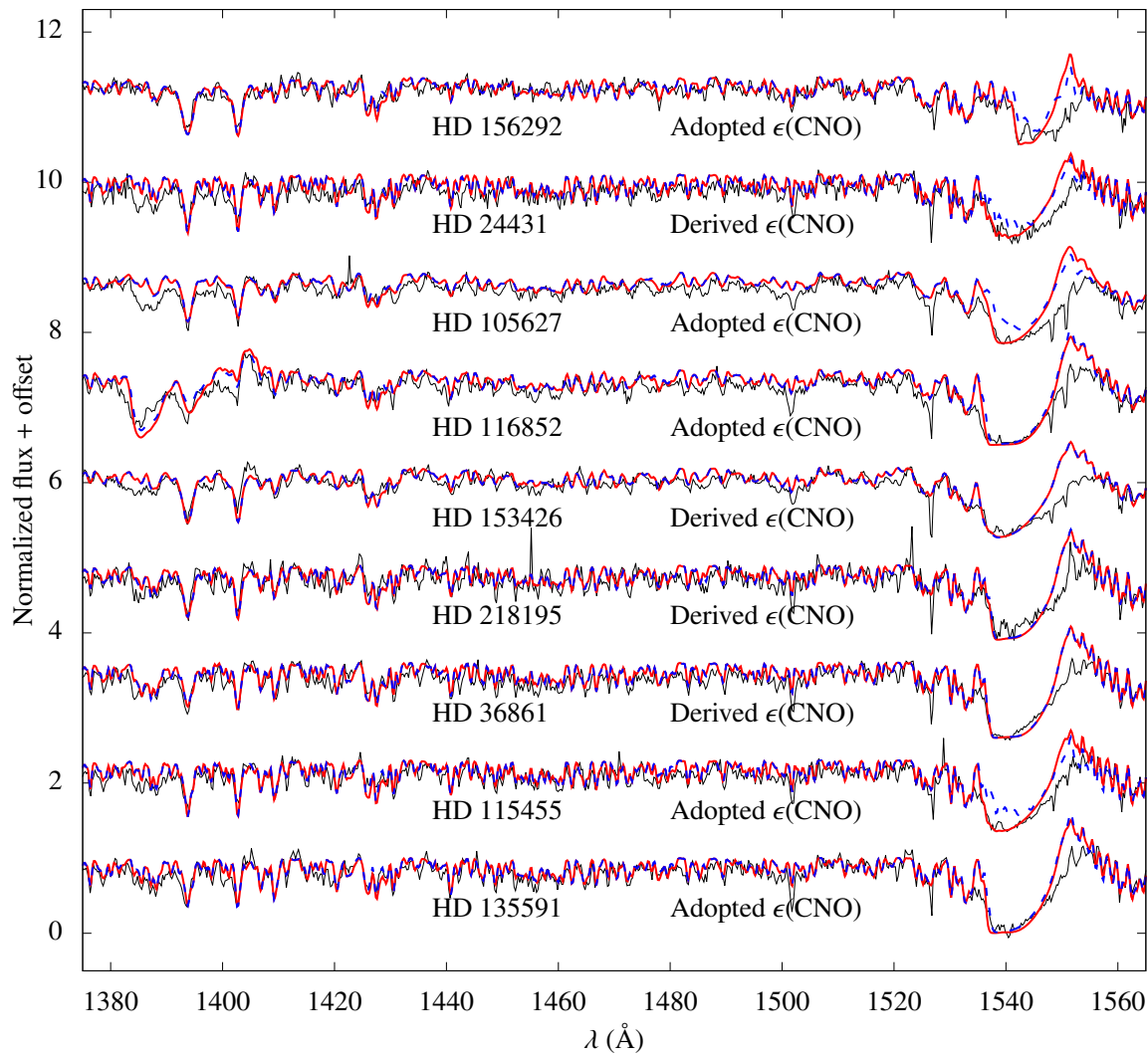}}}
\vspace*{1cm}
\caption{Final models for all our sample considering solar abundances (red). Re-computed models using $\epsilon_{\mathrm{CNO}}$ from \citet{martins_herve15} are in dashed blue. IUE data are in black. We indicate, right above each spectrum, whether $\epsilon_{\mathrm{CNO}}$ is a determination from \citet{martins_herve15} or an adoption based on their results (see text for details). We note that the highest effects in \ion{C}{IV} $\lambda$$\lambda$1548,1551 are due to the adopted $\epsilon_{\mathrm{CNO}}$.}\label{abundanciaCNO_SiIV_CIV}
\end{figure*}

Regarding the wind profiles in the UV region, we cannot perform a similar analysis since there is just one IUE/SWP spectrum for each of these three stars. However, we see that the observed profiles of \ion{Si}{IV} $\lambda$$\lambda$1394,1403 and \ion{C}{IV} $\lambda$$\lambda$1548,1551 are fairly similar in morphology in the overall sample (see, e.g., Fig. \ref{beta_law}). This reflects in the determination of the mass-loss rate from fitting these lines: we obtain a relatively uniform range of $\dot{M}$ values from $\sim 10^{-9}$ to $10^{-8}$ $\mathrm{M_\odot}$ yr\textsuperscript{-1}, with a trend for higher $\dot{M}$ for more luminous objects, as expected. HD 116852 presents a clear morphological exception, with developed P-Cygni profiles in \ion{Si}{IV} $\lambda$$\lambda$1394,1403. However, this object has the highest luminosity of our sample, being of luminosity class II-III, and the highest $\dot{M}$ derived from fitting the UV lines. Additionally, this star is not classified as a binary in the literature. So, this exception in morphology is very unlikely due to binary effects. We stress that our failure to reproduce the emission component of \ion{C}{IV} $\lambda$$\lambda$1548,1551 seems to be independent of the binary status our sample. For example, the final model for HD 156292 overestimates the observed emission component of this line practically as much as in the case of HD 105627 (no binary status). Thus, our results do not support that such an issue arises due to binary effects.\par

We are aware that disentangling of the observed spectra would be the most appropriate method to evaluate quantitatively possible effects of binarity on our results, but this technique is beyond the scope of this paper. Nevertheless, these results indicate that our mass-loss rates, derived from fitting the UV wind lines and the H$\alpha$ line, are unlikely biased by binary effects. Moreover, our models provide reasonable fits to the observed SED, and the effective temperature derived from the UV and the visible regions are in good agreement, showing a trend of higher $T_{\mathrm{eff}}$ towards the earlier stars (O8). This reflects in the analysis of the HR diagram, where our sample's loci are consistent with other results in the literature for O dwarfs, giants, and supergiants.\par

\subsection{Mass-loss rates: Degeneracies}
\label{sec_degeneracies}

Here, we investigate the effects of different stellar parameters on the UV line diagnostics for the mass-loss rate. The computational effort required by CMFGEN hinders degeneracy tests for several parameters simultaneously (e.g., CNO abundances, $T_{\mathrm{eff}}$, and luminosity). Therefore, we calculated the effects of a specific parameter on the $\dot{M}$ diagnostics, leaving all others fixed. We present results separately for CNO abundances ($\epsilon_{\mathrm{CNO}}$), $T_{\mathrm{eff}}$, and $\log (L_\star/\mathrm{L_\odot})$.

\subsubsection{CNO abundances}
\label{sec_deg_CNO_mdot}

Compared with main sequence objects, evolved O stars present an enhanced N/H together with depleted C/H and O/H ratios at their surfaces \citep[see, e.g.,][and references therein]{martins_herve15}. Deviations from the solar CNO abundance have a potential impact on the determination of the mass-loss rates for the stars of our sample. In particular, a very low C/H ratio affects directly the \ion{C}{IV} $\lambda$$\lambda$1548,1551 profile, decreasing its strength\footnote{ The Sobolev optical depth for \ion{C}{IV} is directly proportional to the product $\dot{M} q_{\ion{C}{IV}} \epsilon_{C}$, where $q_{\ion{C}{IV}}$ is the ion fraction of \ion{C}{IV} \citep[see, e.g.,][]{lamers99}. }. Thus, we proceeded as follows to evaluate abundance effects on our results:

\begin{enumerate}[label=(\roman*)]

\begin{table}[!h]
\caption{\label{tabela_mdot_CNO_derivadas}CNO affects on the mass-loss rates of all the sample. Revised mass-loss rates are denoted by $\dot{M}^{\mathrm{CNO}}_{\mathrm{revised}}$. The tick symbol (\cmark) stands for star without changes in $\dot{M}$. In the last column, the asterisk symbol (*) stands for adopted $\epsilon_{\mathrm{CNO}}$. The $\dot{M}_\mathrm{derived}$ were determined using solar $\epsilon_{\mathrm{CNO}}$ from \citet{grevesse10}: C/H = $2.7\e{-4}$, N/H = $0.7\e{-4}$, and O/H = $4.9\e{-4}$.}
\centering
\renewcommand{\arraystretch}{1.4}
\begin{adjustbox}{width=0.5\textwidth}
\begin{tabular}{lccccc}
\toprule
\toprule
Star       & $\dot{M}_\mathrm{derived}$\tablefootmark{a} ($\mathrm{M_\odot}$ yr\textsuperscript{-1}) & $\dot{M}^{\mathrm{CNO}}_{\mathrm{revised}}$ ($\mathrm{M_\odot}$ yr\textsuperscript{-1}) & C/H, N/H, O/H ($\e{-4}$)  \\ 
\midrule
HD 156292  & $1.5^{+4.0}_{-0.75}\e{-9}$ & $3.0^{+2.5}_{-1.5}\e{-9}$ & * 0.7, 1.6, 1.1  \\ 

HD 24431   & $2.5^{+7.5}_{-1.5}\e{-9}$ & $5.0^{+5.0}_{-2.5}\e{-9}$ & 1.3, 0.8, 3.9 \\ 

HD 105627  & $4.0^{+16.0}_{-2.5}\e{-9}$ & $1.0^{+2.0}_{-0.6}\e{-8}$ & * 0.7, 1.6, 1.1 \\ 

HD 116852  & $6.0^{+19.0}_{-4.5}\e{-8}$ & \cmark & * 0.7, 1.6, 1.1 \\ 

HD 153426  & $4.5^{+10.5}_{-2.5}\e{-9}$ & \cmark & 2.9, 1.0, 4.6  \\ 

HD 218195  & $1.0^{+2.5}_{-0.75}\e{-8}$ & \cmark & 2.0, 5.0, 4.6 \\ 

HD 36861   & $2.5^{+3.5}_{-2.2}\e{-8}$ & \cmark & 2.2, 1.5, 4.7 \\ 

HD 115455  & $5.0^{+10.0}_{-3.0}\e{-9}$ & $1.5^{+2.0}_{-1.0}\e{-8}$ & * 0.7, 1.6, 1.1 \\ 

HD 135591  & $2.0^{+6.0}_{-1.85}\e{-8}$ & \cmark & * 0.7, 1.6, 1.1 \\
\bottomrule

\end{tabular}
\end{adjustbox}

\tablefoot{
\tablefoottext{a}{Same notation as in Table \ref{resumo_resultados}.} 
}

\end{table}

\item \citet{martins_herve15} derived CNO abundances for 74 O stars of different luminosity classes through spectral modeling with CMFGEN. Our sample has four stars in common with them: HD 24431, HD 153426, HD 218195, and HD 36861. For these objects, we re-computed our final models (Table \ref{resumo_resultados}) using the values given by \citet{martins_herve15} for C/H, N/H, and O/H. 

\item For HD 156292, HD 105627, HD 116852, HD 115455, and HD 135591, for which there is no detailed abundance analysis, we chose to adopt the lowest C/H, highest N/H, and lowest O/H values among the late O giants found in \citet{martins_herve15}, namely: C/H = 0.7 $\times 10^{-4}$, N/H = 1.6 $\times 10^{-4}$, and O/H = 1.1 $\times 10^{-4}$. This approach is obviously conservative. Our main concern is about C/H, because it is expected to be lower than the solar value, impacting \ion{C}{IV} $\lambda$$\lambda$1548,1551.

\item We then analyzed the effects of non-solar CNO values in the modeling of the UV mass-loss diagnostics, revising, when necessary, the mass-loss rates to re-fit the observed spectra.

\item The revised mass-loss rates were evaluated again in the $\log (\dot{M})$ versus $\log (\dot{M}_{\mathrm{Vink}})$ diagram, allowing us to compare them with the previous $\dot{M}$ using solar abundances.

\end{enumerate}

\begin{figure}[!h]
\centerline{\resizebox{0.43\textwidth}{!}{\includegraphics[angle=90]{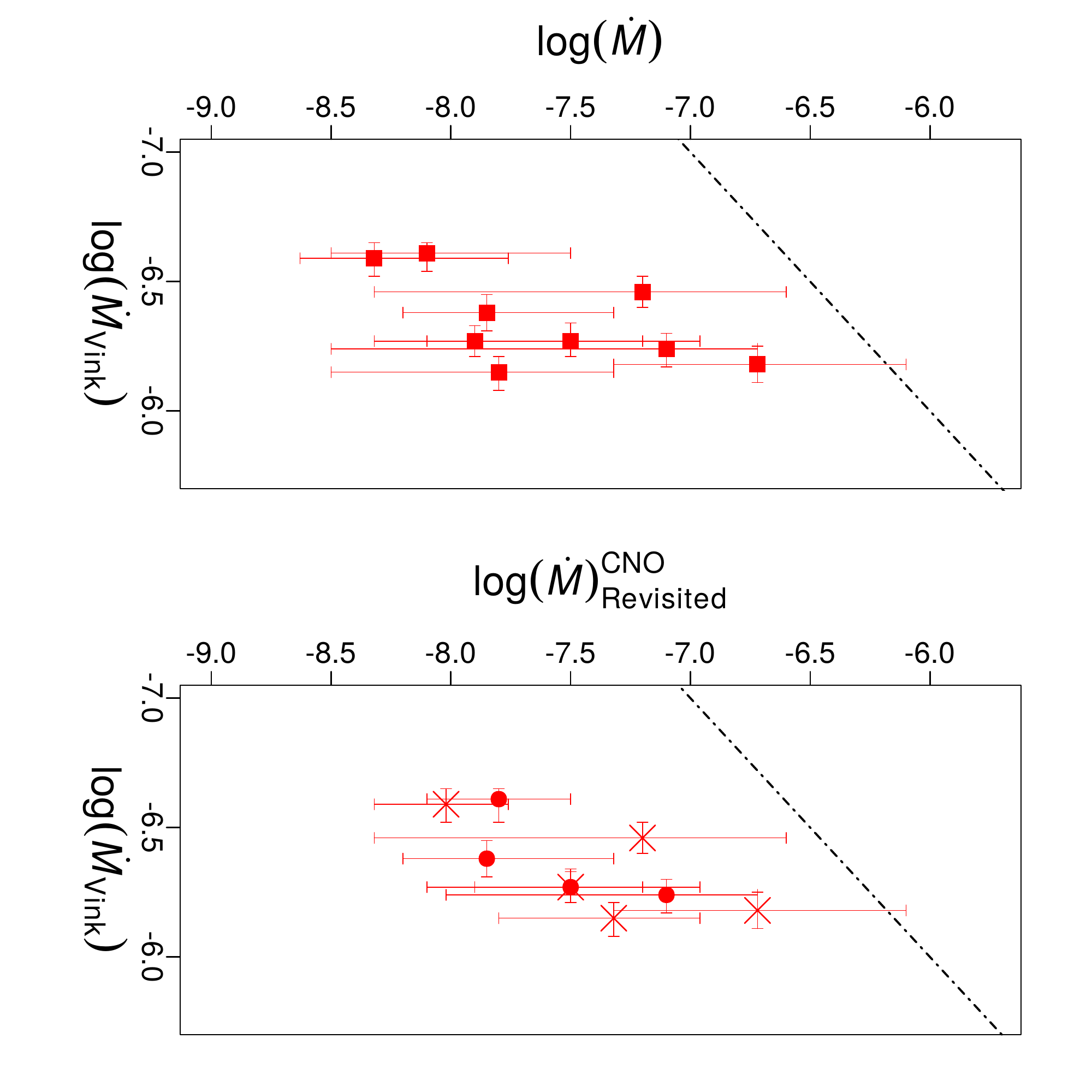}}}
\caption{Effects of $\epsilon_{\mathrm{CNO}}$ in the $\log(\dot{M})$ versus $\log(\dot{M}_{\mathrm{Vink}})$ diagram. On the left are previous results using the solar chemical abundance (shown in Fig. \ref{diagrama_mdot_anas_giants_super}). On the right are revised $\dot{M}$ taking different CNO values into account. Circles indicate stars with exact $\epsilon_{\mathrm{CNO}}$ from \citet{martins_herve15}. Crosses indicate stars for which we adopted extreme $\epsilon_{\mathrm{CNO}}$ based on the results of \citet{martins_herve15}. Our conclusions about weak wind in late O giants are not affected due to CNO abundances.}
\label{abundances_CNO_effect_mdot}
\end{figure}

\begin{figure}[!h]
\centerline{\resizebox{0.42\textwidth}{!}{\input{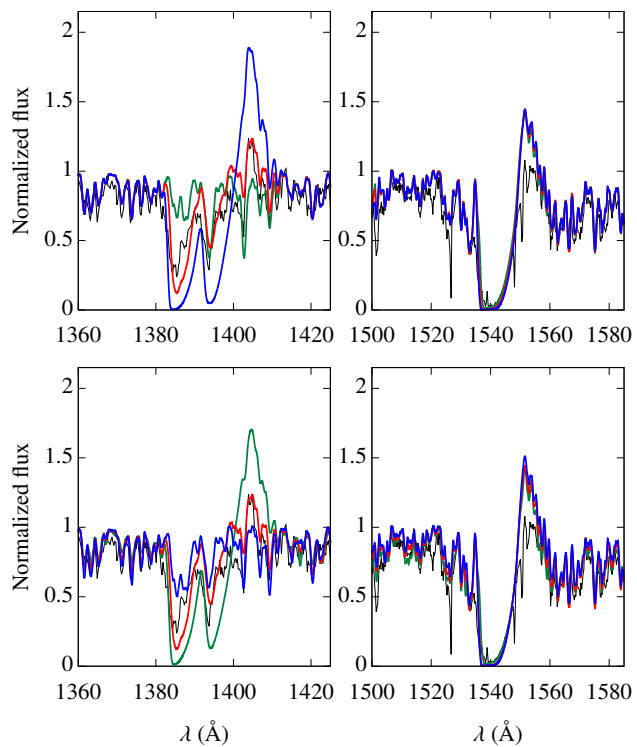}}}
\vspace*{1cm}
\caption{Effect of $T_{\mathrm{eff}}$ on the UV mass-loss diagnostics \ion{Si}{IV} $\lambda$$\lambda$1394,1403 and \ion{C}{IV} $\lambda$$\lambda$1548,1551. We show the analysis for HD 116852 (O8.5II-III) with $T_{\mathrm{eff}}^{\mathrm{UV}}$. In all panels, our final model is shown in red: $\dot{M} = 6.0\e{-8}$ $\mathrm{M_\odot}$ yr\textsuperscript{-1} and $T_{\mathrm{eff}} = 32500$ K. In the top panels are shown models with the limits on $\dot{M}$ ($1.5\e{-8}$ $\mathrm{M_\odot}$ yr\textsuperscript{-1} in green and $2.5\e{-7}$ $\mathrm{M_\odot}$ yr\textsuperscript{-1} in blue): fixed $T_{\mathrm{eff}}$ = 32500 K. In the bottom panels are shown models with limits on $T_{\mathrm{eff}}$ (30000 K in green and 34500 K in blue): fixed $\dot{M} = 6.0\e{-8}$ $\mathrm{M_\odot}$ yr\textsuperscript{-1}. We note that the error bars of $\dot{M}$ are conservative enough to account for the effects due to variation in $T_{\mathrm{eff}}$.}
\label{hd116852_deg_teff_mdot}
\end{figure}

\begin{figure}[!h]
\centerline{\resizebox{0.42\textwidth}{!}{\input{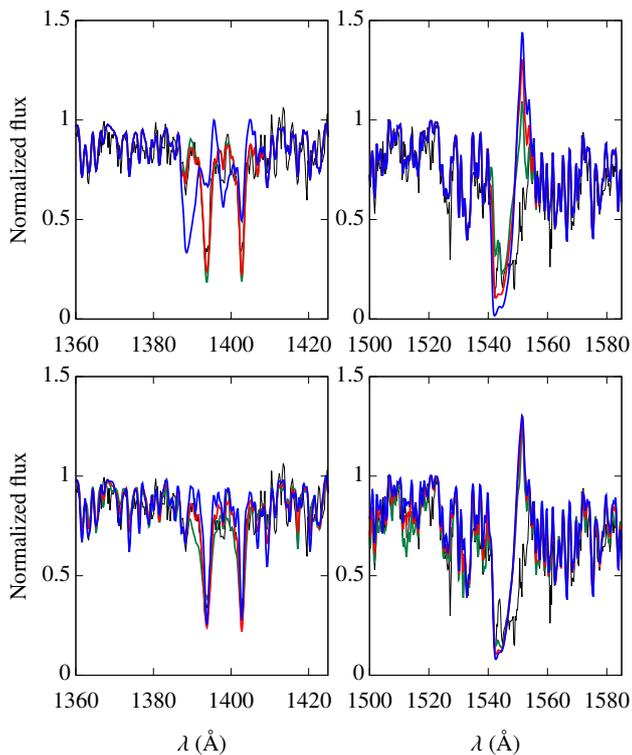}}}
\vspace*{1cm}
\caption{Same analysis as in Fig. \ref{hd116852_deg_teff_mdot}, but for HD 156292 (O9.7III). Final model (red) with $\dot{M} = 1.5\e{-9}$ $\mathrm{M_\odot}$ yr\textsuperscript{-1} and $T_{\mathrm{eff}} = 31000$ K. In the top panels are shown models with the limits on $\dot{M}$ ($7.5\e{-10}$ $\mathrm{M_\odot}$ yr\textsuperscript{-1} in green and $5.5\e{-9}$ $\mathrm{M_\odot}$ yr\textsuperscript{-1} in blue): fixed $T_{\mathrm{eff}}$ = 31000 K. In the bottom panels are shown models with the limits on $T_{\mathrm{eff}}$ (28000 K in green and 33000 K in blue): fixed $\dot{M} = 1.5\e{-9}$ $\mathrm{M_\odot}$ yr\textsuperscript{-1}. We note that the error bars of $T_{\mathrm{eff}}$ do not produce any significant changes in the considered mass-loss diagnostics.}
\label{hd156292_deg_teff_mdot}
\end{figure}

We show in Fig. \ref{abundanciaCNO_SiIV_CIV} the final models with solar abundances in comparison with the ones using the results from \citet{martins_herve15}. The considered $\epsilon_{\mathrm{CNO}}$ are listed in Table \ref{tabela_mdot_CNO_derivadas}. The \ion{C}{IV} $\lambda$$\lambda$1548,1551 profiles of four stars are affected by a lower carbon abundance, becoming weaker than the observed: HD 156292, HD 24431, HD 105627, and HD 115455. For these objects, we re-determined their $\dot{M}$, which are indicated in Table \ref{tabela_mdot_CNO_derivadas}. For three out these stars, we have adopted conservative low values of C/H. That is, these new inferred $\dot{M}$ are likely overestimated due to our adoption. The above results are summarized in Fig. \ref{abundances_CNO_effect_mdot}, where we present again the measured $\dot{M}$ and the ones according to \citet{vink00}. It is clear that our conclusions regarding weak winds in late O giants are not changed due to possible affects created by CNO abundances. Some points shift toward to the expected mass-loss rates, but the changes are minor overall.

\subsubsection{Effective temperature}
\label{sec_deg_teff_mdot}

Reliable effective temperatures are mandatory to derive the wind parameters. Depending on the model parameters range, changes in $T_{\mathrm{eff}}$ can modify the ionization structure in the wind \citep[e.g.,][]{austin11}, directly affecting the wind lines. Here, we analyze the effect of different $T_{\mathrm{eff}}$ values on the mass-loss rate. For simplicity, we present only results for HD 156292 (O9.7III) and HD 116852 (O8.5II-III). These two stars have extreme values of mass-loss rate among the objects of our sample. The $\dot{M}$ of HD 116852 is higher than the $\dot{M}$ of HD 156292 by a factor of 40.\par

In Figs. \ref{hd116852_deg_teff_mdot} (HD 116852) and \ref{hd156292_deg_teff_mdot} (HD 156292), we show the fits to the two main UV diagnostics of $\dot{M}$ considered in this paper. We present the synthetic wind profiles corresponding to the upper and lower limits of the mass-loss rates for these stars, with $T_{\mathrm{eff}}$ fixed (upper panels). For comparison, we also present the synthetic wind profiles corresponding to the upper and lower limits for $T_{\mathrm{eff}}$, with fixed $\dot{M}$ (lower panels). These values for the effective temperature were derived from the UV region, allowing us to develop a more conservative analysis since they have higher error bars than the ones found from fitting the \ion{He}{I-II} lines in the optical.\par

For HD 156292, we do not see significant changes in \ion{Si}{IV} $\lambda$$\lambda$1394,1403 and \ion{C}{IV} $\lambda$$\lambda$1548,1551 with different values for $T_{\mathrm{eff}}$. We note that in the case of HD 116852:

\begin{enumerate}[label=(\roman*)]

\item Variation in $T_{\mathrm{eff}}$ affects the \ion{Si}{IV} $\lambda$$\lambda$1394,1403 transition but not \ion{C}{IV} $\lambda$$\lambda$1548,1551 that is saturated.

\item We need to increase $\dot{M}$ to re-fit the \ion{Si}{IV} $\lambda$$\lambda$1394,1403 profiles.

\item This new (larger) mass-loss rate lies within the error bars already provided by our final $\dot{M}$: this can be seen from the amplitude of the profile variations corresponding to the error bars on $\dot{M}$.

\item The strongest effect on \ion{Si}{IV} $\lambda$$\lambda$1394,1403 is due to the lower limit of $T_{\mathrm{eff}}$. In this case, we need to decrease $\dot{M}$ to fit again the \ion{Si}{IV} $\lambda$$\lambda$1394,1403 profiles. 

\end{enumerate}

Therefore, we do not need to revise upward the UV mass-loss rate due to uncertainties in $T_{\mathrm{eff}}$. The error bars on the mass-loss rates are conservative enough to encompass uncertainties due to $T_{\mathrm{eff}}$.\par

\begin{figure}
\centerline{\resizebox{0.42\textwidth}{!}{\input{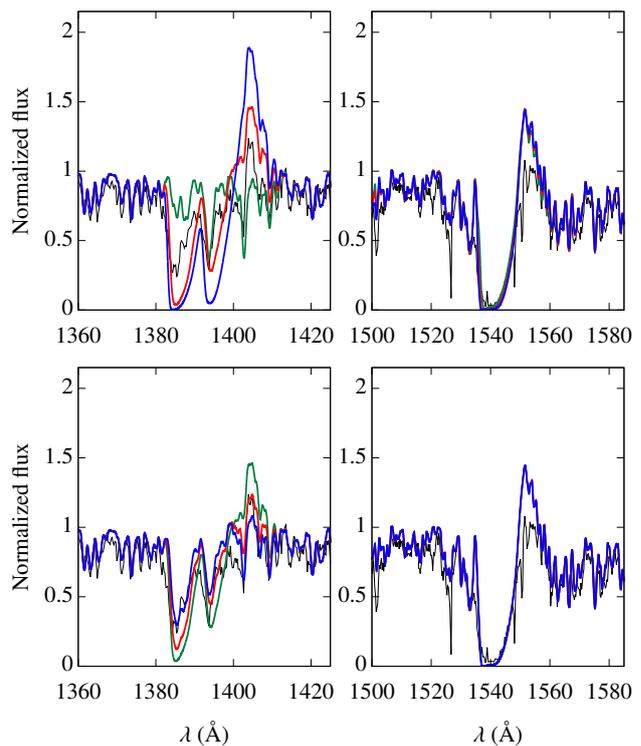}}}
\vspace*{1cm}
\caption{Stellar luminosity effect on the UV mass-loss diagnostics \ion{Si}{IV} $\lambda$$\lambda$1394,1403 and \ion{C}{IV} $\lambda$$\lambda$1548,1551. We show the analysis for HD 116852 with $T_{\mathrm{eff}}^{\mathrm{UV}}$. In all panels, the final model is shown in red: $\dot{M} = 6.0\e{-8}$ $\mathrm{M_\odot}$ yr\textsuperscript{-1} and $\log (L_\star/\mathrm{L_\odot})$ = 5.33. In the top panel are shown models with the limits on $\dot{M}$ ($1.5\e{-8}$ $\mathrm{M_\odot}$ yr\textsuperscript{-1} in green and $2.5\e{-7}$ $\mathrm{M_\odot}$ yr\textsuperscript{-1} in blue): fixed $\log (L_\star/\mathrm{L_\odot})$ = 5.33. In the bottom panel are shown models with the limits on log($L_\star/\mathrm{L_\odot}$) (5.13 in green and 5.53 in blue): fixed $\dot{M} = 6.0\e{-8}$ $\mathrm{M_\odot}$ yr\textsuperscript{-1}. We note that the error bars of $\dot{M}$ are conservative enough to account for effects due to variation in luminosity.}
\label{hd116852_deg_luminosity_mdot}
\end{figure}

\begin{figure}
\centerline{\resizebox{0.42\textwidth}{!}{\input{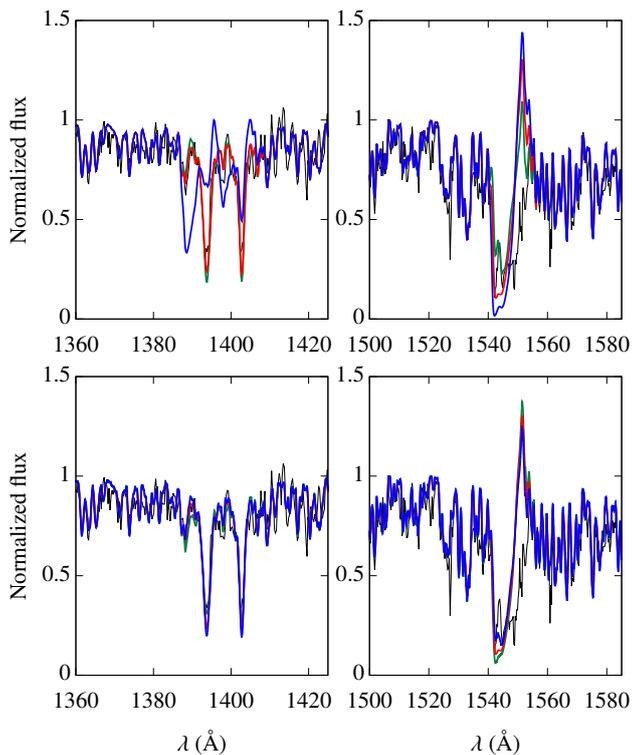}}}
\vspace*{1cm}
\caption{Same analysis as in Fig. \ref{hd116852_deg_luminosity_mdot}, but for HD 156292.  Final model (red) with $\dot{M} = 1.5\e{-9}$ $\mathrm{M_\odot}$ yr\textsuperscript{-1} and $\log (L_\star/\mathrm{L_\odot})$ = 5.12. In the top panel are shown models with the limits on $\dot{M}$ ($7.5\e{-10}$ $\mathrm{M_\odot}$ yr\textsuperscript{-1} in green and $5.5\e{-9}$ $\mathrm{M_\odot}$ yr\textsuperscript{-1} in blue): fixed $\log (L_\star/\mathrm{L_\odot})$ = 5.12. In the bottom panel are shown models with the limits on log($L_\star/\mathrm{L_\odot}$) (4.92 in green and 5.32 in blue): fixed $\dot{M} = 1.5\e{-9}$ $\mathrm{M_\odot}$ yr\textsuperscript{-1}. We note that the error bars of $\log (L_\star/\mathrm{L_\odot})$ do not produce any significant changes on the UV diagnostics for $\dot{M}$.}
\label{hd156292_deg_luminosity_mdot}
\end{figure}

\subsubsection{Stellar luminosity}
\label{sec_deg_luminosity_mdot}

It is well known that the luminosity class affects the \ion{Si}{IV} $\lambda$$\lambda$1394,1403 lines of O-type stars \citep[][]{walborn84,howarth89,pauldrach90}. These lines are almost photospheric in O9.5 dwarfs, while O supergiants show fully P-Cygni profiles in \ion{Si}{IV} $\lambda$$\lambda$1394,1403. Here, we evaluate the impact of our luminosity adoptions, from \citet{martins05_calibration}, on the derivation of $\dot{M}$. As discussed in Sect. \ref{sec_sed}, these values of $\log (L_\star/\mathrm{L_\odot})$ match well the observed SEDs for the most part of our sample, considering astrometric measures from \citet{gaia18}. For this purpose, we analyze the cases of HD 156292 and HD 116852 again. As discussed above, these stars possess, respectively, the lowest and highest density wind in our sample -- $\dot{M} = 1.5\e{-9}$ $\mathrm{M_\odot}$ yr\textsuperscript{-1} and $\dot{M} = 6.0\e{-8}$ $\mathrm{M_\odot}$ yr\textsuperscript{-1}. They are also extreme cases in luminosity, respectively: $\log (L_\star/\mathrm{L_\odot}) = 5.12$  (lowest value) and $\log (L_\star/\mathrm{L_\odot}) = 5.33$ (highest value).

We show the fits to the UV mass-loss diagnostics in Figs. \ref{hd116852_deg_luminosity_mdot} (HD 116852) and \ref{hd156292_deg_luminosity_mdot} (HD 156292). As in Figs. \ref{hd116852_deg_teff_mdot} and \ref{hd156292_deg_teff_mdot}, the models corresponding to the upper and lower limits of the mass-loss rates are shown in the upper panels. Here, they have the luminosity fixed for each case. In the lower panels, we present the synthetic profiles for models with variation in stellar luminosity within our adopted error bars of 0.2 in $\log (L_\star/\mathrm{L_\odot})$ and fixed $\dot{M}$.\par

For HD 156292, we do not see any significant changes in \ion{Si}{IV} $\lambda$$\lambda$1394,1403 and \ion{C}{IV} $\lambda$$\lambda$1548,1551 with different values of luminosity ($\log (L_\star/\mathrm{L_\odot})$ = 4.92 and 5.32). On the other hand, we note that in the case of HD 116852:

\begin{enumerate}[label=(\roman*)]

\item Variation in $\log (L_\star/\mathrm{L_\odot})$, from 5.08 to 5.48, affects the \ion{Si}{IV} $\lambda$$\lambda$1394,1403 transition but not \ion{C}{IV} $\lambda$$\lambda$1548,1551, which is saturated.

\item The strongest effect on \ion{Si}{IV} $\lambda$$\lambda$1394,1403 is due to the lower limit on $\log (L_\star/\mathrm{L_\odot})$. In this case, we need to decrease $\dot{M}$ in order to re-fit the \ion{Si}{IV} $\lambda$$\lambda$1394,1403 profiles. 

\item The upper limit on luminosity provides an acceptable fit to the \ion{Si}{IV} $\lambda$$\lambda$1394,1403 within the error bars provided to $\dot{M}$.

\end{enumerate}

Hence, as in the case of the effective temperature, we do not need to revise our UV mass-loss rates to higher values due to the uncertainties ($\pm$ 0.2 dex) in the luminosity.\par


\section{Conclusions}\label{conclusions}

We presented a quantitative analysis of nine late O giant stars (O8-O9.5III) using non-LTE atmosphere models computed with the code CMFGEN. We used archival high-resolution UV (IUE/SWP) and recent optical data (FEROS, NARVAL, and ESPADONS) to determine the main photospheric and wind parameters for each star of our sample. We analyzed the results in the context of the weak wind phenomenon. Our main findings and conclusions are summarized as follows:

\begin{enumerate}[label=(\roman*)]

\item Overall, we achieved good fits to the UV and optical data. The bulk of our sample has parameters in the following ranges: $\log (L_\star/\mathrm{L_\odot}) \sim 5.10-5.30$, $T_{\mathrm{eff}} \sim 30000-35000$ K, $\log(g) \sim 3.50-3.60$, $v_{\infty} \sim 2000-2400$ km s\textsuperscript{-1}, and $\dot{M} \sim 10^{-9}-10^{-8}$ $\mathrm{M_\odot}$ yr\textsuperscript{-1}.\par

\item It is reliable to derive the main physical properties of late O giants -- $T_{\mathrm{eff}}$, $R_\star$, $M_\star$, $\dot{M}$, and $v_{\infty}$ --  solely from UV spectrum (IUE/SWP data, $\sim 1100-2000$ {\AA}), considering that values for the bolometric luminosity and surface gravity are adopted. The main mass-loss diagnostics for O8-9.5 giants are \ion{Si}{IV} $\lambda$$\lambda$1394,1403 and \ion{C}{IV} $\lambda$$\lambda$1548,1551.

\item Overall, our model SEDs reproduce well the observed ones considering parallaxes from \citet{gaia18} and \citet{hipparcos07}. The highest discrepancies are explained by unreliable distance estimations and uncertainties in the ISM extinction. Thus, the calibrated $\log (L_\star/\mathrm{L_\odot})$ given by \citet{martins05_calibration} is a good assumption for late O giants.\par

\item We determined $T_{\mathrm{eff}}$ separately from the UV (\ion{Fe}{III-V} lines) and the optical (\ion{He}{I-II} lines). These independent measurements for $T_{\mathrm{eff}}$ agree well. For both of them, the expected tendency of higher values of $T_{\mathrm{eff}}$ was obtained, as we go from later (O9.5III) towards earlier (O8III) spectral types. \par

\item The emission component of \ion{C}{IV} $\lambda$$\lambda$1548,1551 is systematically overestimated by our models. We are able to improve the fits with $\xi^{min}_{t}$ up to 20-30 km s\textsuperscript{-1} together with a fuller accounting of atomic species. Our improved model in terms of species is not sufficient to explain this discrepancy. However, such higher values for the photospheric microturbulence prevent a self-consistent $T_{\mathrm{eff}}$ analysis from the UV and optical for the stars of our sample. Considering the adoption of $\xi^{min}_{t} = 10$ km s\textsuperscript{-1}, our tests show that the fits are only improved by using a considerably lower $\beta$ in the velocity law, namely, of $\sim 0.3$. These values are not common among O stars. Dwarfs usually present values close to unity while OB supergiants may present larger values, such as $\beta \gtrsim 3.0$. We consider that this very low $\beta$ value is likely due to the simple $\beta$ parameterization used in this paper, that is, they must be an artifact of our modeling assumption with CMFGEN. A deeper investigation regarding this issue is required.\par

\item We analyzed literature results for O dwarfs, giants, and supergiants with our ones in the HR diagram. The loci of these stars are different in the diagram. We estimated ages for the giants of our sample and confirmed that they are more evolved than dwarfs. The bulk of our sample has ages of $10^{6.7}-10^{6.8}$ yr and evolutionary masses of $M_{\mathrm{ZAM}} \sim 25-28$ $\mathrm{M_\odot}$. These values of mass are consistent with the recent findings of \citet{martins17}, which determined spectroscopic masses around $25-40$ $\mathrm{M_\odot}$ for O7-8 giants. Our spectroscopic masses ($M_{\star} \sim 16-26$ $\mathrm{M_\odot}$) are in marginal agreement with the evolutionary masses.\par

\item The mass-loss rates obtained from the UV analysis were compared with different theoretical works. Considering clumped $\dot{M}$, our values are considerably lower than the values predicted from \citet{vink00} by a factor of $\sim 0.9-2.3$ dex. The discrepancy is reduced to $\sim 0.4-1.7$ dex in comparison with our unclumped $\dot{M}$, still being severe since it reaches up to about two orders of magnitude. That is, late O giants also present weak winds, like late O dwarfs. The disagreement between our results and the predicted values by theory is lessened as we go from O9.5III to O8III objects. Thus, the region of $\log(L_\star/\mathrm{L_\odot}) \sim 5.2$ seems critical for the weak wind phenomenon.\par

\item We also compared our results with the theoretical $\dot{M}$ from \citet{lucy10_grid}. The predictions of \citet{lucy10_grid} are better in comparison with $\dot{M}_{\mathrm{Vink}}$ for low luminosity objects. The weak wind problem is considerably attenuated (where Vink's predictions are worst), with discrepancies up to $\sim$ 1.0 dex. On the other hand, they underestimate the mass-loss rates for OB supergiants (where Vink's predictions work well). The physical reason for these discrepancies should be investigated in detail, but it is beyond the scope of this paper. We verified that the region of $\log(L_\star/\mathrm{L_\odot}) \sim 5.2$ seems to be critical for both the predictions from \citet{vink00} and \citet{lucy10}.\par

\item The mass-loss rates predicted from \citet{vink00} largely overestimate the intensity in \ion{Si}{IV} $\lambda$$\lambda$1394,1403 for all our sample. These values of $\dot{M}$ also overestimate the \ion{C}{IV} $\lambda$$\lambda$1548,1551 for our stars with non-saturated profiles. For H$\alpha$, a good part of our sample (five out of nine stars) shows overestimated profiles using $\dot{M}_{\mathrm{Vink}}$. The mass loss from \citet{vink00} is able to fit the observed H$\alpha$ for three objects: HD 24431, HD 116852, and HD 115455.\par

\item We checked the fits to the H$\alpha$ profiles using our mass-loss rates inferred from the UV diagnostics. We found excellent fits for two objects (HD 156292 and HD 105627) and for another two stars the upper limit on $\dot{M}$ also provide reasonable fits (HD 116852 and HD 135591). However, for five objects, we could not fit H$\alpha$ even considering the UV upper limits on the $\dot{M}$: HD 24431, HD 218195, HD 36861, HD 153426, and HD 115455. Despite our efforts, in these stars our models fail to reproduce simultaneously all the wind diagnostics considered here (UV and optical).

\item Regarding the simultaneous fitting to the UV and H$\alpha$ spectra, weak winds in O8-9.5III are favored in comparison with the values from \citet{vink00}. We fit the UV and optical spectra of four out of nine stars of our sample. Models with $\dot{M}_{\mathrm{Vink}}$ are able to reproduce the H$\alpha$ data for three out of nine objects, but they fail to reproduce the UV wind lines for all our sample. We needed to revise upward the UV $\dot{M}$ to model the H$\alpha$ profiles in about half of our sample. This issue regarding the UV and optical modeling with CMFGEN is also found in works about O dwarfs. It is still an open question in the literature.

\item Our results show that effects of binarity on the UV and optical analyses are unlikely to change our conclusions about the weak wind phenomenon in late O giants. Moreover, we performed different tests to address the impact of stellar parameters on the mass-loss rates derivation from the UV wind lines. We also found that the depletion of C/H, caused by evolutionary effects in late O giants, is not enough to modify our conclusions regarding the weak winds. The same is valid for the effects of $T_{\mathrm{eff}}$ and $L_\star$ on the UV wind diagnostics.

\end{enumerate}

In conclusion, our results indicate the weak wind phenomenon in O8-9.5III stars. It is the first time that weak winds are found for spectral types other than O8-9.5V. Despite our efforts, we are not able to model at the same time both the UV wind diagnostic lines and the H$\alpha$ profile for all the stars of our sample. This issue could be solved by investigations regarding macroclumping implementation in the modeling with CMFGEN and potential H$\alpha$ variability (as observed in late OB supergiants) among late O giants. Apart from this problem, low $\dot{M}$ (weak winds) are favored to model the spectra (UV + optical regions) of late O dwarfs and giants in comparison with values provided by theory. In other words, the measured mass-loss rates of these stars are systematically lower than the predictions of \citet{vink00}. This is important as they are low luminosity O stars (latter spectral types), implying that the majority of the O-type stars must undergo a weak wind phase. Therefore, we suggest that the mass-loss recipe in the majority of modern stellar evolution codes must severely overestimate $\dot{M}$ during the H-burning phase. Further investigations are needed to evaluate the consequences of this in terms of physical parameters for massive stars (e.g., angular momentum and CNO surface abundances).\par


\bibliographystyle{aa} 

\begin{acknowledgements}

We would like to thank the anonymous referee for comments that helped to improve the paper. E. S. G. de Almeida thanks the ``Ville de Nice'' (Nice, France) for the financial support to this work through the ``Bourse Doctorale Olivier Chesneau'' during the period of 2016-2019. E. S. G. de Almeida thanks the CAPES Agency (Brazil) for the financial support to this work through MSc grant during the period of 2015-2016. E. S. G. de Almeida thanks A. Domiciano de Souza, A. Meilland, and P. Stee for reading the original manuscript. W. L. F. Marcolino acknowledges CNPq for the PQ grant (307152/2016-2). Based on observations collected at the European Southern Observatory under ESO programme 096.A-9027(A). Based on data obtained from the ESO Science Archive Facility under request numbers 389364, 389364, and 389372. Some of the data presented in this paper were obtained from the Mikulski Archive for Space Telescopes (MAST). STScI is operated by the Association of Universities for Research in Astronomy, Inc., under NASA contract NAS5-26555. Support for MAST for non-HST data is provided by the NASA Office of Space Science via grant NNX13AC07G and by other grants and contracts. This work has made use of data from the European Space Agency (ESA) mission {\it Gaia} (\url{https://www.cosmos.esa.int/gaia}), processed by the {\it Gaia} Data Processing and Analysis Consortium (DPAC, \url{https://www.cosmos.esa.int/web/gaia/dpac/consortium}). Funding for the DPAC has been provided by national institutions, in particular the institutions participating in the {\it Gaia} Multilateral Agreement. This research has made use of the SIMBAD database, operated at CDS, Strasbourg, France.

\end{acknowledgements}

\bibliography{./references} 

\begin{thebibliography}{86}
\expandafter\ifx\csname natexlab\endcsname\relax\def\natexlab#1{#1}\fi

\bibitem[{{Abbott} {et~al.}(2016){Abbott}, {Abbott}, {Abbott}, {Abernathy},
  {Acernese}, {Ackley}, {Adams}, {Adams}, {Addesso}, {Adhikari}, \&
  et~al.}]{abbott16}
{Abbott}, B.~P., {Abbott}, R., {Abbott}, T.~D., {et~al.} 2016, \apjl, 818, L22

\bibitem[{{Abbott}(1982)}]{abbott82}
{Abbott}, D.~C. 1982, \apj, 263, 723

\bibitem[{{Austin} \& {Prinja}(2011)}]{austin11}
{Austin}, M.~J. \& {Prinja}, R.~K. 2011, in IAU Symposium, Vol. 272, Active OB
  Stars: Structure, Evolution, Mass Loss, and Critical Limits, ed. C.~{Neiner},
  G.~{Wade}, G.~{Meynet}, \& G.~{Peters}, 600--601

\bibitem[{{Bailer-Jones} {et~al.}(2018){Bailer-Jones}, {Rybizki}, {Fouesneau},
  {Mantelet}, \& {Andrae}}]{bailer18}
{Bailer-Jones}, C.~A.~L., {Rybizki}, J., {Fouesneau}, M., {Mantelet}, G., \&
  {Andrae}, R. 2018, \aj, 156, 58

\bibitem[{{Berghoefer} {et~al.}(1996){Berghoefer}, {Schmitt}, \&
  {Cassinelli}}]{berghoefer96}
{Berghoefer}, T.~W., {Schmitt}, J.~H.~M.~M., \& {Cassinelli}, J.~P. 1996,
  \aaps, 118, 481

\bibitem[{{Bouret} {et~al.}(2012){Bouret}, {Hillier}, {Lanz}, \&
  {Fullerton}}]{bouret12}
{Bouret}, J.-C., {Hillier}, D.~J., {Lanz}, T., \& {Fullerton}, A.~W. 2012,
  \aap, 544, A67

\bibitem[{{Bouret} {et~al.}(2003){Bouret}, {Lanz}, {Hillier}, {Heap}, {Hubeny},
  {Lennon}, {Smith}, \& {Evans}}]{bouret03}
{Bouret}, J.-C., {Lanz}, T., {Hillier}, D.~J., {et~al.} 2003, \apj, 595, 1182

\bibitem[{{Bouret} {et~al.}(2013){Bouret}, {Lanz}, {Martins}, {Marcolino},
  {Hillier}, {Depagne}, \& {Hubeny}}]{bouret13}
{Bouret}, J.-C., {Lanz}, T., {Martins}, F., {et~al.} 2013, \aap, 555, A1

\bibitem[{{Cardelli} {et~al.}(1989){Cardelli}, {Clayton}, \&
  {Mathis}}]{cardelli89}
{Cardelli}, J.~A., {Clayton}, G.~C., \& {Mathis}, J.~S. 1989, \apj, 345, 245

\bibitem[{{Carneiro} {et~al.}(2018){Carneiro}, {Puls}, \&
  {Hoffmann}}]{carneiro18}
{Carneiro}, L.~P., {Puls}, J., \& {Hoffmann}, T.~L. 2018, \aap, 615, A4

\bibitem[{{Chlebowski} \& {Garmany}(1991)}]{chlebowski91}
{Chlebowski}, T. \& {Garmany}, C.~D. 1991, \apj, 368, 241

\bibitem[{{Crowther} {et~al.}(2006){Crowther}, {Lennon}, \&
  {Walborn}}]{crowther06}
{Crowther}, P.~A., {Lennon}, D.~J., \& {Walborn}, N.~R. 2006, \aap, 446, 279

\bibitem[{{Cutri} {et~al.}(2003){Cutri}, {Skrutskie}, {van Dyk}, {Beichman},
  {Carpenter}, {Chester}, {Cambresy}, {Evans}, {Fowler}, {Gizis}, {Howard},
  {Huchra}, {Jarrett}, {Kopan}, {Kirkpatrick}, {Light}, {Marsh}, {McCallon},
  {Schneider}, {Stiening}, {Sykes}, {Weinberg}, {Wheaton}, {Wheelock}, \&
  {Zacarias}}]{cutri03}
{Cutri}, R.~M., {Skrutskie}, M.~F., {van Dyk}, S., {et~al.} 2003, VizieR Online
  Data Catalog, 2246

\bibitem[{{de Mink} {et~al.}(2014){de Mink}, {Sana}, {Langer}, {Izzard}, \&
  {Schneider}}]{demink14}
{de Mink}, S.~E., {Sana}, H., {Langer}, N., {Izzard}, R.~G., \& {Schneider},
  F.~R.~N. 2014, \apj, 782, 7

\bibitem[{{Dean} \& {Bruhweiler}(1985)}]{dean85}
{Dean}, C.~A. \& {Bruhweiler}, F.~C. 1985, \apjs, 57, 133

\bibitem[{{Ducati}(2002)}]{ducati02}
{Ducati}, J.~R. 2002, VizieR Online Data Catalog, 2237

\bibitem[{{Ekstr{\"o}m} {et~al.}(2012){Ekstr{\"o}m}, {Georgy}, {Eggenberger},
  {Meynet}, {Mowlavi}, {Wyttenbach}, {Granada}, {Decressin}, {Hirschi},
  {Frischknecht}, {Charbonnel}, \& {Maeder}}]{ekstrom12}
{Ekstr{\"o}m}, S., {Georgy}, C., {Eggenberger}, P., {et~al.} 2012, \aap, 537,
  A146

\bibitem[{{ESA}(1997)}]{hipparcos97}
{ESA}, ed. 1997, ESA Special Publication, Vol. 1200, {The HIPPARCOS and TYCHO
  catalogues. Astrometric and photometric star catalogues derived from the ESA
  HIPPARCOS Space Astrometry Mission}

\bibitem[{{Gaia Collaboration} {et~al.}(2018){Gaia Collaboration}, {Brown},
  {Vallenari}, {Prusti}, {de Bruijne}, {Babusiaux}, \& {Bailer-Jones}}]{gaia18}
{Gaia Collaboration}, {Brown}, A.~G.~A., {Vallenari}, A., {et~al.} 2018, ArXiv
  e-prints [\eprint[arXiv]{1804.09365}]

\bibitem[{{Gaia Collaboration} {et~al.}(2016){Gaia Collaboration}, {Prusti},
  {de Bruijne}, {Brown}, {Vallenari}, {Babusiaux}, {Bailer-Jones}, {Bastian},
  {Biermann}, {Evans}, \& et~al.}]{gaia16}
{Gaia Collaboration}, {Prusti}, T., {de Bruijne}, J.~H.~J., {et~al.} 2016,
  \aap, 595, A1

\bibitem[{{Gehrels} \& {Razzaque}(2013)}]{gehrels13}
{Gehrels}, N. \& {Razzaque}, S. 2013, Frontiers of Physics, 8, 661

\bibitem[{{Gordon} {et~al.}(2018){Gordon}, {Gies}, {Schaefer}, {Huber},
  {Ireland}, \& {Hillier}}]{gordon18}
{Gordon}, K.~D., {Gies}, D.~R., {Schaefer}, G.~H., {et~al.} 2018, \apj, 869, 37

\bibitem[{{Grevesse} {et~al.}(2010){Grevesse}, {Asplund}, {Sauval}, \&
  {Scott}}]{grevesse10}
{Grevesse}, N., {Asplund}, M., {Sauval}, A.~J., \& {Scott}, P. 2010, \apss,
  328, 179

\bibitem[{{Grunhut} {et~al.}(2017){Grunhut}, {Wade}, {Neiner}, {Oksala},
  {Petit}, {Alecian}, {Bohlender}, {Bouret}, {Henrichs}, {Hussain},
  {Kochukhov}, \& {MiMeS Collaboration}}]{grunhut17}
{Grunhut}, J.~H., {Wade}, G.~A., {Neiner}, C., {et~al.} 2017, \mnras, 465, 2432

\bibitem[{{Hillier} {et~al.}(2003){Hillier}, {Lanz}, {Heap}, {Hubeny}, {Smith},
  {Evans}, {Lennon}, \& {Bouret}}]{hillier03}
{Hillier}, D.~J., {Lanz}, T., {Heap}, S.~R., {et~al.} 2003, \apj, 588, 1039

\bibitem[{{Hillier} \& {Miller}(1998)}]{hillier98}
{Hillier}, D.~J. \& {Miller}, D.~L. 1998, \apj, 496, 407

\bibitem[{{Hillier} \& {Miller}(1999)}]{hillier99}
{Hillier}, D.~J. \& {Miller}, D.~L. 1999, \apj, 519, 354

\bibitem[{{Holgado} {et~al.}(2018){Holgado}, {Sim{\'o}n-D{\'{\i}}az},
  {Barb{\'a}}, {Puls}, {Herrero}, {Castro}, {Garcia}, {Ma{\'{\i}}z
  Apell{\'a}niz}, {Negueruela}, \& {Sab{\'{\i}}n-Sanjuli{\'a}n}}]{holgado18}
{Holgado}, G., {Sim{\'o}n-D{\'{\i}}az}, S., {Barb{\'a}}, R.~H., {et~al.} 2018,
  \aap, 613, A65

\bibitem[{{Howarth} \& {Prinja}(1989)}]{howarth89}
{Howarth}, I.~D. \& {Prinja}, R.~K. 1989, \apjs, 69, 527

\bibitem[{{Howarth} {et~al.}(1997){Howarth}, {Siebert}, {Hussain}, \&
  {Prinja}}]{howarth97}
{Howarth}, I.~D., {Siebert}, K.~W., {Hussain}, G.~A.~J., \& {Prinja}, R.~K.
  1997, \mnras, 284, 265

\bibitem[{{Hubeny} \& {Lanz}(1995)}]{hubeny95}
{Hubeny}, I. \& {Lanz}, T. 1995, \apj, 439, 875

\bibitem[{{Kaufer} {et~al.}(1999){Kaufer}, {Stahl}, {Tubbesing},
  {N{\o}rregaard}, {Avila}, {Francois}, {Pasquini}, \& {Pizzella}}]{kaufer99}
{Kaufer}, A., {Stahl}, O., {Tubbesing}, S., {et~al.} 1999, The Messenger, 95, 8

\bibitem[{{Kaufer} {et~al.}(2000){Kaufer}, {Stahl}, {Tubbesing}, {Norregaard},
  {Avila}, {Francois}, {Pasquini}, \& {Pizzella}}]{kaufer00}
{Kaufer}, A., {Stahl}, O., {Tubbesing}, S., {et~al.} 2000, in \procspie, Vol.
  4008, Optical and IR Telescope Instrumentation and Detectors, ed. M.~{Iye} \&
  A.~F. {Moorwood}, 459--466

\bibitem[{{Keszthelyi} {et~al.}(2017){Keszthelyi}, {Wade}, \&
  {Petit}}]{keszthelyi17}
{Keszthelyi}, Z., {Wade}, G.~A., \& {Petit}, V. 2017, in IAU Symposium, Vol.
  329, The Lives and Death-Throes of Massive Stars, ed. J.~J. {Eldridge}, J.~C.
  {Bray}, L.~A.~S. {McClelland}, \& L.~{Xiao}, 250--254

\bibitem[{{Lamers} {et~al.}(1999){Lamers}, {Haser}, {de Koter}, \&
  {Leitherer}}]{lamers99}
{Lamers}, H.~J.~G.~L.~M., {Haser}, S., {de Koter}, A., \& {Leitherer}, C. 1999,
  \apj, 516, 872

\bibitem[{{Lanz} \& {Hubeny}(2003)}]{lanz03}
{Lanz}, T. \& {Hubeny}, I. 2003, \apjs, 146, 417

\bibitem[{{Lucy}(2010{\natexlab{a}})}]{lucy10_grid}
{Lucy}, L.~B. 2010{\natexlab{a}}, \aap, 524, A41

\bibitem[{{Lucy}(2010{\natexlab{b}})}]{lucy10}
{Lucy}, L.~B. 2010{\natexlab{b}}, \aap, 512, A33

\bibitem[{{Lucy} \& {Solomon}(1970)}]{lucy70}
{Lucy}, L.~B. \& {Solomon}, P.~M. 1970, \apj, 159, 879

\bibitem[{{Mahy} {et~al.}(2015){Mahy}, {Rauw}, {De Becker}, {Eenens}, \&
  {Flores}}]{mahy15}
{Mahy}, L., {Rauw}, G., {De Becker}, M., {Eenens}, P., \& {Flores}, C.~A. 2015,
  \aap, 577, A23

\bibitem[{{Ma{\'{\i}}z Apell{\'a}niz} {et~al.}(2008){Ma{\'{\i}}z
  Apell{\'a}niz}, {Alfaro}, \& {Sota}}]{maiz08}
{Ma{\'{\i}}z Apell{\'a}niz}, J., {Alfaro}, E.~J., \& {Sota}, A. 2008, arXiv
  e-prints [\eprint[arXiv]{0804.2553}]

\bibitem[{{Ma{\'{\i}}z Apell{\'a}niz} \& {Barb{\'a}}(2018)}]{maiz18}
{Ma{\'{\i}}z Apell{\'a}niz}, J. \& {Barb{\'a}}, R.~H. 2018, \aap, 613, A9

\bibitem[{{Ma{\'{\i}}z Apell{\'a}niz} {et~al.}(2013){Ma{\'{\i}}z
  Apell{\'a}niz}, {Sota}, {Morrell}, {Barb{\'a}}, {Walborn}, {Alfaro}, {Gamen},
  {Arias}, \& {Gallego Calvente}}]{maiz13}
{Ma{\'{\i}}z Apell{\'a}niz}, J., {Sota}, A., {Morrell}, N.~I., {et~al.} 2013,
  in Massive Stars: From alpha to Omega, 198

\bibitem[{{Ma{\'{\i}}z-Apell{\'a}niz}
  {et~al.}(2004){Ma{\'{\i}}z-Apell{\'a}niz}, {Walborn}, {Galu{\'e}}, \&
  {Wei}}]{maiz04}
{Ma{\'{\i}}z-Apell{\'a}niz}, J., {Walborn}, N.~R., {Galu{\'e}}, H.~{\'A}., \&
  {Wei}, L.~H. 2004, \apjs, 151, 103

\bibitem[{{Majaess} {et~al.}(2016){Majaess}, {Turner}, {D{\'e}k{\'a}ny},
  {Minniti}, \& {Gieren}}]{majaess16}
{Majaess}, D., {Turner}, D., {D{\'e}k{\'a}ny}, I., {Minniti}, D., \& {Gieren},
  W. 2016, \aap, 593, A124

\bibitem[{{Marcolino} {et~al.}(2009){Marcolino}, {Bouret}, {Martins},
  {Hillier}, {Lanz}, \& {Escolano}}]{marcolino09}
{Marcolino}, W.~L.~F., {Bouret}, J.-C., {Martins}, F., {et~al.} 2009, \aap,
  498, 837

\bibitem[{{Martins}(2011)}]{martins11}
{Martins}, F. 2011, Bulletin de la Societe Royale des Sciences de Liege, 80, 29

\bibitem[{{Martins} {et~al.}(2015{\natexlab{a}}){Martins}, {Herv{\'e}},
  {Bouret}, {Marcolino}, {Wade}, {Neiner}, {Alecian}, {Grunhut}, \&
  {Petit}}]{martins_herve15}
{Martins}, F., {Herv{\'e}}, A., {Bouret}, J.-C., {et~al.} 2015{\natexlab{a}},
  \aap, 575, A34

\bibitem[{{Martins} \& {Hillier}(2012)}]{martins12}
{Martins}, F. \& {Hillier}, D.~J. 2012, \aap, 545, A95

\bibitem[{{Martins} {et~al.}(2012){Martins}, {Mahy}, {Hillier}, \&
  {Rauw}}]{martins_mahy12}
{Martins}, F., {Mahy}, L., {Hillier}, D.~J., \& {Rauw}, G. 2012, \aap, 538, A39

\bibitem[{{Martins} {et~al.}(2015{\natexlab{b}}){Martins}, {Marcolino},
  {Hillier}, {Donati}, \& {Bouret}}]{martins_marcolino15}
{Martins}, F., {Marcolino}, W., {Hillier}, D.~J., {Donati}, J.-F., \& {Bouret},
  J.-C. 2015{\natexlab{b}}, \aap, 574, A142

\bibitem[{{Martins} \& {Plez}(2006)}]{martins06}
{Martins}, F. \& {Plez}, B. 2006, \aap, 457, 637

\bibitem[{{Martins} {et~al.}(2005{\natexlab{a}}){Martins}, {Schaerer}, \&
  {Hillier}}]{martins05_calibration}
{Martins}, F., {Schaerer}, D., \& {Hillier}, D.~J. 2005{\natexlab{a}}, \aap,
  436, 1049

\bibitem[{{Martins} {et~al.}(2004){Martins}, {Schaerer}, {Hillier}, \&
  {Heydari-Malayeri}}]{martins04}
{Martins}, F., {Schaerer}, D., {Hillier}, D.~J., \& {Heydari-Malayeri}, M.
  2004, \aap, 420, 1087

\bibitem[{{Martins} {et~al.}(2005{\natexlab{b}}){Martins}, {Schaerer},
  {Hillier}, {Meynadier}, {Heydari-Malayeri}, \&
  {Walborn}}]{martins05_weakwinds}
{Martins}, F., {Schaerer}, D., {Hillier}, D.~J., {et~al.} 2005{\natexlab{b}},
  \aap, 441, 735

\bibitem[{{Martins} {et~al.}(2017){Martins}, {Sim{\'o}n-D{\'{\i}}az},
  {Barb{\'a}}, {Gamen}, \& {Ekstr{\"o}m}}]{martins17}
{Martins}, F., {Sim{\'o}n-D{\'{\i}}az}, S., {Barb{\'a}}, R.~H., {Gamen}, R.~C.,
  \& {Ekstr{\"o}m}, S. 2017, \aap, 599, A30

\bibitem[{{Mayne} \& {Naylor}(2008)}]{mayne08}
{Mayne}, N.~J. \& {Naylor}, T. 2008, \mnras, 386, 261

\bibitem[{{Meynet} {et~al.}(2015){Meynet}, {Chomienne}, {Ekstr{\"o}m},
  {Georgy}, {Granada}, {Groh}, {Maeder}, {Eggenberger}, {Levesque}, \&
  {Massey}}]{meynet15}
{Meynet}, G., {Chomienne}, V., {Ekstr{\"o}m}, S., {et~al.} 2015, \aap, 575, A60

\bibitem[{{Meynet} {et~al.}(2011){Meynet}, {Georgy}, {Hirschi}, {Maeder},
  {Massey}, {Przybilla}, \& {Nieva}}]{meynet11}
{Meynet}, G., {Georgy}, C., {Hirschi}, R., {et~al.} 2011, Bulletin de la
  Societe Royale des Sciences de Liege, 80, 266

\bibitem[{{Mokiem} {et~al.}(2005){Mokiem}, {de Koter}, {Puls}, {Herrero},
  {Najarro}, \& {Villamariz}}]{mokiem05}
{Mokiem}, M.~R., {de Koter}, A., {Puls}, J., {et~al.} 2005, \aap, 441, 711

\bibitem[{{Muijres} {et~al.}(2012){Muijres}, {Vink}, {de Koter}, {M{\"u}ller},
  \& {Langer}}]{muijres12}
{Muijres}, L.~E., {Vink}, J.~S., {de Koter}, A., {M{\"u}ller}, P.~E., \&
  {Langer}, N. 2012, \aap, 537, A37

\bibitem[{{Naz{\'e}}(2009)}]{naze09}
{Naz{\'e}}, Y. 2009, \aap, 506, 1055

\bibitem[{{Oskinova} {et~al.}(2007){Oskinova}, {Hamann}, \&
  {Feldmeier}}]{oskinova07}
{Oskinova}, L.~M., {Hamann}, W.-R., \& {Feldmeier}, A. 2007, \aap, 476, 1331

\bibitem[{{Patriarchi} {et~al.}(2001){Patriarchi}, {Morbidelli}, {Perinotto},
  \& {Barbaro}}]{patriarchi01}
{Patriarchi}, P., {Morbidelli}, L., {Perinotto}, M., \& {Barbaro}, G. 2001,
  \aap, 372, 644

\bibitem[{{Pauldrach} {et~al.}(1990){Pauldrach}, {Kudritzki}, {Puls}, \&
  {Butler}}]{pauldrach90}
{Pauldrach}, A.~W.~A., {Kudritzki}, R.~P., {Puls}, J., \& {Butler}, K. 1990,
  \aap, 228, 125

\bibitem[{{Puls} {et~al.}(2009){Puls}, {Sundqvist}, {Najarro}, \&
  {Hanson}}]{puls09}
{Puls}, J., {Sundqvist}, J.~O., {Najarro}, F., \& {Hanson}, M.~M. 2009, in
  American Institute of Physics Conference Series, Vol. 1171, American
  Institute of Physics Conference Series, ed. I.~{Hubeny}, J.~M. {Stone},
  K.~{MacGregor}, \& K.~{Werner}, 123--135

\bibitem[{{Puls} {et~al.}(2005){Puls}, {Urbaneja}, {Venero}, {Repolust},
  {Springmann}, {Jokuthy}, \& {Mokiem}}]{puls05}
{Puls}, J., {Urbaneja}, M.~A., {Venero}, R., {et~al.} 2005, \aap, 435, 669

\bibitem[{{Puls} {et~al.}(2008){Puls}, {Vink}, \& {Najarro}}]{puls08}
{Puls}, J., {Vink}, J.~S., \& {Najarro}, F. 2008, \aapr, 16, 209

\bibitem[{{Rauw} {et~al.}(2015){Rauw}, {Naz{\'e}}, {Wright}, {Drake},
  {Guarcello}, {Prinja}, {Peck}, {Albacete Colombo}, {Herrero}, {Kobulnicky},
  {Sciortino}, \& {Vink}}]{rauw15}
{Rauw}, G., {Naz{\'e}}, Y., {Wright}, N.~J., {et~al.} 2015, \apjs, 221, 1

\bibitem[{{Repolust} {et~al.}(2004){Repolust}, {Puls}, \&
  {Herrero}}]{repolust04}
{Repolust}, T., {Puls}, J., \& {Herrero}, A. 2004, \aap, 415, 349

\bibitem[{{Sana} {et~al.}(2014){Sana}, {Le Bouquin}, {Lacour}, {Berger},
  {Duvert}, {Gauchet}, {Norris}, {Olofsson}, {Pickel}, {Zins}, {Absil}, {de
  Koter}, {Kratter}, {Schnurr}, \& {Zinnecker}}]{sana14}
{Sana}, H., {Le Bouquin}, J.-B., {Lacour}, S., {et~al.} 2014, \apjs, 215, 15

\bibitem[{{Sana} {et~al.}(2006){Sana}, {Rauw}, {Naz{\'e}}, {Gosset}, \&
  {Vreux}}]{sana06}
{Sana}, H., {Rauw}, G., {Naz{\'e}}, Y., {Gosset}, E., \& {Vreux}, J.-M. 2006,
  \mnras, 372, 661

\bibitem[{{Schr{\"o}der} {et~al.}(2004){Schr{\"o}der}, {Kaper}, {Lamers}, \&
  {Brown}}]{schroder04}
{Schr{\"o}der}, S.~E., {Kaper}, L., {Lamers}, H.~J.~G.~L.~M., \& {Brown},
  A.~G.~A. 2004, \aap, 428, 149

\bibitem[{{Sembach} \& {Savage}(1994)}]{sembach94}
{Sembach}, K.~R. \& {Savage}, B.~D. 1994, \apj, 431, 201

\bibitem[{{Sota} {et~al.}(2014){Sota}, {Ma{\'{\i}}z Apell{\'a}niz}, {Morrell},
  {Barb{\'a}}, {Walborn}, {Gamen}, {Arias}, \& {Alfaro}}]{sota14}
{Sota}, A., {Ma{\'{\i}}z Apell{\'a}niz}, J., {Morrell}, N.~I., {et~al.} 2014,
  \apjs, 211, 10

\bibitem[{{Sundqvist} \& {Puls}(2018)}]{sundqvist18}
{Sundqvist}, J.~O. \& {Puls}, J. 2018, \aap, 619, A59

\bibitem[{{Sundqvist} {et~al.}(2011){Sundqvist}, {Puls}, {Feldmeier}, \&
  {Owocki}}]{sundqvist11}
{Sundqvist}, J.~O., {Puls}, J., {Feldmeier}, A., \& {Owocki}, S.~P. 2011, \aap,
  528, A64

\bibitem[{{Sundqvist} {et~al.}(2014){Sundqvist}, {Puls}, \&
  {Owocki}}]{sundqvist14}
{Sundqvist}, J.~O., {Puls}, J., \& {Owocki}, S.~P. 2014, \aap, 568, A59

\bibitem[{{van Leeuwen}(2007)}]{hipparcos07}
{van Leeuwen}, F. 2007, \aap, 474, 653

\bibitem[{{Vink} {et~al.}(1999){Vink}, {de Koter}, \& {Lamers}}]{vink99}
{Vink}, J.~S., {de Koter}, A., \& {Lamers}, H.~J.~G.~L.~M. 1999, \aap, 350, 181

\bibitem[{{Vink} {et~al.}(2000){Vink}, {de Koter}, \& {Lamers}}]{vink00}
{Vink}, J.~S., {de Koter}, A., \& {Lamers}, H.~J.~G.~L.~M. 2000, \aap, 362, 295

\bibitem[{{Vink} {et~al.}(2001){Vink}, {de Koter}, \& {Lamers}}]{vink01}
{Vink}, J.~S., {de Koter}, A., \& {Lamers}, H.~J.~G.~L.~M. 2001, \aap, 369, 574

\bibitem[{{Wade} {et~al.}(2016){Wade}, {Neiner}, {Alecian}, {Grunhut}, {Petit},
  {de Batz}, {Bohlender}, {Cohen}, {Henrichs}, {Kochukhov}, {Landstreet},
  {Manset}, {Martins}, {Mathis}, {Oksala}, {Owocki}, {Rivinius}, {Shultz},
  {Sundqvist}, {Townsend}, {ud-Doula}, {Bouret}, {Braithwaite}, {Briquet},
  {Carciofi}, {David-Uraz}, {Folsom}, {Fullerton}, {Leroy}, {Marcolino},
  {Moffat}, {Naz{\'e}}, {Louis}, {Auri{\`e}re}, {Bagnulo}, {Bailey},
  {Barb{\'a}}, {Blaz{\`e}re}, {B{\"o}hm}, {Catala}, {Donati}, {Ferrario},
  {Harrington}, {Howarth}, {Ignace}, {Kaper}, {L{\"u}ftinger}, {Prinja},
  {Vink}, {Weiss}, \& {Yakunin}}]{wade16}
{Wade}, G.~A., {Neiner}, C., {Alecian}, E., {et~al.} 2016, \mnras, 456, 2

\bibitem[{{Walborn} \& {Panek}(1984)}]{walborn84}
{Walborn}, N.~R. \& {Panek}, R.~J. 1984, \apjl, 280, L27

\bibitem[{{Wegner}(1994)}]{wegner94}
{Wegner}, W. 1994, \mnras, 270, 229

\bibitem[{{Wegner}(2003)}]{wegner03}
{Wegner}, W. 2003, Astronomische Nachrichten, 324, 219

\end{thebibliography}


\begin{appendix}

\onecolumn

\section{Final models: Ultraviolet}\label{appendix_uv}

\begin{figure}[!h]
\centerline{\resizebox{1.0\textwidth}{!}{\input{./fig_hd156292_uv.tex}}}
\caption{Final CMFGEN model with $T_{\mathrm{eff}}^{\mathrm{UV}}$ (red) for HD 156292 in the UV. The IUE/SWP spectrum is shown in black. Model parameters are listed in Table \ref{resumo_resultados}.}
\label{final_model_uv_hd156292}
\end{figure}

\begin{figure}[!h]
\centerline{\resizebox{1.0\textwidth}{!}{\input{./fig_hd24431_uv.tex}}}
\caption{Final CMFGEN model with $T_{\mathrm{eff}}^{\mathrm{UV}}$ (red) for HD 24431 in the UV. The IUE/SWP spectrum is shown in black. Model parameters are listed in Table \ref{resumo_resultados}.}
\label{final_model_uv_hd24431}
\end{figure}

\begin{figure}[!h]
\centerline{\resizebox{1.0\textwidth}{!}{\input{./fig_hd105627_uv.tex}}}
\caption{Final CMFGEN model with $T_{\mathrm{eff}}^{\mathrm{UV}}$ (red) for HD 105627 in the UV. The IUE/SWP spectrum is shown in black. Model parameters are listed in Table \ref{resumo_resultados}.}
\label{final_model_uv_hd105627}
\end{figure}

\begin{figure}[!h]
\centerline{\resizebox{1.0\textwidth}{!}{\input{./fig_hd116852_uv.tex}}}
\caption{Final CMFGEN model with $T_{\mathrm{eff}}^{\mathrm{UV}}$ (red) for HD 116852 in the UV. The IUE/SWP spectrum is shown in black. Model parameters are listed in Table \ref{resumo_resultados}.}
\label{final_model_uv_hd116852}
\end{figure}

\begin{figure}[!h]
\centerline{\resizebox{1.0\textwidth}{!}{\input{./fig_hd153426_uv.tex}}}
\caption{Final CMFGEN model with $T_{\mathrm{eff}}^{\mathrm{UV}}$ (red) for HD 153426 in the UV. The IUE/SWP spectrum is shown in black. Model parameters are listed in Table \ref{resumo_resultados}.}
\label{final_model_uv_hd153426}
\end{figure}

\begin{figure}[!h]
\centerline{\resizebox{1.0\textwidth}{!}{\input{./fig_hd218195_uv.tex}}}
\caption{Final CMFGEN model with $T_{\mathrm{eff}}^{\mathrm{UV}}$ (red) for HD 218195 in the UV. The IUE/SWP spectrum is shown in black. Model parameters are listed in Table \ref{resumo_resultados}.}
\label{final_model_uv_hd218195}
\end{figure}

\begin{figure}[!h]
\centerline{\resizebox{1.0\textwidth}{!}{\input{./fig_hd36861_uv.tex}}}
\caption{Final CMFGEN model with $T_{\mathrm{eff}}^{\mathrm{UV}}$ (red) for HD 36861 in the UV. The IUE/SWP spectrum is shown in black. Model parameters are listed in Table \ref{resumo_resultados}.}
\label{final_model_uv_hd36861}
\end{figure}

\begin{figure}[!h]
\centerline{\resizebox{1.0\textwidth}{!}{\input{./fig_hd115455_uv.tex}}}
\caption{Final CMFGEN model with $T_{\mathrm{eff}}^{\mathrm{UV}}$ (red) for HD 115455 in the UV. The IUE/SWP spectrum is shown in black. Model parameters are listed in Table \ref{resumo_resultados}.}
\label{final_model_uv_hd115455}
\end{figure}

\begin{figure}[!h]
\centerline{\resizebox{1.0\textwidth}{!}{\input{./fig_hd135591_uv.tex}}}
\caption{Final CMFGEN model with $T_{\mathrm{eff}}^{\mathrm{UV}}$ (red) for HD 135591 in the UV. The IUE/SWP spectrum is shown in black. Model parameters are listed in Table \ref{resumo_resultados}.}
\label{final_model_uv_hd135591}
\end{figure}

\clearpage

\section{Final models: Optical}\label{appendix_optical}

\begin{figure}[!h]
\centerline{\resizebox{1.0\textwidth}{!}{\input{./fig_hd156292_optical.tex}}}
\caption{Final CMFGEN model with $T_{\mathrm{eff}}^{\mathrm{UV}}$ (red) for HD 156292 in the optical. The FEROS spectrum is shown in black. Model parameters are listed in Table \ref{resumo_resultados}.}
\label{final_model_optical_hd156292}
\end{figure}

\begin{figure}[!h]
\centerline{\resizebox{1.0\textwidth}{!}{\input{./fig_hd24431_optical.tex}}}
\caption{Final CMFGEN model with $T_{\mathrm{eff}}^{\mathrm{UV}}$ (red) for HD 24431 in the optical. The ESPADONS spectrum is shown in black. Model parameters are listed in Table \ref{resumo_resultados}.}
\label{final_model_optical_hd24431}
\end{figure}

\begin{figure}[!h]
\centerline{\resizebox{1.0\textwidth}{!}{\input{./fig_hd105627_optical.tex}}}
\caption{Final CMFGEN model with $T_{\mathrm{eff}}^{\mathrm{UV}}$ (red) for HD 105627 in the optical. The FEROS spectrum is shown in black. Model parameters are listed in Table \ref{resumo_resultados}.}
\label{final_model_optical_hd105627}
\end{figure}

\begin{figure}[!h]
\centerline{\resizebox{1.0\textwidth}{!}{\input{./fig_hd116852_optical.tex}}}
\caption{Final CMFGEN model with $T_{\mathrm{eff}}^{\mathrm{UV}}$ (red) for HD 116852 in the optical. The FEROS spectrum is shown in black. Model parameters are listed in Table \ref{resumo_resultados}.}
\label{final_model_optical_hd116852}
\end{figure}

\begin{figure}[!h]
\centerline{\resizebox{1.0\textwidth}{!}{\input{./fig_hd153426_optical.tex}}}
\caption{Final CMFGEN model with $T_{\mathrm{eff}}^{\mathrm{UV}}$ (red) for HD 153426 in the optical. The FEROS spectrum is shown in black. Model parameters are listed in Table \ref{resumo_resultados}.}
\label{final_model_optical_hd153426}
\end{figure}

\begin{figure}[!h]
\centerline{\resizebox{1.0\textwidth}{!}{\input{./fig_hd218195_optical.tex}}}
\caption{Final CMFGEN model with $T_{\mathrm{eff}}^{\mathrm{UV}}$ (red) for HD 218195 in the optical. The ESPADONS spectrum is shown in black. Model parameters are listed in Table \ref{resumo_resultados}.}
\label{final_model_optical_hd218195}
\end{figure}

\begin{figure}[!h]
\centerline{\resizebox{1.0\textwidth}{!}{\input{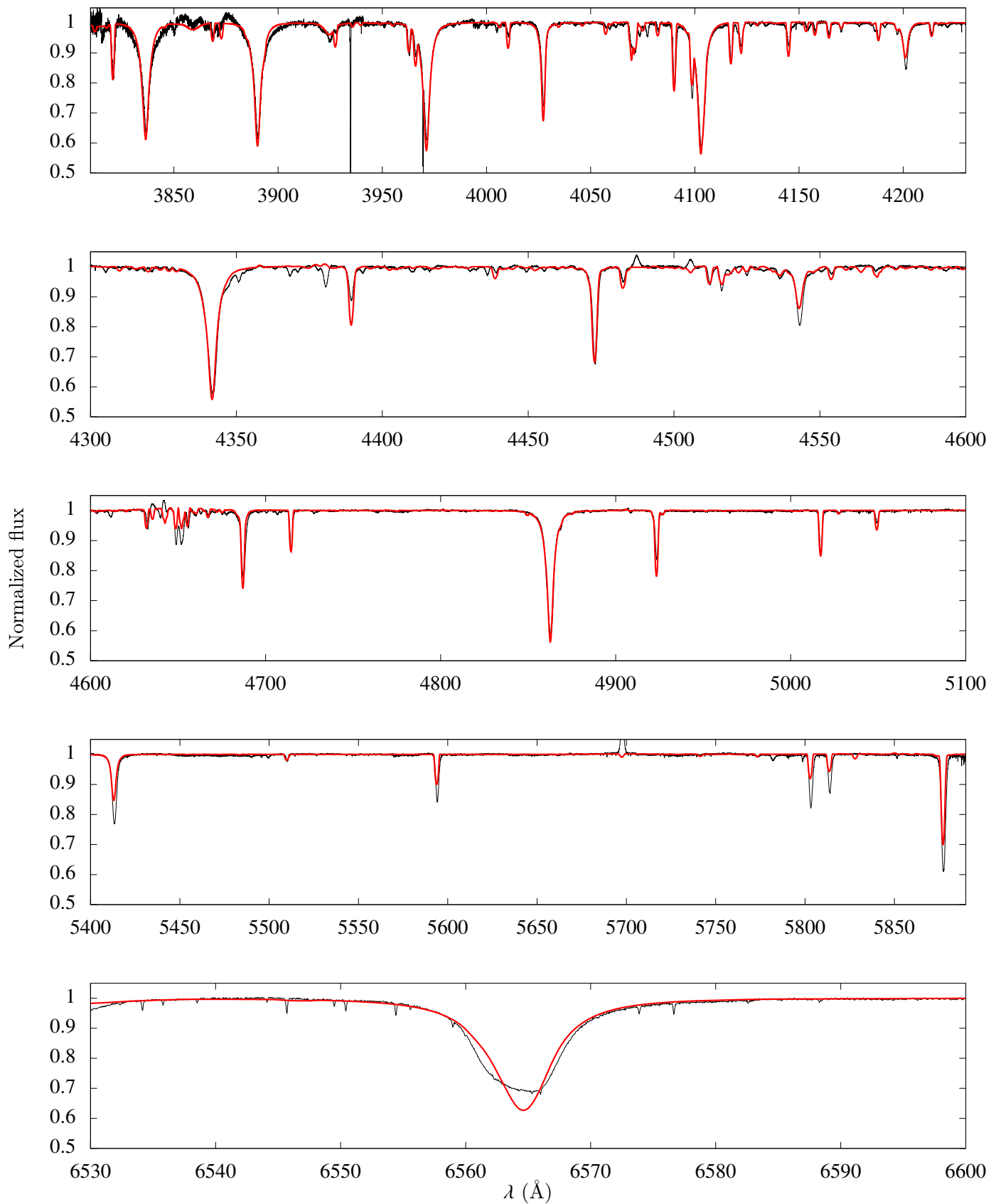}}}
\caption{Final CMFGEN model with $T_{\mathrm{eff}}^{\mathrm{UV}}$ (red) for HD 36861 in the optical. The NARVAL spectrum is shown in black. Model parameters are listed in Table \ref{resumo_resultados}.}
\label{final_model_optical_hd36861}
\end{figure}

\begin{figure}[!h]
\centerline{\resizebox{1.0\textwidth}{!}{\input{./fig_hd115455_optical.tex}}}
\caption{Final CMFGEN model with $T_{\mathrm{eff}}^{\mathrm{UV}}$ (red) for HD 115455 in the optical. The FEROS spectrum is shown in black. Model parameters are listed in Table \ref{resumo_resultados}.}
\label{final_model_optical_hd115455}
\end{figure}

\begin{figure}[!h]
\centerline{\resizebox{1.0\textwidth}{!}{\input{./fig_hd135591_optical.tex}}}
\caption{Final CMFGEN model with $T_{\mathrm{eff}}^{\mathrm{UV}}$ (red) for HD 135591 in the optical. The FEROS spectrum is shown in black. Model parameters are listed in Table \ref{resumo_resultados}.}
\label{final_model_optical_hd135591}
\end{figure}

\end{appendix}


\end{document}